\documentclass{aa}
\usepackage{natbib,epsfig,txfonts,graphicx}

\def\HeII{He{\sc ii}}
\def\HeIII{He{\sc iii}}

\def\Pv{P{\sc v}}
\def\Teff{$T_{\rm eff}$}
\def\logg{$\log g$}
\def\YHe   {$Y_{\rm He}$}
\def\vmic{$v_{\rm t}$}

\def\Rsun {$R_{\odot}$}
\def\Rsune {R_{\odot}}
\def\Msun {${\rm M}_{\odot}$}

\def\Rstar{$R_{\ast}$}
\def\Rmax{$R_{\rm max}$}

\def\Mdot{${\dot M}$}

\def\MV {M$_{\rm v}$}

\def\vinf {$v_{\rm \infty}$}

\def \Trad {$T_{\rm rad}$}

\def\Hg {H$_{\rm \gamma}$}

\def\Ha {H$_{\rm \alpha}$}

\def\kms {km~s$^{-1}$}
\def\Mdu{$\cdot 10^{-6}\, {\rm M_{\odot}/yr}$}

\def\rarrow{$\rightarrow$}
\def\dd{{\rm d}}

\def\fcl{f_{\rm cl}}

\def\beq{\begin{equation}}
\def\eeq{\end{equation}}
\def\beqa{\begin{eqnarray}}
\def\eeqa{\end{eqnarray}}

\def\Rv{R$_V$}
\def \Lsune {L_{\odot}}

\def \Rstare {R_\star}

\def \Teffe {T_{\rm eff}}

\def \vinfe {v_\infty}
\def \Mdote {\dot M}

\def\rin{r_{\rm in}}
\def\rmid{r_{\rm mid}}
\def\rout{r_{\rm out}}
\def\rfar{r_{\rm far}}
\def\fin{f_{\rm cl}^{\rm in}}
\def\fmid{f_{\rm cl}^{\rm mid}}
\def\fmidm{f_{\rm max}^{\rm mid}}
\def\fout{f_{\rm cl}^{\rm out}}
\def\foutm{f_{\rm max}^{\rm out}}
\def\ffar{f_{\rm cl}^{\rm far}}

\def\kmax{$\kappa_{\rm max}$}

\newcommand{\rb}[1]{\raisebox{1.5ex}[-1.5ex]{#1}}

\begin{document}

\title{Bright OB  stars in the Galaxy}
\subtitle{III. Constraints on the radial stratification of the clumping
factor in hot star winds from a combined \Ha, IR and radio
analysis\thanks{based in part on observations obtained with the VLA operated
by the National Radio Astronomy Observatory (NRAO)}}

\author{J. Puls\inst{1}, N. Markova\inst{2}, S. Scuderi\inst{3}, 
C. Stanghellini\inst{4}, O.~G.~Taranova\inst{5}, 
A.~W.~Burnley\inst{6} and I.~D.~Howarth\inst{6}}

\offprints{J. Puls}

\institute{Universit\"ats-Sternwarte M\"unchen, Scheinerstr. 1, 
  D-81679 M\"unchen, Germany, \email{uh101aw@usm.uni-muenchen.de}
\and
Institute of Astronomy, Bulgarian National Astronomical Observatory,
  P.O. Box 136, 4700 Smoljan, Bulgaria,\\
  \email{nmarkova@astro.bas.bg}
\and
INAF - Osservatorio Astrofisico di Catania, Via S. Sofia 78, I-95123 Catania, 
Italy, \email{scuderi@oact.inaf.it}
\and
INAF - Istituto di Radioastronomia, Via P. Gobetti 101, I-40129 Bologna, 
Italy, \email{c.stanghellini@ira.inaf.it}
\and
Sternberg Astronomical Institute, Universitetski  pr. 13, Moscow, 119992,
Russia, \email{taranova@sai.msu.ru}
\and
Department of Physics and Astronomy, University College London, Gower
Street, London WC1E 6BT, UK,\\
\email{awxb@star.ucl.ac.uk, idh@star.ucl.ac.uk}
}

\date{Received; accepted }

\abstract{Recent results strongly challenge the canonical picture of
massive star winds: various evidence indicates that currently accepted
mass-loss rates, \Mdot, may need to be revised downwards, by factors
extending to one magnitude or even more. This is because the most commonly
used mass-loss diagnostics are affected by ``clumping'' (small-scale density
inhomogeneities), influencing our interpretation of observed spectra and
fluxes.\\ 
Such downward revisions would have dramatic consequences for the
evolution of, and feedback from, massive stars, and thus robust
determinations of the clumping properties and mass-loss rates are urgently
needed. We present a first attempt concerning this
objective, by means of constraining the radial stratification of the
so-called clumping factor.\\
To this end, we have analyzed a sample of 19 Galactic O-type
supergiants/giants, by combining our own and archival data for \Ha, IR, mm and
radio fluxes, and using approximate methods, calibrated to more
sophisticated models. Clumping has been included into our analysis in the
``conventional'' way, by assuming the inter-clump matter to be void. Because
(almost) all our diagnostics depends on the square of density, we cannot
derive absolute clumping factors, but only factors normalized to a certain
minimum.\\
This minimum was usually found to be located in the outermost, radio-emitting
region, i.e., the radio mass-loss rates are the lowest ones,
compared to \Mdot\ derived from \Ha\ and the IR. The radio rates agree well
with those predicted by theory, but are only upper limits, due to unknown
clumping in the outer wind. \Ha\ turned out to be a useful tool to derive
the clumping properties inside $r < 3{\ldots}5\, \Rstare$. Our most
important result concerns a (physical) difference between denser and thinner
winds: for denser winds, the innermost region is more strongly clumped than
the outermost one (with a normalized clumping factor of $4.1 \pm 1.4$),
whereas thinner winds have similar clumping properties in the inner and
outer regions.\\
Our findings are compared with theoretical
predictions, and the implications are discussed in detail, by assuming
different scenarios regarding the still unknown clumping properties of the
outer wind.

}

\titlerunning{Constraints on the clumping factors in hot star winds}
\authorrunning{J. Puls et al.}

\maketitle

\section{Introduction}
\label{intro}
In the last few years, massive stars ($M_{\rm ZAMS} \ga$ 10 \Msun) have
(re-)gained considerable interest among the astrophysical community, in
particular because of their role in the development of the early Universe
(e.g., its chemical evolution and re-ionization; \citealt{bromm01}, but also
\citealt{mc05}). Unfortunately, however, our knowledge of these objects is
not as complete as we would like it to be, and present efforts concentrate
on modeling various {\it dynamical} processes in the stellar interior, as
well as in the stellar atmosphere (mass loss, rotation, magnetic fields,
convection, and pulsation).

Most important in this regard is the {\it mass loss} that occurs through
supersonic winds, which modifies evolutionary time-scales, chemical
profiles, surface abundances and luminosities. As shown by numerous
stellar-evolution calculations, changing the mass-loss rates of massive
stars by even a factor of two has a dramatic effect on their evolution
(\citealt{meynet94}).

The winds from massive stars in their O-, B- and A-supergiant phase are well
described by radiation-driven wind theory (\citealt{cak, ppk}); the even
stronger mass outflows observed during their Wolf-Rayet (WR) and Luminous
Blue Variable (LBV) phases are also thought to be driven by radiation
pressure (for recent progress, see \citealt{graefener05} for the case of WRs
and \citealt{owo05} for LBVs).

Notwithstanding its considerable successes (e.g., \citealt{vink00, kud02, 
puls03}),
the theory is certainly over-simplified. Stellar rotation (e.g.,
\citealt{ocg96, puls99} and references therein), and the intrinsic
instability of the line-driving mechanism (see below), produce non-spherical
and inhomogeneous structure, observationally evident from, e.g., X-ray
emission and line-profile variability (for summaries, see \citealt{kp00} and
\citealt{oskinova04} regarding the present status of X-ray line emission).  
As long as the time-dependent structuring of stellar winds is not well
understood, we cannot be sure about even their ``average'' properties, such
as mass-loss rates and emergent ionizing fluxes. Even worse, most
spectroscopic analyses of hot stars aiming at {\it deriving} stellar and
wind parameters have been performed by relying on the assumption of a
globally stationary wind with a {\it smooth} density/velocity
stratification. Consequently, the underlying models are incapable in
principle of describing the aforementioned features, and the derived results
(including the verification of the theory) may depend strongly on this
assumption.

Theoretical effort to understand the nature and origin of these
observational findings have generally focused on the line-driving mechanism 
itself; the first linear-stability analyses showed the line force to be
inherently unstable (\citealt{or84} and references therein).  Subsequent
numerical simulations of the non-linear evolution of the line-driven flow
instability (for a review, see \citealt{owo94}), with various degrees of
approximation concerning the stabilizing diffuse, scattered radiation field
(\citealt{op96, op99}), have shown that the outer wind (typically, from 1.3
\Rstar\ on) develops extensive structure, consisting of strong reverse
shocks separating slower, dense shells from high-speed rarefied regions.
Only a very small fraction of material is accelerated to high speed and then
shocked; for most of the flow the major effect is a compression into narrow,
dense ``clumps'', separated by large regions of much lower density.

At first glance, these models appear to be in strong contrast with our
assumptions for the ``standard model'' for wind diagnostics based on
stationarity and homogeneity, especially when viewed with respect to the
spatial variation of velocity and density. However, when viewed with respect
to the {\em mass} distribution of these quantities, the models are not so
very different (e.g., \citealt{ocr, pof}). Given the intrinsic
mass-weighting of spectral formation, and the extensive temporal and spatial
averaging involved, the observational properties of such structured models
are quite similar to what is derived from the ``conventional'' diagnostics,
in an average sense.  Structured winds also explain `steady-state'
characteristics like X-rays, and the black absorption troughs observed in
saturated UV resonance lines (\citealt{lucy82, lucy83}, later confirmed by
\citealt{pof} on the basis of hydrodynamical simulations).

Recent time-dependent simulations have aimed at investigating two specific
problems. First, \citet{ro02, ro05} have introduced new methods to
numerically resolve even the outermost wind. In particular, they provide
theoretical predictions for the radial stratification of the so-called
clumping factor,
\beq
\label{defcl}
\fcl = \frac{<\rho^2>}{<\rho>^2} \quad \ge 1,
\eeq
where angle brackets denote (temporal) average quantities. For self-excited
instabilities (e.g., without any photospheric disturbances such as
pulsations or sound-waves), they find that, beginning with an unclumped
wind in the lowermost part ($f_{\rm cl} = 1$), the clumping becomes
significant ($f_{\rm cl} \simeq 4$) at wind speeds of a few hundreds of
\kms, reaches a maximum ($f_{\rm cl} \simeq 15{\ldots}20$), and thereafter
decays, settling at a factor of roughly four again.

On the other hand, \citet{desowo03, desowo05}, building on a pilot
investigation by \citet{owo99}, have taken the first steps towards including
2-D effects of the radiation field into a higher-dimensional hydrodynamical
description, to obtain constraints on the {\it lateral} extent of clumping.

Taken together, and with respect to NLTE modeling and spectral analysis,
the above scenario has the following major implications, related to radiation field
and density/velocity structure:-
\begin{enumerate}
\item X-ray emission arising from the formation and interaction of
clumps and shocks, in concert with an enhanced EUV flux (e.g.,
\citealt{feld97}), can have a strong influence on the
ionization/excitation balance in the wind (e.g., \citealt{paul01}).
\item Clumping introduces depth-dependent deviations from a smooth
density structure, which particularly affects common observational
mass-loss indicators, such as \Ha\ emission and the IR/radio excess,
since these diagnostics directly depend on $<\rho^2>$ (being larger
than $<\rho>^2$). Furthermore, the ionization balance becomes
modified, primarily because of the additional $<\rho^2>$-dependence of
radiative recombination rates (see also \citealt{bouret05}).
\item Not only the modified density stratification, but also the
highly perturbed velocity field can affect the spectral line formation,
because of its multiple non-monotonic nature. This gives rise to 
modified escape probabilities and multiple resonance zones for certain
frequencies. A major example of such an influence is the formation of
black absorption troughs in saturated UV-resonance lines (see above).
Optical lines (e.g., \Ha) can also be affected, though to a lesser extent
(e.g., \citealt{pulsetal93}).

\end{enumerate}

Although the potential effects of clumping were first discussed some time
ago (e.g., \citealt{abbott81, lamerswaters84b, pulsetal93}), and have been
accounted for in the diagnostics of Wolf-Rayet stars since pioneering work
by \citet{hil91} and \citet{schmutz95}, this problem has been reconsidered
by the ``OB-star community'' only recently, mostly because of improvements
in the diagnostic tools, and particularly the inclusion of
line-blocking/blanketing in NLTE atmospheric models.

\citet{repo04} presented results of a re-analysis of Galactic O-stars,
previously modeled using unblanketed model atmospheres \citep{puls96}, with
new line-blanketed calculations. As a result of the line blanketing, the
derived effective temperatures were significantly lower than previously
found, whereas the modified wind-momentum rates remained roughly at their
former values. Based on this investigation, and deriving new
spectral-type--\Teff\ and spectral-type--\logg\ calibrations (see also
\citealt{martins05}), \citet[``Paper I'']{markova04} extended this sample
considerably and obtained a ``new'' empirical wind-momentum--luminosity
relationship\footnote{The presence of such a relationship is explained by
the radiation-driven wind theory, namely that the modified wind-momentum
rate, \Mdot \vinf (\Rstar/\Rsun)$^{0.5}$, should depend almost exclusively
on the stellar luminosity, $L/\Lsune$, to some power.}(WLR; \citealt{kud95,
kp00}) for Galactic O~stars, based on \Ha\ mass-loss rates.

\begin{table*}
\caption{Sample stars and stellar/reddening parameters as used in this
study. Note that radii, mass-loss rates (assuming an unclumped medium) 
absolute visual magnitudes, \MV, and reddening parameters have been modified
with respect to the original values (from ``ref1'' and ``ref2'') by a combined
$V/J/H/K$-band de-reddening procedure (see Sect.~\ref{dered}), using the
indicated distances. Gravitational accelerations, \logg, are ``effective''
values, i.e., without centrifugal correction, derived from \Hg\ or
calibrations; \vinf\ is in \kms; \Mdot\ is in $10^{-6}~{\rm M_{\odot}/yr}$;
and distances are in kpc. ``pt'' denotes the \Ha\ profile type
(emission/absorption/intermediate). ``ref1'' and ``ref2'' refer to the
sources of the original stellar and magnitude/color/reddening parameters,
respectively, where the extensions given for reference ``1'' denote the
``preferred'' model chosen in Paper~I (see text). For $\zeta$~Pup (HD\,66811),
we provide two entries, based on different distances (see Paper~I).}
\tabcolsep1.65mm
\begin{tabular}{l l c c r c c c r c c c c c c l}
\hline
Star & Sp.Type & \Teff & \logg & \Rstar & \YHe & \vinf & pt & \Mdot(opt) &
$\beta$(opt) & \MV & E(B-V) & \Rv   & dist & ref1 & ref2 \\
\hline
  Cyg\,OB2\#7 &           O3If* & 45800 &  3.93 & 15.0 & 0.21 & 3080 & e & 10.61 & 0.77 & -5.98 &  1.77 &  3.00 & 1.71 & 2 &   5\\
 HD190429A &           O4If+ & 39200 &  3.65 & 22.7 & 0.14 & 2400 & e & 16.19 & 0.95 & -6.63 &  0.47 &  3.10 & 2.29 & 1 & 1-0\\
   HD15570 &           O4If+ & 38000 &  3.50 & 24.0 & 0.18 & 2600 & e & 17.32 & 1.05 & -6.69 &  1.00 &  3.10 & 2.19 & 4 &   6\\
   HD66811 &         O4I(n)f & 39000 &  3.60 & 29.7 & 0.20 & 2250 & e & 16.67 & 0.90 & -7.23 &  0.04 &  3.10 & 0.73 & 3 & 1-4\\
           &                 &       &       & 18.6 &      &      &   &  8.26 &      & -6.23 &  0.04 &  3.10 & 0.46 & 3 & 1-0\\
   HD14947 &           O5If+ & 37500 &  3.45 & 26.6 & 0.20 & 2350 & e & 16.97 & 0.95 & -6.90 &  0.71 &  3.10 & 3.52 & 3 & 1-2\\
 Cyg\,OB2\#11 &           O5If+ & 36500 &  3.62 & 23.6 & 0.10 & 2300 & e &  8.12 & 1.03 & -6.67 &  1.76 &  3.15 & 1.71 & 2 &   5\\
 Cyg\,OB2\#8C &            O5If & 41800 &  3.73 & 15.6 & 0.13 & 2650 & a &  4.28 & 0.85 & -5.94 &  1.62 &  3.00 & 1.71 & 2 &   5\\
 Cyg\,OB2\#8A &        O5.5I(f) & 38200 &  3.56 & 27.0 & 0.14 & 2650 & i & 11.26 & 0.74 & -6.99 &  1.63 &  3.00 & 1.71 & 2 &   5\\
  HD210839 &         O6I(n)f & 36000 &  3.55 & 23.3 & 0.10 & 2250 & e &  7.95 & 1.00 & -6.61 &  0.49 &  3.10 & 1.08 & 3 & 1-2\\
  HD192639 &         O7Ib(f) & 35000 &  3.45 & 18.5 & 0.20 & 2150 & e &  6.22 & 0.90 & -6.07 &  0.61 &  3.10 & 1.82 & 3 & 1-0\\
   HD34656 &         O7II(f) & 34700 &  3.50 & 25.5 & 0.12 & 2150 & a &  2.61 & 1.09 & -6.79 &  0.31 &  3.40 & 3.20 & 1 & 1-6\\
   HD24912 & O7.5III(n)((f)) & 35000 &  3.50 & 24.2 & 0.15 & 2450 & a &  2.45 & 0.80 & -6.70 &  0.33 &  3.10 & 0.85 & 3 & 1-2\\
  HD203064 &         O7.5III & 34500 &  3.50 & 12.4 & 0.10 & 2550 & a &  0.98 & 0.80 & -5.23 &  0.23 &  3.10 & 0.79 & 3 &   6\\
   HD36861 &      O8III((f)) & 33600 &  3.56 & 14.4 & 0.10 & 2400 & a &  0.74 & 0.80 & -5.52 &  0.08 &  5.00 & 0.50 & 1 & 1-1\\
  HD207198 &         O9Ib/II & 36000 &  3.50 & 11.6 & 0.15 & 2150 & a &  1.05 & 0.80 & -5.15 &  0.58 &  2.56 & 0.83 & 3 & 1-1\\
   HD37043 &           O9III & 31400 &  3.50 & 17.9 & 0.12 & 2300 & a &  1.03 & 0.85 & -5.92 &  0.04 &  5.00 & 0.50 & 1 & 1-1\\
   HD30614 &          O9.5Ia & 29000 &  3.00 & 20.7 & 0.10 & 1550 & e &  3.07 & 1.15 & -6.00 &  0.25 &  3.10 & 0.79 & 3 & 1-2\\
 Cyg\,OB2\#10 &           O9.5I & 29700 &  3.23 & 30.7 & 0.08 & 1650 & i &  2.74 & 1.05 & -6.95 &  1.80 &  3.15 & 1.71 & 2 &   5\\
  HD209975 &          O9.5Ib & 32000 &  3.20 & 14.7 & 0.10 & 2050 & a &  1.11 & 0.80 & -5.45 &  0.35 &  2.76 & 0.83 & 3 & 1-1\\
\hline
\end{tabular}
\smallskip
\newline
References: 1. \citet{markova04}, 2. \citet{mokiem05}, 3. \citet{repo04},
4. \citet{repo05}, 5. \citet{hanson03} (distance from \citet{massey91}), 
6. \citet{maiz04}.\\
\label{sample}
\end{table*}

A comparison of the ``observed'' wind-momentum rates with theoretical
predictions from \citet{vink00} and independent calculations performed by 
our group (\citealt{puls03})\footnote{which proved to be almost identical,
though the two approaches are rather different; see also \citet{kud02}.},
revealed that objects with \Ha\ in emission and those with \Ha\ in
absorption form two distinct WLRs. The latter is in agreement with theory,
whilst the former appears to be located in parallel, but above the
theoretical relation. This difference was interpreted as being a consequence
of wind clumping, with the contribution of wind emission to the total
profile being significantly different for objects with \Ha\ in absorption
compared to those with \Ha\ in emission (since for the former group only
contributions from the lowermost wind can be seen, whereas for the latter
the emission is due to a significantly more extended volume). Thus, there is
the possibility that for these objects one {\it sees directly} the effects
of a clumped wind, which would mimic a higher mass-loss rate (as is most
probably the case for Wolf-Rayet winds). With this interpretation, the
presence of clumping in the winds of objects with \Ha\ in absorption is not
excluded; owing to the low optical depth, however, one simply cannot see it,
and corresponding mass-loss rates would remain unaffected.

The ``actual'' mass-loss rates for objects with \Ha\ in emission can then be
estimated by shifting the observed wind-momentum rates onto the theoretical
predictions, with a typical reduction in \Mdot\ by factors of 2--2.5,
corresponding to clumping factors of the order of 4--6. 

Though factors between two and three seem reasonable when compared to
results from Wolf-Rayet stars (also factors of $\sim$3; e.g.,
\citealt{mr94}), there is increasing evidence that the situation might be
even more extreme. From an analysis of the ultraviolet \Pv\ resonance doublet 
(which is unsaturated, and can therefore be used as a mass-loss indicator),
\citet{massa03} and \citet{fulli04, fulli06} conclude that typical O-star
mass-loss rates derived from \Ha\ or radio emission might overestimate the
actual values by factors of up to 100 (with a median of 20, if \Pv\ were the
dominant ion for spectral types between O4 to O7; see \citealt{fulli06}).
\citet{bouret05}, from a combined UV and optical analysis, obtained factors
between 3 and 7, though from only two stars. In addition, the latter work
suggests that the medium is clumped from the wind base on, in strong
contrast with typical hydrodynamical simulations (see above). If this were
true, presently accepted mass-loss rates for non-supergiant stars also need
to be revised, and even the analysis of quasi-photospheric lines (i.e.,
stellar parameters and abundances) might be affected, since the cores of
important lines are formed in the transonic region.

In this paper, we attempt to undertake a first step towards a clarification
of the present puzzling situation. From a simultaneous analysis of \Ha, IR
and radio observations, we obtain constraints on the radial stratification
of the clumping factor, and test how far the results meet the predictions
given by \citet{ro02, ro05}. Since all these diagnostics depend on
$<\rho^2>$, however, we are able to derive only relative, not absolute,
values, as detailed in Sect.~\ref{analysis}. Let us point out here that our
analysis is based upon the assumption of {\it small-scale} inhomogeneities
redistributing the matter into overdense clumps and an (almost) void
inter-clump medium, in accordance with (but not necessarily related to) the
basic effects of the line-driven instability. Indeed, the question of
whether the wind material is predominantly redistributed on such small
scales and not on larger spatial scales (e.g., in the form of co-rotating
interaction regions; \citealt{mullan84, mullan86, co96}) has not yet been
resolved, but unexplained residuals from the results of our analysis might
help to clarify this issue.

Investigations such as we perform here are not new. Indeed, a number of
similar studies have been presented during recent years, e.g.,
\citet{leitherer82,abbott84,ll93,RB96, BR97,scu98,blomme02,blomme03}. 
The improvements
underpinning our study, which hopefully will allow us to obtain more
conclusive results, are related to the following facts. First, and in
contrast to earlier work, the uncertainties concerning the adopted stellar
parameters have been greatly reduced, since they have been derived by means
of state-of-the-art, line-blanketed models. Secondly, we do not derive
(different) mass-loss rates from the different wavelength domains based on a
homogeneous wind model, but aim at a unique solution by explicitly allowing
for clumping as a function of radius, at least in a simplified way. Thirdly,
we use recent radio observations obtained with the Very Large Array (VLA),
which, because of its gain in sensitivity due to improved performance
(mostly at 6cm, where the system temperature improved from 60 to 45~K)
allows us to measure the radio fluxes for stars with only moderate wind
densities, which produce \Ha\ in absorption. In this way we are able to test
the above hypothesis concerning the differences of \Ha\ mass-loss rates from
stars with \Ha\ emission and absorption. Lastly, our IR analysis does not
depend on assumptions used in previous standard methods exploiting the IR
excess (e.g., \citealt{lamerswaters84a}), since we calibrate against results
from line-blanketed NLTE models.  (Note that uncertainties in the stellar
radii due to distance errors cancel out as far as the derived run of the
clumping factors is concerned, and affects ``only'' the absolute mass-loss
rates.)

The plan of this paper is as follows. In Sect.~\ref{obs}, we describe our
stellar sample and the observational material used in this study. We also
comment on some problems related to reddening. In Sect.~\ref{simul}, we
present the methods used to analyze the different wavelength regimes, and
discuss how we deal with a clumped wind medium. Applying these methods, we
derive constraints for the radial stratification of the clumping factor in
Sect.~\ref{analysis}, and give a discussion and summary of our findings in
Sects.~\ref{discussion} and \ref{summary}.

\section{Stellar sample and observational material}
\label{obs}

The stellar sample consists of 19 Galactic supergiants/giants, covering
spectral types O3 to O9.5. These stars have been analyzed in the optical
and, to a large part, (re-)observed by us with the VLA.  To our knowledge,
the only {\it confirmed} non-thermal radio emitter included in our sample is
Cyg\,OB2\#8A \citep{bieging89}, which was recently detected as an
O6I/O5.5III, colliding-wind binary system by \citet{deBecker04}. Somewhat
inconsistently, we will use corresponding stellar parameters resulting from
an analysis assuming a single star. Note also that HD\,37043 is listed
as an SB2 binary in the recent Galactic O-star catalogue of \citet{maiz04}.

Most of the optical analyses were performed by either \citet{repo04} or
\citet{mokiem05} (Cyg OB2 objects), using the NLTE line-blanketed
model-atmosphere code {\sc fastwind} \citep{puls05}. For a few stars (those
denoted by ``1'' in Table~\ref{sample}, column ``ref1''), stellar parameters
have been derived from calibrations only, as outlined in Paper~I. At least
for HD\,190429A, an independent re-analysis by \citet{bouret05}, by means of
the alternative model-atmosphere code {\sc cmfgen} \citep{hil98}, confirms
the corresponding calibration.  Finally, for HD\,15570, we use parameters
derived from $H$- \& $K$-band spectroscopy by \citet{repo05}.

The parameters adopted in this study are presented in Table~\ref{sample}.
For those objects which have been analyzed exclusively in Paper~I, and for
which more than one choice concerning distance, reddening or luminosity has
been discussed, we have used the ``preferred'' parameter set (Paper~I,
Table~2), denoted by the corresponding extension in entry ``ref 2''. Only
for HD\,66811 ($\zeta$~Pup) do we provide two entries, referring to its 
``conventional'' distance, $d=460$~pc (2nd entry), and the assumption that
this star is a runaway star, located at $d=730$~pc (see \citealt{Sahu93} and
Paper~I, Sect.~5). Unless stated explicitly, we will use the latter
parameter set in our further discussion. 

Note that due to minor revisions with respect to reddening, the stellar
radii and \Ha\ mass-loss rates (rescaled by assuming $\Mdote/\Rstare^{1.5}$
= const, e.g., \citealt{puls96}) for most objects are (slightly) different
from the original sources. In Sect~\ref{dered}, we will discuss why
these revisions were necessary, and how they have been obtained.

\subsection{Variability of the diagnostics used}
\label{variab}

Before we discuss the observations obtained in the individual bands (\Ha, IR
and radio), let us first give some important comments on the variability
of the different diagnostics. Stellar winds are known to be variable on
different timescales and in all wavelength ranges in which they are
observed. Thus, the use of non-simultaneous measurements, as in our
analysis, can be an issue. 

Regarding \Ha, line profile variations in early-type stars have been
observed for years. Since the first extensive surveys by \citet{rosen73a,
rosen73b}, a large number of investigations have been conducted to establish
the properties of the \Ha\ variability and also its origin (e.g.,
\citealt{ebbets82, scuderi92, kaufer96, kaper97, morel04}). Although the
variations in the \Ha\ profile in some cases look very dramatic, they
indicate, when interpreted in terms of a variable mass-loss rate, only
moderate changes in \Mdot, usually not exceeding the uncertainties on the
corresponding estimates. Recently, for a sample of 15 O-type supergiants,
\citet{markova05} constrained the \Mdot\ variability to about $\pm 4\%$ of
the corresponding mean value for stars with stronger winds, and to about
$\pm 16\%$ for stars with weaker winds. These estimates are in remarkably
good agreement with those from previous studies \citep{ebbets82, scuderi92}
who report variations in \Mdot\ of about 10 to 30\%. 

In the case of IR and radio continua, and assuming the emission to be
thermal, the timescales of variability (due to variations of micro- or
macro-structure, i.e., of the local density or mass-loss rate\footnote{Note
that variations in the ionization can also induce temporal variability,
e.g., \citet{panagia91}.}) can differ by orders of magnitude in the two
wavelength regimes. Considering variations in \Mdot, the transit time of a
front would be of the order of hours in the near-IR forming region (given
typical sizes of the emitting region and velocity of the expanding
material), and as much as months, or even years, in the radio domain. This
implies that whilst the IR emission would display short-term variability,
following the mass-loss rate variations very closely, variations in the
radio would be averaged out if they occurred on timescales much shorter than
the transit time. 

Different considerations apply when the variability is of non-thermal
origin. In this case, only the radio emission is affected. The process
responsible is usually cited as being synchrotron emission \citep{white85},
most probably produced in colliding-wind binaries \citep{vanLoo06}. The main
observed characteristics are variability over timescales of up to months,
and a power-law spectrum increasing with wavelength and with a variable
spectral index \citep{bieging89}. In such a case, which is met at least by
one of our objects, Cyg\,OB2\#8A, the measured radio-flux(es) can still be
used as an upper limit of the thermal free-free emission, by analyzing the
lowermost flux measured at the shortest radio wavelength.

Regarding the {\it amplitude} of variability, no clear evidence of IR
continuum variability has been reported up to now. Amongst the IR
observations we have obtained from the literature, there are some studies
(e.g., \citealt{castor83} or \citealt{abbott84}) with data sampled on
timescales ranging from a few hours up to a few months, but no variation of
the observed IR fluxes above the errors was reported. If, on the other hand,
we compare sets of measurements of the same object, obtained by different
authors with different instruments, we do observe differences in the
measured fluxes, more likely related to calibration problems than to genuine
IR variability (see also Sect.~\ref{absolute_fluxes}). 

With respect to radio emission, there are several pieces of evidence for
variability, both in the observed fluxes and in the spectral index. Again,
we have to distinguish between thermal and non-thermal emission. In the case
of non-thermal origin, variability is {\it always} present (e.g.,
\citealt{bieging89}). This has to be accounted for whenever we have no clear
indication about the thermal origin of the observed emission, but where we
do see variations. Of our targets, in addition to \#8A, this might be a
problem only for HD\,190429A (and for HD\,34656 and HD\,37043 for other
reasons).

\begin{table*}
\caption{VLA radio fluxes (in $\mu$Jy), with 1-$\sigma$ errors in
brackets. Data without superscripts are new observations (see
Sect.~\ref{radio}), whilst data with superscripts correspond to either (a)
unpublished measurements by Scuderi et al.\ or literature values used to
complement our database: (b) \citet{scu98}; (c) \citet{bieging89}; (d)
3.6~cm observations from \citet{ll93} (concerning HD\,15570, see text); (e)
\citet[including 20~cm data for $\zeta$~Pup, at
760$\pm$90~$\mu$Jy]{blomme03}.
\newline
Also indicated are the adopted IR to mm fluxes and the sources from which
they have been drawn (see foot of table). Data denoted by ``own'' refer to
$JHKLM$-band observations performed by OGT at the Crimean 1.25~m telescope
(see Sect.~\ref{irobs}). {\sc scuba} data (at 1.35\,mm) were obtained by AWB
and IDH (see Sect.~\ref{mmobs}), and {\sc scuba} 0.85\,mm data are from \citet{blomme03}.}
\begin{tabular}{lllllll}
\hline
\multicolumn{1}{c}{Star}
&\multicolumn{1}{c}{4.86\,GHz}
&\multicolumn{1}{c}{8.46\,GHz}
&\multicolumn{1}{c}{14.94\,GHz}
&\multicolumn{1}{c}{43.34\,GHz}
&\multicolumn{1}{l}{IR- and mm-}
&\multicolumn{1}{l}{references}\\
\multicolumn{1}{c}{}
&\multicolumn{1}{c}{(6\,cm)}
&\multicolumn{1}{c}{(3.5\,cm)}
&\multicolumn{1}{c}{(2\,cm)}
&\multicolumn{1}{c}{(0.7\,cm)}
&\multicolumn{1}{l}{bands used}
&\multicolumn{1}{l}{(IR and mm)}\\
\hline                
Cyg\,OB2\#7      & $<$112          & $<$100       &              &      &$HKLMN$    & 1,14   \\
Cyg\,OB2\#8A     & $<$540$^{a}$    & 920(70)$^{a}$&              &      &$JHKLMNQ$  & 1,5,14,19,20\\
                 & 1000(200)$^{c}$ &              & 500(200)$^{c}$ &      \\
                 & 800(100)$^{c}$  &              &              &        \\
                 & 700(100)$^{c}$  &              &              &        \\
                 & 400(100)$^{c}$  &              &              &        \\
Cyg\,OB2\#8C     & $<$200$^{c}$    &              &              &      &$HKLMN$    & 14      \\
Cyg\,OB2\#10     & 134(29)         & 155(26)      & 300(100)     &      &$JHKLMN$   & 5,14,19 \\
Cyg\,OB2\#11     & 182(33)         & 228(28)      & $<$400       &      &$JHKLMN$   & 5,14   \\
HD\,14947        & $<$110          & $<$135       & $<$700       &      &$JHKLMN$   & 2,5,15,own\\
                 &$<90^{a}$        &  90(30)$^{a}$&              &      & \\
                 &                 & 120(30)$^{a}$&              &      & \\
                 &                 & 110(30)$^{b}$&              &      & \\
HD\,15570        & 100(40)$^{a}$   & 220(40)$^{a}$&              &      &$JHKLMNQ$,1.35\,mm & 1,5,8,11,15,18,{\sc scuba}\\
                 &                 & 125(25)$^{d}$&              &        \\
HD\,24912        & $<$200          & $<$120       & $<$390       & $<$840 &$JHKLMN$,IRAS & 3,5,7,16   \\
HD\,30614        & 230(50)$^{b}$   & 440(40)$^{b}$& 650(100)$^{b}$ &      &$JHKLMN$ & 5,7,own        \\
HD\,34656        & $<$132          & 119(24)      & $<$510       &        &$JHKL$   & 17,own     \\
HD\,36861        & $<$112          & $<$90        & $<$1000      &        &$JHKLMN$ & 2,5,7    \\
HD\,37043        & 203(38)         & $<$90        & $<$330       &        &$JHKLMN$ & 4,5,16,21,22 \\
                 &                 & 46(15)$^{d}$ &              &        \\
HD\,66811        & 1640(70)$^{e}$  & 2380(90)$^{e}$& 2900(300)$^{c}$ &    &$JHKLM$,IRAS,0.85\,mm,1.3\,mm & 6,9,10,12,13,22,23,24        \\
                 & 1490(110)$^{c}$ &              &              &        &        \\
HD\,190429A      & 250(37)         & 199(36)      & $<$420       & $<$540 &$JHKLM$  & 5,20,own  \\
                 &                 & 280(30)$^{b}$ &             &        \\
HD\,192639       &                 & $<$90$^{a}$  &              &        &$JHKLM$  & 5,15,own  \\
HD\,203064       & 114(27)         & 126(20)      & $<$330       &        &$JHKLM$,IRAS & 3,5,own        \\ 
HD\,207198       & 105(25)         & 101(21)      & 249(82)      &        &$JHKLM$,IRAS & 3,own \\
HD\,209975       & 165(36)         & 184(28)      & 422(120)     &        &$JHKLM$,IRAS & 3,own \\
HD\,210839       & 238(34)         & 428(26)      & 465(120) & 790(190)   &$JHKLMNQ$,IRAS,1.35\,mm & 1,2,3,5,14,15,own,{\sc scuba}  \\
\hline
\end{tabular}
\smallskip
\newline
References for IR and mm data: 
1. \citet{abbott84}, 2. \citet{barlow77}, 3. \citet{beichman88},
4. \citet{breger81}, 5. \citet{castor83}, 6. \citet{dachs82}, 
7. \citet{gehrz74},  8. \citet{guetter89}, 9. \citet{johnson63}, 
10. \citet{johnson64}, 11. \citet{johnson66a}, 12. \citet{johnson66b}, 
13. \citet{lamers84}, 14. \citet{leitherer82}, 15. \citet{leitherer84},
16. \citet{ney73}, 17. \citet{polcaro90}, 18. \citet{sagar90},
19. \citet{sneden78}, 20. \citet{tapia81}, 21. \citet{the86},
22. \citet{whittet80}, 23. \citet{leitherer91}, 24. \citet{blomme03}.
\label{radiotable}
\end{table*}

Among thermal emitters, on the other hand, the situation is less clear.
There are very few studies which have observed one object several times and
at more than one frequency. In the sample studied by \cite{bieging89}, two
from six definite thermal emitters showed variability, both of which are B
supergiants (Cyg\,OB2\#12 and $\zeta$~Sco). In these cases, the flux
variation reached values of up to 70\%. Interpreted in terms of \Mdot, this
would mean a change of 50\% (see Eq.~\ref{radioflux_approx}). In
\cite{scu98}, one out of six objects (again Cyg\,OB2\#12) showed variability
whilst having a spectral index compatible with thermal emission. 
\citet{blomme02} studied the variability of $\epsilon$~Ori (B0Ia) and found
no evidence for variability, both on shorter and longer timescales. The
best-studied object with regard to thermal radio variability is $\zeta$~Pup
(O4If+), as a result of the work by \citet{blomme03}, who investigated both
new and various archival data. Again, short-term variability could be ruled
out, and long-term variability (with observations beginning in 1978)
appeared to be low or even negligible. 

\smallskip \noindent The major hypothesis underlying our present
investigation (being in agreement with most other investigations performed
thus far) is that the clumping properties of a specific wind are controlled by
small-scale structures. Further comments on this hypothesis, in connection
with the outcome of this analysis, will be given in Sect.~\ref{summary}. If
related to any intrinsic wind property (e.g., the instability of radiative
line-driving, even if externally triggered by short wavelength/short period
modulations), the derived clumping properties should be (almost) independent
of time, as long as the major wind characteristics remain largely constant.
Accounting for the observational facts above, this assumption
seems to be reasonable, and justifies our approach of using observational
diagnostics from different epochs. 
Note also
that the observed X-ray variability (where the X-rays are thought to arise
mostly from clump-clump collisions) is low as well \citep{bergetal96}, due
to the cancellation effects of the large number of participating
clumps being accelerated in laterally independent cones (of not too large
angular extent, \citealt{feld97}). 

If we had analyzed only one object, the derived results might be considered
as spurious, of course. However, due to the significant size of our sample,
any global property (if present) should become visible. Let us already
mention here that our findings, on average, indicate rather similar
behaviour for similar objects, and thus we are confident that these results
remain largely unaffected by issues related to {\it strong} temporal
variability.

\subsection{\Ha\ observations}
For our analysis of \Ha\ by means of clumped wind models, we have used the
same observational material as described in Paper~I, i.e., \Ha\ spectra
obtained at the Coud\'e spectrograph of the 2m RCC telescope at the National
Astronomical Observatory, Bulgaria, with a typical resolution of 15\,000.
For further information concerning technical details and reduction, see
Paper~I and references therein.

\subsection{Radio observations}
\label{radio}

New radio observations for 13 stars have been carried out at the VLA (in CnB
and C configuration), in several sessions between February and April 2004,
for a total of about 36 hours. Some of these stars were already known to be
radio emitters, but for many of them only upper limits for their radio
emission were available. Exploiting the gain in sensitivity of the VLA, and
guided by the requirement of using consistent data at {\it all} radio
frequencies for our analysis, we decided to (re-)observe them. Note
particularly that it was possible to observe not only stars with strong
winds, but also those with weaker winds (i.e., with \Ha\ in absorption).

The journal of observations is given in Table~\ref{radioobs}, with dates,
observing frequencies, time on targets, calibrators for flux-density
bootstrapping and VLA configuration. The observations were performed with a
total bandwidth of 100 MHz at all frequencies. The target stars were
observed for several scans of about 10 minutes, interleaved with a proper
phase calibrator at 4.86, 8.46, and 14.94 GHz. A faster switching between
the target star and the phase calibrator was used at 43 GHz, to remove the
rapid phase fluctuations introduced between the antenna elements by the
troposphere at this frequency. The data at 15 and 43 GHz have been
corrected for atmospheric opacity using a combination of a seasonal model
and the surface weather conditions during the experiment. The Astronomical
Image Processing System (AIPS) developed by the National Radio Astronomy
Observatory (NRAO) was used for editing, calibrating and imaging the data.

Table~\ref{radiotable} (left part) displays the corresponding fluxes, 
together with data from other sources (in most cases, at 2, 3.5 and 6 cm)
used to complement our sample. For four objects, partly overlapping with the
13 stars mentioned above, we relied on data (denoted by superscript ``a'' in
Table~\ref{radiotable}) derived by Scuderi et al. (in preparation), using
the VLA in B, BnC and C configuration, in different sessions between 1998
and 1999. The reduction and analysis of these data has been performed in a
similar way as outlined above for our new observations. The quoted flux 
limits for both datasets refer to 3 times the RMS noise on the images,
whereas the errors are 1-$\sigma$ errors, again measured on the images. Note
that at these low flux densities the contribution of errors introduced by
the calibration procedure is negligible.

For the remaining stars, literature values have been used, in particular
from \citet{scu98} and \citet{bieging89}, together with 3.6\,cm observations
from \citet{ll93}; these are denoted by superscripts ``b'', ``c'' and ``d'',
respectively. Note that the indicated 3.6\,cm flux for HD\,15570 deviates
from the ``original'' value of 110$\,\pm\,30$~$\mu$Jy provided by
\citet{ll93}, as a result of a recent recalibration of the original Howarth
\& Brown VLA data, performed by IDH. For $\zeta$~Pup, finally, we used the
data obtained by \citet[ denoted by superscript ``e'']{blomme03}, at 3.6 and
6~cm (Australia Telescope Compact Array, ATCA) and 20~cm (VLA), in
combination with the 2~cm data from Bieging et al. (recalibrated, see Blomme
et al.)

For those objects which have been observed both by us and by others, or where
multiple observations have been obtained (particularly for the non-thermal
emitter Cyg\,OB2\#8A), we have added these values to our database.  In almost
all cases, the different values are consistent with each other, especially
for the weaker radio sources when comparing with the upper limits derived by
\citet{bieging89}.

\subsection{IR observations}
\label{irobs}

\begin{table}
\caption{Near IR magnitudes and errors (last two digits) for program stars
as observed with the Crimean 1.25 m telescope.} 
\tabcolsep1.6mm
\begin{tabular}{lcccccc}
\hline
Star       & JD      &      $J$  &       $H$  &  $K$       &   $L$  \\
           & (2453+) &         &          &          &   $M$  \\
\hline
HD\,14947  & 067.204 & 7.17  02 & 6.95 02 & 6.88 01 & 6.85 04  \\ 
           & 307.456 & 7.10  01 & 6.98 01 & 6.85 04 & 6.67 08  \\
HD\,30614  & 073.238 & 4.32  02 & 4.25 02 & 4.20 02 & 4.20 01  \\
           & 100.224 & 4.30  01 & 4.26 01 & 4.27 01 & 4.23 01  \\
HD\,34656  & 072.356 & 6.67  01 & 6.71 01 & 6.61 01 & 6.60 04  \\
           & 100.244 & 6.69  02 & 6.71 01 & 6.69 01 & 6.65 03  \\
HD\,190429A& 216.414 & 6.28  01 & 6.12 01 & 6.14 01 & 6.13 04  \\
           & 225.439 & 6.18  01 & 6.01 01 & 6.19 01 & 5.98 08  \\
HD\,192639 & 216.439 & 6.45  01 & 6.24 01 & 6.22 01 & 6.26 04  \\
           & 307.254 & 6.44  01 & 6.23 01 & 6.17 01 & 6.24 04  \\
HD\,203064 & 223.459 & 5.17  01 & 5.12 01 & 5.13 01 & 5.13 03 \\
           &         &         &         &         & 4.98 10 \\
           & 311.301 & 5.19  01 & 5.17 01 & 5.17 00 & 5.02 02 \\
           &         &         &         &         & 5.02 05 \\
HD\,207198 & 223.496 & 5.51  01 & 5.35 02 & 5.39 01 & 5.37 04 \\
           &         &         &         &         & 5.58 10 \\
           & 309.167 & 5.48  02 & 5.42 01 & 5.45 01 & 5.58 03 \\
           &         &         &         &         & 5.57 20 \\
HD\,209975 & 223.535 & 5.01  01 & 4.97 01 & 5.00 01 & 5.12 05 \\
           &         &         &         &         & 5.00 07 \\
HD\,210839 & 223.567 & 4.62  01 & 4.52 01 & 4.54 01 & 4.57 02 \\
           &         &         &         &         & 4.44 05 \\
           & 309.193 & 4.61  01 & 4.51 01 & 4.58 01 & 4.62 02 \\
           &         &         &         &         &4.37 06  \\
\hline                
\end{tabular}
\label{irtable}
\end{table}

In the right part of Table~\ref{radiotable} we have summarized the IR data 
used, which are to a large part drawn from the literature. For a few
objects, IRAS data for 12, 25, 60 and 100 $\mu$m are also available
\citep{beichman88}, unfortunately mostly as upper limits for $\lambda \ge$
25 $\mu$m. For $\zeta$ Pup (HD\,66811), however, actual values
are present at all but the last wavelength (100 $\mu$m); see \citet{lamers84}. 

For nine objects (denoted by ``own'' in the ``references'' column of
Table~\ref{radiotable}), new $JHKLM$ fluxes (see Table~\ref{irtable}) have
been obtained at the 1.25 m telescope of the Crimean Station of the
Sternberg Astronomical Institute (Cassegrain focus, with an exit aperture of
12\arcsec), using a photometer with an InSb detector cooled with
liquid nitrogen. Appropriate stars from the Johnson catalog
\citep{johnson66b} were selected and used as photometric standards. Where
necessary, the $HLM$ magnitudes of the standards have been estimated from
their spectral types using relations from \citet{Korneef}.

\subsubsection{Absolute flux calibration}
\label{absolute_fluxes}

In order to convert the various IR magnitudes from the literature and our own
observations into {\it meaningful} (i.e., {\it internally consistent}) physical
units, we have to perform an adequate absolute flux calibration. For such a
purpose, at least three different methods can be applied:-
\begin{enumerate}
\item Calibration by means of the solar absolute flux, using analogous stars.
\item Direct comparison of the observed Vega flux with a blackbody.
\item Extrapolation of the visual absolute flux calibration of Vega, 
using suitable model atmospheres.
\end{enumerate}
Although the first two methods are more precise, the latter one provides
the opportunity to interpolate in wavelength, allowing the derivation of different 
sets of IR-band Vega fluxes for various photometric systems. Thus, such an
approach is advantageous in the case encountered here (observational
datasets obtained in different photometric systems), and we have 
elected to follow this strategy.

\paragraph{Atmospheric model for Vega.} To this end, we used the latest
Kurucz models\footnote{from {\tt http://kurucz.harvard.edu/stars/vega}} to
derive a set of absolute IR fluxes for Vega in a given photometric system,
by convolving the model flux distribution (normalized to the Vega absolute
flux at a specific wavelength; see below) with the corresponding filter
transmission functions.  In particular, we used a model with \Teff\ = 9550
K, \logg\ = 3.95, [M/H] = -0.5 and \vmic\ = 2.0 \kms \citep{CK94}. In order
to account for the possibility that the metallicity of Vega might differ
from that adopted by us, an alternative model with [M/H] = -1.0 (cf.
\citealt{GG05}) was used to check for the influence of a different
metallicity on the derived calibration. At least for the Johnson photometric
system, the differences in the corresponding fluxes turned out to lie always
below 1\%.

\paragraph{Visual flux calibration.}
The most commonly used visual flux calibration for Vega is based on the
compilation by \citet{Hayes85}, which has since been questioned by
\citet{Megessier95}, who recommends a value being 0.6\% larger than the
value provided by Hayes (3540 Jy),
and equals 3560 Jy (i.e., $3.46 \cdot 10^{-9}$ erg
cm$^{-2}$ s$^{-1}$ \AA$^{-1}$) at $\lambda$ = 5556 \AA.  
This value has been used when normalizing the Kurucz model fluxes to the
monochromatic flux at $\lambda$ = 5556 \AA. Since the standard error of the
Megessier calibration is about one percent, this error is also inherent in 
our absolute flux distribution.

\paragraph{Vega $V$-band magnitude.} 
The available $V$-band magnitudes of Vega range from 0\,\fm026
\citep{Bohlin04} to 0\,\fm035 \citep{Colina94}, while in the present
investigation we adopt $V$ = 0\,\fm03 mag in agreement with
\citet{johnson66b}. With this value, the monochromatic flux for a Vega-like
star at the effective wavelength of the $V$ filter is
F$_{5500} $(m$_V$ = 0\,\fm0) = 3693 Jy.

\paragraph{Filter transmission functions.} To calculate the absolute fluxes of
Vega in a given photometric system, we have to know the
corresponding filter transmission functions, for each band of this system. 
In those cases where such functions were explicitly available we used them,
while for the rest (including our own IR data) we used trapezoidal
transmission curves based on the published effective wavelength and FWHM of
the filters.\footnote{For more detailed information about the shape of the
filter transmission functions used to convert the literature data, see 
\citet[their Table~3]{RB96}.} The use of trapezoidal instead of actual
response functions might, of course, lead to some error in the derived
absolute fluxes. Indeed, in the particular case of the ESO filter system,
this error was estimated to be less than 5\% \citep{SM93}, with typical
values of about 2\% systematically larger fluxes from the trapezoidal
approximation. 

\paragraph{Vega IR magnitudes.}
To convert stellar magnitudes into absolute fluxes using Vega as a
standard, the magnitudes of Vega in the different filters for the
various photometric system have to be known. In our case, these data
have been  taken from the corresponding literature, and the errors inherent
to these measurements are usually very small.

Finally, let us mention that we are aware of the problem that the use of
(simplified) model atmospheres for calculating the IR flux distribution of
Vega might lead to some uncertainties, as discussed by \citet{Bohlin04}
(e.g., the possibility that Vega is a pole-on rapid rotator, \citealt{GHA94,
Peterson04}).  Note, however, that \citet{Tokunaga05} have recently shown
that the near-IR (1 to 5 $\mu$m) absolute flux densities of Vega derived by
means of atmospheric models (e.g., \citealt{Cohen92}) and by means of direct
measurements (e.g., \citealt{Megessier95}) are actually indistinguishable
within the corresponding uncertainties, which, in these specific cases, are
of the order of 1.45\% and 2\%, respectively.

On the other hand, given the fact that Vega has a dust and gas disk
\citep{Wilner02} which produces an IR excess, one cannot exclude the
possibility that a flux calibration based on a comparison of Vega observed
magnitudes and model fluxes might lead to systematic errors, at least for
$\lambda\, > 5\, \mu$m, as discussed also by \citet{Megessier95}. There are
12 stars in our sample for which we have ground based mid-IR photometry
obtained in the $N$- and $Q$-bands. In the case that Vega indeed displays a
mid-IR flux excess (as compared to the models), one might expect that the
observed fluxes of our targets (based on this calibration) are somewhat
underestimated in these bands. Such a systematic error can be easily
detected, however, and we shall keep this possibility in mind when
performing our analysis.

\subsection{Mm observations}
\label{mmobs}

For three objects, we were also able to use 1.3/1.35\,mm fluxes, acquired
either with the Swedish ESO Submillimeter Telescope (SEST) at La Silla
($\zeta$ Pup; see \citealt{leitherer91}) or with the Submillimetre Common
User Bolometer Array ({\sc scuba}; \citealt{holland99}) at the James Clerk
Maxwell Telescope (HD\,15570 and HD\,210839). (For Cyg\,OB2\#8A, which was
also observed with {\sc scuba}, only badly defined upper limits were obtained.)

The {\sc scuba} observations were obtained in the instrument's photometry
mode (the standard mode employed for point-like sources), using the single,
1.35\,mm photometric pixel, located at the outer edge of the long-wavelength
(LW) array. The data were acquired in service mode over the period
May--July, 1998. Table~\ref{scubatable} lists the observation dates,
integration times and measured fluxes. Data reduction was performed using
the {\sc scuba} User Reduction Facility ({\sc surf}; \citealt{jenness00}).

Additional 0.85\,mm {\sc scuba} data have been taken from the literature 
\citep{blomme03}, again for $\zeta$~Pup.

\begin{table}
\caption{1.35\,mm fluxes and errors for program stars observed with {\sc scuba}.}
\begin{tabular}{lccc}
\hline
Star        & date of obs.\  &   integration (s) &  flux (mJy) \\
\hline
Cyg\,OB2\#8A  & May 7, 1998 &   3600    &  $-2.50\pm5.95\phantom{-}$  \\
HD\,15570   & Jul 3, 1998 &   2160    &   $4.76\pm2.43$  \\
HD\,210839  & May 4, 1998 &   4500    &   $4.25\pm1.92$  \\
            & Jun 1, 1998 &   2340    &   $8.87\pm3.58$  \\
\hline                
\end{tabular}
\label{scubatable}
\end{table}

\subsection{De-reddening and stellar radii}
\label{dered}

Since we are aiming for a combined optical/IR/radio study, all parameters
used have to be consistent in order to allow for a meaningful analysis of
the observed fluxes, in particular the excesses caused by the (clumped) wind
alone. To compare the observed with the theoretical fluxes, we have ($i$) to
de-redden the observed fluxes and ($ii$) to derive a consistent stellar radius
for a given distance $d$ (or vice versa, see below), which has been drawn
from the literature cited or recalculated from the assumed value of \MV\
(for models ``1-2'' in column ``ref2'' of Table~\ref{sample}).

For this purpose, we have used our (simplified) model as described in
Sect.~\ref{irfluxes} to synthesize theoretical $VJHK$ fluxes.\footnote{Only
near-IR fluxes were used to ensure that the flux excess due to the wind
remains low, i.e., rather unaffected by clumping.} Note that this model has
been calibrated to reproduce the corresponding predictions obtained from a
large OB-model grid calculated by {\sc fastwind} \citep{puls05}.

By comparing the observed IR fluxes (from the various sources given in
Table~\ref{radiotable}) with the theoretical predictions, we derive
``empirical'' values for the color excess E(B-V) and/or the extinction ratio
\Rv, by requiring the ratio between de-reddened observed (plus/minus error)
and distance-diluted theoretical fluxes to be constant within the $V$- to
$K$-bands. For this purpose, we adopt the reddening law provided by
\citet{cardelli89}. Visual fluxes have been calculated using $V$-magnitudes
from Paper~I or from \citet{maiz04}.

\begin{figure}
\resizebox{\hsize}{!} {\includegraphics{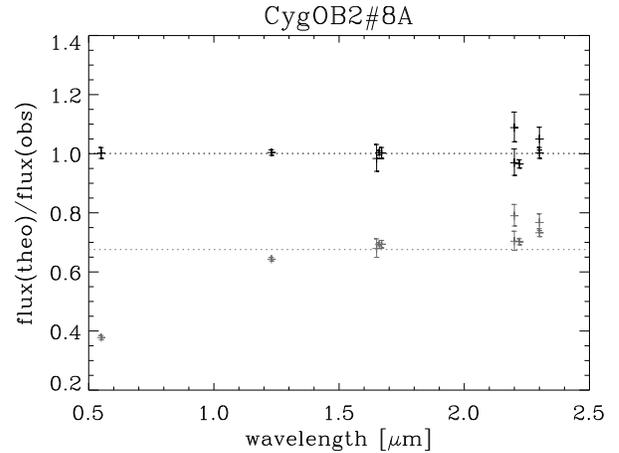}}
\caption{De-reddening procedure, for the example of Cyg\,OB2\#8A. Displayed
is the ratio of distance-diluted, theoretical fluxes and de-reddened,
observed $VJHK$ fluxes, as a function of wavelength, with bars accounting
for the observational errors. Grey entries correspond to ``fit-parameters'' 
E(B-V) = 1.9, \Rv\ = 3.0 and \Rstar\ = 24 \Rsun. Obviously, the extinction
is too large: the ratio of theoretical to de-reddened fluxes is much smaller
at shorter than at larger wavelengths (extinction decreasing with
wavelength). Moreover, the assumed radius is too small, since the {\it mean}
flux ratio (indicated by a dotted line) is well below unity (too small an
angular diameter). The black entries show our final solution, for E(B-V) =
1.63, \Rv = 3.0 and \Rstar = 27 \Rsun. Any curvature has vanished, the
optical flux corresponds to the mean, and the mean ratio itself (again
indicated by a dotted line) is located at unity.} 
\label{dered_radius}
\end{figure}

In a second step, we adapt the stellar radius (for a given distance) in such
a way that the {\it mean} ratio becomes unity. This procedure ensures the
correct {\it ratio} between radius and distance, i.e., angular diameter,
which is the only quantity which can be specified from a comparison between
synthetic and observed fluxes. Of course, we could have also chosen to
modify the distance for a given radius; however, in order to be consistent
with previous mass-loss estimates from radio observations, which rely on
certain distances, we have followed the former approach.
Fig.~\ref{dered_radius} gives an impression of this procedure, for the
example of Cyg\,OB2\#8A.

Note that in parallel with re-defining the stellar radius, the mass-loss
rate used to calculate the $V$- to $K$-band model fluxes has to be modified
as well, because the latter depend on the assumed value of \Mdot\ (see
below). Since we do not know the actual mass-loss rate in advance, we follow
a simplified approach and use a value equal or related to the \Ha\
mass-loss rate provided by previous investigations (entry ``ref1''). This
mass-loss rate, however, had been derived for a certain stellar radius,
which we claim to improve by our procedure. Consequently, we also have to
modify our ``input value'' of \Mdot, to maintain the \Ha\ fit-quality of the
former investigations. As outlined already above, this can be obtained by
keeping the ratio $Q'=\Mdote/\Rstare^{1.5}$ constant. 

This scaling has a further advantage, namely that not only \Ha\, but {\it
all} $\rho^2$-dependent diagnostics (i.e., \Ha\ profile shape, IR and radio
fluxes) and the finally {\it derived run of the clumping factor} remain
almost unaffected if a different radius or distance (though identical
angular diameter) are chosen.

This notion follows from the fact that the \Ha\ profile shape depends on $Q'$
alone (for given \vinf\ and assuming that the NLTE departure coefficients do
not vary), and that the IR and radio optical depths scale with this quantity
as well, whereas the corresponding fluxes are additionally diluted by
$(\Rstare/d)^2$. As an example, remember that under certain conditions (see 
Sect~\ref{radiofluxes}) the radio fluxes scale according to
\beq 
\label{radioflux_approx}
F_\nu \propto \frac{\Mdote^{4/3}}{d^2}\, = \, 
\bigr(\frac{\Mdote}{\Rstare^{3/2}}\bigl)^{4/3} \,
\bigr(\frac{\Rstare}{d}\bigl)^2. 
\eeq 

In other words, as long as $Q'$ and the angular diameter (``measured'' from
aligning synthetic and observed, de-reddened fluxes; see
Fig.~\ref{dered_radius}) remain conserved, almost all further results become
independent of the individual choice of \Rstar\ or $d$, and a translation of
our results to different assumptions, e.g., due to future improvements
concerning distance measurements ({\sc gaia}), becomes easily possible. The
only quantities which depend directly on these values are the mass-loss and
wind-momentum rate (e.g., Paper~I), which are of minor
importance regarding the objectives of this paper. 

One problem inherent to our approach is the fact that the derivation of
reddening parameters and \Rstar\ requires an a priori knowledge of \Mdot\
(and clumping properties), since, as stated already above, the model fluxes
depend on this quantity.

First note that the flux excess increases as a function of \Mdot.
Consequently, the average {\it slope} of the model fluxes decreases, which
affects our de-reddening procedure (operating in the $V$- to $K$-band). This
dependence, however, is only moderate, due to the rather low excess in this
wavelength region for typical OB-star winds. Moreover, it is predicted
correctly by our models if $Q'$ is of the correct order.

The {\it absolute} flux level in the optical and near IR, on the other hand,
is much more affected by our choice of \Mdot, thus influencing our
derivation of \Rstar. For {\it identical} stellar parameters, the $V$-flux
is a (monotonically) decreasing function of \Mdot.\footnote{More precisely:
for those wavelength bands where the wind is not optically thick, i.e.,
where the fluxes depend on both the photospheric radiation and the wind
absorption/emission, there is an additional dependence on the wind density,
$\propto \Mdote/\Rstare^2$, which scales somewhat differently than $Q'$.} To
a large extent, this behaviour is induced by a decreasing source function at
bf-continuum formation depth, related to the decrease in electron
temperature (at $\tau(\Mdote) \approx 2/3$) when \Mdot\ is increasing, and
increasing electron scattering. Both effects apply to blanketed {\it and}
unblanketed models; the ``only'' difference concerns the absolute flux level
at optical and (N)IR bands, which is larger for blanketed models, due to
flux-conservation arguments (compensation of the blocked (E)UV radiation
field).

Since a precise knowledge of the ``real'' wind density and the
near-photospheric clumping properties is not possible at this stage, only an
iteration cycle exploiting the results of our following mass-loss/clumping
analysis could solve the problem ``exactly''. 

In order to avoid such a cycle, we follow a simplified approach, in
accordance with our findings from Paper~I and anticipating our results from
Sect.~\ref{analysis} (cf. column ``ratio'' in Table~\ref{table_clf}). To
calculate the theoretical fluxes required for our de-reddening procedure,
for objects with \Ha\ in absorption we have used the actual, $Q'$-scaled,
\Ha\ mass-loss rate, whereas for objects with \Ha\ in emission we have
reduced the corresponding value by a factor of 0.48. This approach is based
on our hypothesis that the lowermost wind is unclumped (see
Sect.~\ref{clumping}), and that the previously derived \Ha\ mass-loss rates
for objects with \Ha\ in emission are contaminated by clumping, with average
clumping factors of the order of $\bigl(\frac{1}{0.48}\bigr)^2$.

From the almost perfect agreement of the theoretical $V$-to-$K$ fluxes with
the observations for our final, clumped models, this assumption seems to be
fairly justified. In any case (i.e., even if the lowermost wind were to be
clumped as well), the most important quantity is the {\it effective}
mass-loss rate (i.e., the actual, unclumped \Mdot\ times square root of
local clumping factor), so any reasonable error regarding this quantity
would barely affect the corresponding theoretical fluxes and thus our
de-reddening procedure.

We will now comment, where appropriate, on the results of our procedure for
a few individual objects. For the majority of stars, only small
modifications of the E(B-V) values resulting from optical photometry,
$(B-V)$, and intrinsic colors, $(B-V)_0$, were necessary, while keeping the
total-to-selective extinction ratio, \Rv, at its ``normal'' value of 3.1, or
at a value suggested from other investigations. The intrinsic colors used
here have been adapted from \citet{wegner94}, particularly because of their
extension towards hotter spectral types. However, since this calibration
deviates considerably from the widely used alternative provided by
\citet{fitzgerald70} at the cool end (-0.24 mag vs. -0.28 mag for O9.5
supergiants), we adopt, as a compromise, only values $\le -0.27$, and
$-0.27$ if Wegner's calibration exceeds this threshold.

Concerning the Cyg\,OB2 stars, for three objects (\#7, \#8A and \#8C), our
procedure results in rather similar reddening parameters to those presented by
\citet[based on UBV photometry by Massey, priv. comm., and IR-photometry from
2MASS]{hanson03}. Only for stars \#10 and \#11 did we find larger
discrepancies, which were corrected for by using \Rv\ = 3.15 instead of 
\Rv\ = 3.0, as suggested by Hanson and previous work in the optical
(\citealt{massey91, torres91}). Note that ``our'' value is consistent with
the values provided by \citet[see below]{patetal03}: \Rv\ = 3.17 and 3.18,
respectively.

\begin{figure}
\resizebox{\hsize}{!}
   {\includegraphics{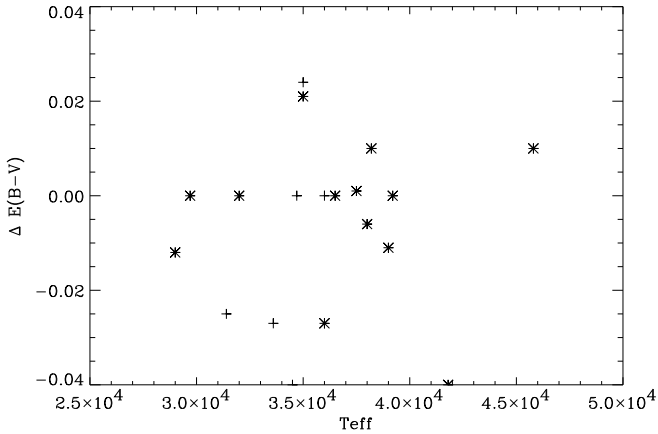}}
\caption{
Differences between {\it derived} color excess, E(B-V), and
corresponding literature value, $(B-V)-(B-V)_0$ (Table~\ref{sample}, entry
``ref 2''; for intrinsic colors see text), as a function of \Teff. Asterisks
denote supergiants, and crosses bright giants and giants, respectively. The mean
deviation for supergiants is $-0.004\,\pm\,0.016$ mag, and for l.c. II/III stars
$-0.01\,\pm\,0.023$ mag.
}
\label{compred}
\end{figure}

In disagreement with the work by Hanson, however, we still used the
canonical distance of $d=1.71$~kpc for the Cyg\,OB2 stars, as determined
by \citet{massey91}. In our opinion, the alternative, lower value(s)
claimed by Hanson would result in too low luminosities. Most probably,
though, the ``real'' distance {\it is} smaller than the value used here. As
pointed out, this would imply ``only'' a down-scaling of radii and
mass-loss rates, and would not affect our conclusions concerning the wind
clumping.

The only other objects worthy of closer inspection are those with extinction
ratios \Rv\ $\ne 3.1$ (cf. Paper~I). Unfortunately, the recent catalogue of
\Rv\ values for Galactic O-stars by \citet{patetal03} covers only few stars
in our sample (in particular, Cyg\,OB2\#8A, \#10, \#11 and HD\,34656), such
that a comparison is not possible for the majority of objects. Due to the
degeneracy of E(B-V) and \Rv\ (different combinations can result in rather
similar extinction laws, if the considered wavelength range is not too
large), we applied the following philosophy: in those cases with peculiar
extinction ratios, \Rv\ $\ne 3.1$, (obtained from Paper~I and references
therein), we firstly checked, by using these values, whether the derived
color excess is consistent (within small errors; see below) with measured
and intrinsic colors in the optical. If so, we adopted these values here. In
this way, we confirmed the values of \Rv\ = 5.0, 5.0 and 2.76 for the stars
HD\,36861, HD\,37043 (both belonging to Ori\,OB1) and HD\,209975,
respectively.

For HD\,207198, with a literature value of \Rv\ =2.76, the derived color
excess would have fallen 0.04 mag below the ``optical'' value (a deviation
which we considered to be too large if \Rv\ $\ne$3.1 anyway). Therefore, for
this object, we kept the optical E(B-V) value and fitted \Rv\ using our
procedure, resulting in \Rv\ = 2.56. This star is the only one for which our
procedure showed significant deviations from previous work. Given the
difficulties in deriving reliable \Rv\ values, however, we consider this
deviation as not too troublesome. 

For the last object in this group, HD\,34656, we could check for the 
consistency of our results with the work by Patriarchi et al. By keeping
\Rv\ = 3.1, as suggested in Paper~I, the derived E(B-V) would lie 0.03 mag
above the ``optical'' value, which, compared to the other objects (see
below), is rather large. On the other hand, by keeping our value of E(B-V),
we derived \Rv\ = 3.4, which is consistent with the value claimed by
Partriarchi et al. (\Rv\ = 3.5), and we adopted this solution.

Fig~\ref{compred} summarizes the results of our de-reddening procedure, by
comparing the {\it derived} values of E(B-V) for our complete sample with
the corresponding ``optical'' values, $(B-V)-(B-V)_0$, as a function of
\Teff\ (with (B-V) given by the references in Table~\ref{sample}, entry
``ref 2'', and the intrinsic colors as discussed above). 

From this figure, we find no obvious trend of the difference in E(B-V) as a
function of \Teff\ (the average differences being almost exactly zero for
supergiants and $-0.01$~mag for the remaining objects), which is also true
if we plot this quantity as a function of \MV\ (not shown). The majority of
these differences are less than $0.02$~mag, which seems to be a reasonable
value when accounting for the inaccuracy in the observed $(B-V)$ colors, the
uncertainties in the intrinsic ones, the errors resulting from our flux
calibration and the typical errors on the theoretical fluxes (cf.
Sect.~\ref{irfluxes}).

\section{Simulations}
\label{simul}

In this section, we will describe our approach to calculating the various
energy distributions required for our analysis, and our approximate treatment
of wind clumping, which is based upon the assumption of {\it small-scale}
inhomogeneities. Since this treatment consists of a simple manipulation
of our homogeneous models, we will start with a description of these.

Because of the large number of parameters to be varied (\Mdot, $\beta$,
clumping factors), and accounting for the rather large sample size, an
``exact'' treatment by means of NLTE atmospheres is (almost) prohibitive.
Thus, we follow our previous philosophy of using {\it approximate} methods,
which are calibrated by means of our available NLTE model grids
\citep{puls05}, to provide reliable results. Note that these grids have
been calculated without the inclusion of X-rays; the influence
of X-rays on the occupation numbers and IR/radio opacities of hydrogen
is negligible (e.g., \citealt{paul01}), whilst their
effect on helium (through their EUV tail) has not been
investigated in detail. From a comparison of models with and
without X-rays though, any effect seems to be small.

We have been able to design {\it interactive} procedures (written in {\sc
idl} acting as a wrapper around {\sc fortran}-programs), which allow for a
real-time treatment of the problem, where all required fits and manipulation
of \Ha\ spectra and IR/radio fluxes are obtained in parallel.

\subsection{\Ha}
\label{halpha}

\begin{table}
\caption{Consistency check for \Ha\ mass-loss rates and velocity field
exponents, for those objects with stellar and wind parameters derived from a
complete NLTE analysis (cf. Table~\ref{sample}). All mass-loss rates are in
units of of $10^{-6}{\rm M_{\odot}/yr}$. $\Mdote_1$ is the mass-loss rate as
derived from our approximate method, adopting $\beta_1=\beta$(in), where
possible. In some cases, a second solution ($\Mdote_2, \beta_2$) is
possible, mostly for objects with \Ha\ in absorption (see text).}
\begin{tabular}{lrr|rr|rr}
\hline
Star &  \Mdot(in) & $\beta$(in) & $\Mdote_1$ &$\beta_1$ & $\Mdote_2$ &$\beta_2$\\
\hline
  Cyg\,OB2\#7    & 10.61 & 0.77 &   \Mdot(in) & $\beta$(in) &        9.5  &        0.90\\
   HD\,15570     & 17.32 & 1.05 &       16.00 & $\beta$(in) &   \Mdot(in) &        0.95\\
   HD\,66811     & 16.67 & 0.90 &       13.50 & $\beta$(in) &             &            \\
                 &  8.26 & 0.90 &        6.69 & $\beta$(in) &             &            \\
   HD\,14947     & 16.97 & 0.95 &   \Mdot(in) & $\beta$(in) &             &            \\
 Cyg\,OB2\#11    &  8.12 & 1.03 &        9.50 &        1.10 &             &            \\
 Cyg\,OB2\#8C    &  4.28 & 0.85 &        3.50 &        1.00 &             &            \\
 Cyg\,OB2\#8A    & 11.26 & 0.74 &       13.00 & $\beta$(in) &   \Mdot(in) &        0.95\\
  HD\,210839     &  7.95 & 1.00 &   \Mdot(in) & $\beta$(in) &             &            \\
  HD\,192639     &  6.22 & 0.90 &        5.70 &        1.14 &   \Mdot(in) &        1.05\\
   HD\,24912     &  2.45 & 0.80 &        4.00 & $\beta$(in) &   \Mdot(in) &        1.05\\
  HD\,203064     &  0.98 & 0.80 &        1.30 & $\beta$(in) &   \Mdot(in) &        0.92\\
  HD\,207198     &  1.05 & 0.80 &        1.30 & $\beta$(in) &   \Mdot(in) & 0.90\\
   HD\,30614     &  3.07 & 1.15 &        2.40 & $\beta$(in) &             &            \\
 Cyg\,OB2\#10    &  2.74 & 1.05 &        3.30 & $\beta$(in) &             &            \\
  HD\,209975     &  1.11 & 0.80 &        1.20 &        0.90 &             &            \\
\hline                
\end{tabular}
\label{haconsist_table}
\end{table}

\begin{figure*}
\resizebox{\hsize}{!}
   {\includegraphics[angle=90]{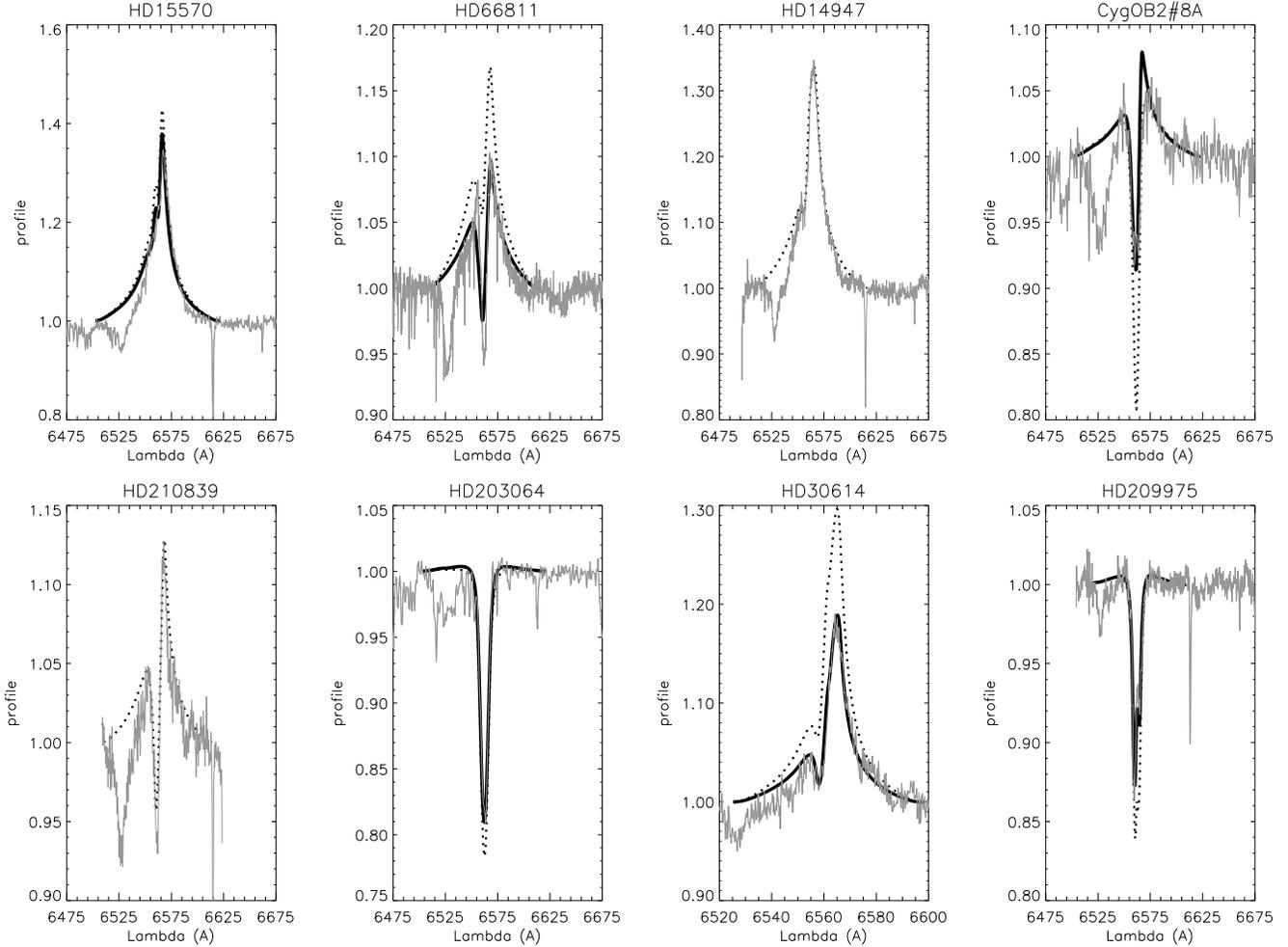}}
\caption{Consistency check for \Ha\ profiles: results of our approximate \Ha\
line synthesis, for some representative cases from
Table~\ref{haconsist_table}. Dotted: \Ha\ line profiles with parameters
\Mdot(in) and $\beta$(in) as derived from a complete NLTE analysis (cf.
Table~\ref{sample}); bold: corresponding profiles with \Mdot\ = $\Mdote_1$ or
$\beta=\beta_2$ (see Table~\ref{haconsist_table}).}
\label{haconsist}
\end{figure*}

In the present study, synthetic \Ha\ profiles are calculated as described in
Paper~I. This approach bases on the approximate treatment as introduced by
\citet{puls96}, updated to account for line-blanketing effects. Except for
the inclusion of clumping, no further modifications have been applied; note
in particular that we have used the same \Ha\ observations and H/He departure
coefficients as adopted in Paper~I. 

On the other hand, for most of our sample stars we have quoted (and used,
within our de-reddening procedure) wind parameters from a complete NLTE
analysis, which do not rely exclusively on \Ha, but also on \HeII\ 4686 and
other diagnostics. Furthermore, the observed \Ha\ profiles used here are
different to those in the corresponding sources, because of the variability
of \Ha\ (cf. Sect.~\ref{variab}). Thus, we have to check how far the
values from the complete analysis (denoted by \Mdot(in) and $\beta$(in))
might deviate from solutions resulting from our simplified method, used in
combination with our different \Ha\ data, to obtain consistent initial
numbers for the following investigations and to re-check the reliability of
our approach.\footnote{Concerning those (four) objects with wind parameters
taken from Paper~I, we have convinced ourselves that the corresponding fits
could be reproduced.} To this end, we have re-determined mass-loss rates and
velocity exponents, using our observational material, the stellar parameters
from Table~\ref{sample} and the approximate \Ha\ line synthesis as outlined
above. Table~\ref{haconsist_table} summarizes the results from this
exercise.

For three objects (Cyg\,OB2\#7\footnote{The second solution with $\beta=0.9$
gives a better fit for the absorption trough.}, HD\,14947 and HD\,210839),
no modifications were required at all, whereas for the other stars small
variations of \Mdot\ were sufficient to reproduce our observational data,
mostly by keeping the nominal velocity exponent. The average ratio between
modified and input mass-loss rates was 1.07\,$\pm$\,0.22. 

In some cases (particularly for objects with \Ha\ in absorption), a second
solution is possible, and in all but one case, we kept the nominal mass-loss
rate constant, while varying $\beta$ (entry $\beta_2$ in
Table~\ref{haconsist_table}). All derived velocity exponents still lie in
the expected range. For representative cases, Fig.~\ref{haconsist} displays
the results of our line synthesis, both for models with the nominal values,
\Mdot(in) and $\beta$(in), and for the best-fitting models from
Table~\ref{haconsist_table}, with $\Mdote_1$ or $\beta_2$.

In conclusion, our simplified routine delivers reliable numbers and thus can
be used in our further approach to derive constraints on the clumping
factors.

\subsection{Infrared fluxes}
\label{irfluxes}

For the calculation of the infrared fluxes, we closely followed the
approximations as outlined by \citet{lamerswaters84a}, with Gaunt factors
from \citet{waterslamers84}. The major difference concerns the fact that 
the radiative transfer is solved by means of the ``Rybicki algorithm''
\citep{ryb71}, to account for electron scattering in a convenient way. 

A further modification regards the photospheric input fluxes which were
chosen in such a way as to assure that the emergent fluxes, on average, comply
with the results from our detailed NLTE model grids. 

After some experiments, 
it turned out that the best choice for the various parameters is the
following:-

\paragraph{The velocity law} is specified by
\beq
v(r)=\vinfe(1-b/r)^\beta, \qquad  b=1 - (v_{\rm min}/\vinfe)^{1/\beta},
\eeq
where $r$ is calculated in units of \Rstar, and
the minimum velocity, $v_{\rm min}$, is set to 10~\kms. 

\paragraph{Electron temperature.} All Gaunt factors are calculated at a
temperature of 0.9 \Teff, and the electron temperature is calculated using
Lucy's temperature law for spherical atmospheres \citep[his Eq. 12, and
using grey opacities]{lucy71}, with an optical depth scale accounting for
electron scattering only and a temperature cut-off at 0.5 \Teff. Remember
that the radio fluxes are almost independent of the temperature, and a
number of tests have shown that different (reasonable) temperature
stratifications have negligible effects on the derived IR fluxes as well.

\paragraph{Ionization equilibrium.} Hydrogen is assumed to be (almost)
completely ionized, helium as singly ionized outside the recombination
radius (see below) and the CNO metals as either two or three times ionized. 

Throughout the parameter range considered here, helium is singly ionized in
the {\it radio} emitting region (for $\lambda > 2$ cm; concerning mm fluxes
see below), as we have convinced ourselves by an inspection of our model
grid.  (Only for O3/4 dwarfs and earlier types -- which are missing in our
sample -- does helium remain completely ionized throughout the entire wind).

\begin{figure}
\resizebox{\hsize}{!}
   {\includegraphics{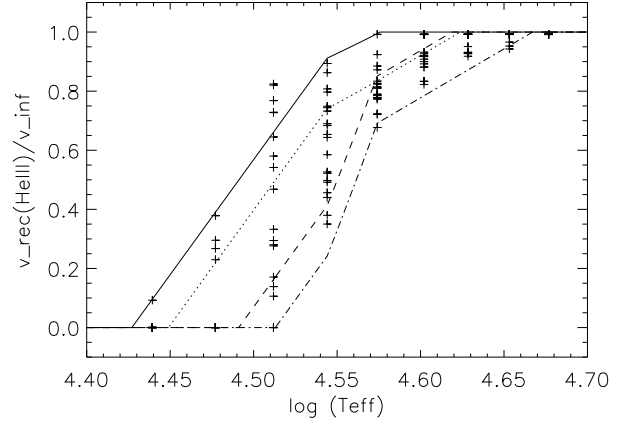}}
\caption{Location of \HeIII\ recombination in velocity space (in units of
\vinf), as a function of \Teff. Crosses display this location as derived from 
our model grid, with 27.5~kK $\le \Teffe \le$ 47.5~kK, different gravities
and wind densities. Curves indicate the results of our linear regression, 
Eq.~\ref{herec_para},
for the (limiting) cases (\logg=3.0, $\bar\rho = -13$, solid), (\logg=3.0, 
$\bar\rho = -11$, dotted), (\logg=4.5, $\bar\rho = -13$, dashed) and
(\logg=4.5, $\bar\rho = -11$, dashed-dotted). For units of $\bar\rho$, see
text. The only region which is not matched by our regression is the low
gravity, low wind density region around \Teff = 33,000~K, where the regression
yields too low recombination velocities.}
\label{herec}
\end{figure}

With respect to the mid- and far-IR emitting region, this statement is no
longer justified, and one would have to calculate a consistent ionization
structure, which is beyond the scope of this paper. In order to obtain 
an approximate solution of this problem, we have parameterized the
recombination radius, in dependence of \Teff, \logg\ and mean wind density,
$\bar \rho = \Mdote/(\Rstare^2 \vinfe)$, from a linear regression to
corresponding results of our model grid (cf. Fig.~\ref{herec}, crosses). It
turned out that a best fit could be obtained for the recombination {\it
velocity} (defined as the velocity where the ionization fraction of \HeIII\
becomes larger than the fraction of \HeII\ when proceeding from outside to
inside), which can expressed (in units of \vinf) as
\beqa
\label{herec_para}
\lefteqn{v_{\rm rec} = c + a_1 \log \Teffe + a_2 \log g + a_3 \log \bar\rho,}\\
\lefteqn{{\rm min}(v_{\rm rec}) = 0, \qquad {\rm max}(v_{\rm rec}) = 1,} \nonumber
\eeqa
with 
\begin{eqnarray*}
\lefteqn{27,500\, {\rm K} < \Teffe < 35,000\, {\rm K}:}\\
\lefteqn{c = -34.60\,\, a_1=7.79\,\, a_2=-0.3325\,\, a_3=-0.0854} \\
\lefteqn{37,500\, {\rm K} < \Teffe < 47,500\, {\rm K}:}\\
\lefteqn{c = -14.90\,\, a_1=3.31\,\, a_2=-0.0956\,\, a_3=-0.0798,}
\end{eqnarray*}
where \Teff\ is measured in K, \Mdot\ in \Msun/yr, \Rstar\ in \Rsun\ and \vinf\
in \kms. For $35 {\rm k}K < \Teffe < 37.5 {\rm k}K$, we have applied a linear
interpolation. 

Concerning the models of our grid, this relation results in a mean
difference, $\langle v_{\rm rec}$(Eq.~\ref{herec_para})-$v_{\rm
rec}$(model)$\rangle = 0.011\pm0.079$ (in units of \vinf), where the largest
discrepancies are found in the low gravity, low wind density region around
\Teff\ = 33,000~K (cf. Fig.~\ref{herec}). Note that for high gravity, \logg\
= 4.5, and low wind density, $\log \bar\rho = -13.0$ (dashed line), the {\it
complete} wind contains solely \HeII\ for $\Teffe \la$ 31,000~K, whereas for
low gravity, \logg\ = 3.0, and high wind density, $\log \bar\rho = -11.0$
(dotted line), it remains completely ionized for $\Teffe \ga$ 42,000~K. For
our final model of $\zeta$~Pup (HD\,66811), our approximation yields $v_{\rm
rec} = 0.87$, which is in good agreement with the value of $v_{\rm rec} =
0.83$ found by \citet{hil93} in their paper on the X-ray emission of this
object.

Mostly affected by the presence of \HeIII\ (compared to the assumption that
helium is singly ionized throughout the wind) is the mid and far IR-band,
where the {\it effective} photosphere might be located below the
recombination radius. (In the near-IR, the emitted flux is still dominated
by the ``real'' photosphere.) Except for a few objects, the former
wavelength range has not been observed so far, so that our predictions
remain to be verified in the future. Note finally that from the scaling
relations provided by, e.g., \citet{lamerswaters84a}, the difference in the
derived mass-loss rates (using \HeIII\ instead of \HeII\ as the major ion,
i.e., no recombination at all) would result in a factor of roughly 0.85 for
solar helium content. Further comments on the influence of the helium
ionization balance will be given in Sect.~\ref{proto}.

\paragraph{Photospheric input fluxes} were chosen as follows: For $\lambda <
1\, \mu{\rm m}$, we used Kurucz fluxes, whereas for higher wavelengths we
used Planck functions with \Trad = 0.87 \Teff\ for $1\,\mu{\rm m} \le
\lambda \le 2\, \mu{\rm m}$, \Trad = 0.85 \Teff\ for $2\,\mu{\rm m} \le
\lambda \le 5\, \mu{\rm m}$ and \Trad = 0.9 \Teff\ elsewhere. Note that for
considerably larger wavelengths, the emergent fluxes become independent of
the input fluxes, due to increasing optical depths.

\medskip \noindent We have compared the fluxes resulting from
this simplified model with those from our NLTE model grid as calculated by
{\sc fastwind}, for the wavelength bands $V$ to $Q$. (A comparison beyond 30
$\mu{\rm m}$ is not possible, since this is the maximum wavelength
considered in {\sc fastwind}, which follows from the constraint that, for
all IR wavelengths and all wind densities, the wind plasma should become
optically thick only well inside the outermost radius point, \Rmax\ =
100~\Rstar.)

For this comparison, 204 models within the range 30~kK $\le \Teffe \le$
45~kK, with different gravities and wind densities (corresponding to $\log Q
= \log (Q'/\vinfe^{1.5})$ = -13.15{\ldots}-12.1, if \vinf\ is calculated in
\kms, see \citealt{puls05}, Sect.~10) have been used. As a result, the mean
ratio of IR fluxes from our simplified model to those from {\sc fastwind} is
of the order of 0.99{\ldots}1.01 (different for different wavelengths), 
and the typical standard deviation for each wavelength band is below 5\%.

\subsection{Radio fluxes}
\label{radiofluxes}

Radio fluxes are calculated in analogy to the IR fluxes (with identical
parameters), but neglecting electron scattering. We use a numerical 
integration, with \Rmax\ = 10,000~\Rstar\footnote{Within our procedure, we
always check that the plasma remains optically thin until well inside the 
outermost grid point.}, for the following reasons: first, the analytical
expression by analogy to Eq.~\ref{radioflux_approx}, as provided by
\citet{panagia75} and \citet{wright75}, is valid only under the condition
that the plasma is already optically thick at $v(r) \approx \vinfe$, which
is not the case for objects with thin winds. Secondly, the inclusion of
depth-dependent clumping factors requires a numerical integration anyway. Of
course, we have checked that for constant clumping factors and large wind
densities, the analytical results are recovered by our approach. Remember
that the emitted fluxes are almost independent of the assumed electron
temperature. From our final results, it turned out that except for the mm
fluxes of our hottest objects, Cyg\,OB2\#7 and HD\,15570, the radio
photospheres of the complete sample (even if sometimes below \vinf) are well
above the corresponding recombination radius (cf. Table \ref{table_clf}).
Thus, unless explicitly stated otherwise, helium is adopted to be singly
ionized in our radio simulations.\footnote{Concerning the influence of the
adopted He ionization on derived mass-loss rates, see also
\citealt{schmutz86}.} In the following figures, the radio range is indicated
to start at 400 $\mu{\rm m}$ = 0.4 mm (end of IR treatment at 200 $\mu{\rm
m}$), but this serves only as a guideline, since at these wavelengths helium
might still not be completely recombined. 

\subsection{Inclusion of wind clumping}
\label{clumping}

To account for the influence of wind clumping, we follow the approach as
described by \citet{abbott81}. Modified by one additional assumption (see
below), this approach has been implemented into NLTE model atmospheres
already by \citet{schmutz95}, and is presently also used by the alternative
NLTE code {\sc cmfgen}. In the following, we will recapitulate the
method and give some important caveats.

Regarding the hydrodynamical simulations of radiatively driven winds, the
term ``clumping factor'' has been introduced by \citet{ocr}, as defined from
the temporal averages in Eq.~\ref{defcl}. To allow for a translation to
stationary model atmospheres, one usually assumes that the wind plasma is
made up of two components, namely dense clumps and rarefied inter-clump
material, in analogy to snapshots obtained from the hydrodynamics. The
volume filling factor, $f$, is then defined as the fractional volume of the
{\it dense} gas, and one can define appropriate {\it spatial} averages for
densities and density-squares (cf. \citealt{abbott81}),
\beqa
<\rho> & = & \frac{1}{\Delta V} \int \bigl[f \rho^+\,+\,(1-f)\,\rho^-\bigr]\dd V \\
<\rho^2> & = & \frac{1}{\Delta V} \int \bigl[f
(\rho^+)^2\,+\,(1-f)\,(\rho^-)^2 \bigr]\dd V,
\eeqa
where $\rho^+$ and $\rho^-$ denote the overdense and rarefied material,
respectively. Here, and in the following, we have suppressed in our notation
any spatial dependence, both of these quantities and of $f$. The actual
mass-loss rate (still assumed to be spatially constant, in analogy to the
temporal averaged mass-loss rate resulting from hydrodynamics) is then
defined from the mean density,
\beq
\label{defmdot}
\Mdote= 4 \pi r^2 <\rho> v,
\eeq
and {\it any} disturbance of the velocity field (e.g., influencing the
line-transfer escape probabilities; see \citealt{pof}) is neglected. 

The modification introduced by \citet{schmutz95} relates to the results from
all hydrodynamical simulations collected so far, namely that the inter-clump
medium becomes almost void {\it after the instability is fully grown}, i.e,
outside a certain radius. In this case then, $\rho^-$ \rarrow\, 0, and we
find, assuming sufficiently small length scales (see below),
\beqa
<\rho> & = & \frac{1}{\Delta V} \int \bigl[f \rho^+\bigr] \dd V 
\, = \,f \rho^+ \\
\nonumber \\
<\rho^2> & = & \frac{1}{\Delta V} \int \bigl[f (\rho^+)^2 \bigr]\dd V
\, = \,f (\rho^+)^2 \,=\, \frac{<\rho>^2}{f}.
\eeqa
Comparing with Eq.~\ref{defcl} and identifying temporal with spatial averages,
we obtain
\beq
\fcl=\frac{1}{f} \qquad {\rm and} \qquad \rho^+ =  \frac{<\rho>}{f}\,
=\,\fcl <\rho>,
\eeq
i.e., the clumping factor describes the overdensity of the clumps, if the 
inter-clump densities are negligible.

Concerning model atmospheres and (N)LTE treatment, this averaging process
has the following consequences:-
\begin{itemize}
\item Since, according to our model, matter is present only inside the 
clumps, the actual (over-)density entering the rate equations is $\rho^+ = \fcl
<\rho>$ (where the latter quantity is defined by Eq.~\ref{defmdot}). 
Since both ion and electron densities become larger, the recombination rates
grow, and the ionization balance changes. As a simple example, under LTE
conditions (Saha equation), and for hot stars, we would find an increased
fraction of neutral hydrogen {\it inside the clumps}, being larger by a
factor of $\fcl^2$ compared to an unclumped model of the same mass-loss
rate. Further, more realistic, examples for important ions have been given by
\citet{bouret05}. 
\item The overall effect of this increase, however, is somewhat
compensated for by the ``holes'' in the wind plasma, since the radiative
transfer (and, consequently, the ionization and excitation rates) is 
affected by the averaging process as well, at least for processes which
depend non-linearly on the density. Note that for processes which are
linearly dependent on the density (e.g., resonance lines of major ions), the
optical depth is similar in clumped and unclumped models, provided that the
scales of the clump/inter-clump matter are significantly smaller than the
domain of integration. For $\rho^2$-dependent processes, on the other hand, 
the optical depth is proportional to the integral over $<\rho^2> = \fcl
<\rho>^2 \approx \fcl (\rho^{\rm uncl})^2$, i.e., the optical depths are
larger by just the clumping factor. Consequently, mass-loss rates
derived from such diagnostics become lower by the square root of this
factor, compared to an analysis performed by means of unclumped models.
\end{itemize}

Before we now comment on the implementation of this process into our models,
let us give two important caveats. Implicit to the assumption of small 
length scales, the simple approach as described above breaks down (at least
to some extent) if the clumps become optically thick. In this case, the
so-called ``porosity length'' becomes important, and the distribution and
shape of the clumps has to be specified to allow for more quantitative 
conclusions. For opacities scaling linearly with density, \citet{owo05}
have provided a suitable formalism to describe the effects of
clumping/porosity in this context, whereas for $\rho^2$-dependent opacities
such an analysis is still missing.

Besides the questions of the length scales involved, related optical depth
effects and the neglect of velocity disturbances, the other important
assumption concerns the treatment of the inter-clump matter as being void.
This approximation is legitimate as long as clumping is decisive only in
those parts of the wind which are significantly separated from the base.
Under this condition, the line-driven instability has already passed its
linear phase and shocks have developed, compressing the material into clumps
and rarefying the medium in between.

As has been discussed in Sect.~\ref{intro}, recent evidence indicates that
clumping becomes important from close to the wind base on
(\citealt{bouret05}). In this region, however, the instability is still in
its linear phase and resembles more a fluctuation (with similar
positive and negative density amplitudes) than a clumped structure.\footnote
{This should be true, even if a different, unknown instability were responsible
for the development of an inhomogeneous structure.}
Consequently, the assumption that the inter-clump medium is void becomes
questionable. In such a case, it might be more appropriate to follow the
original approach by \citet{abbott81}, namely to account explicitly for
the ``under-dense'' medium.

\begin{figure*}
\begin{minipage}{8.8cm}
\resizebox{\hsize}{!}
   {\includegraphics{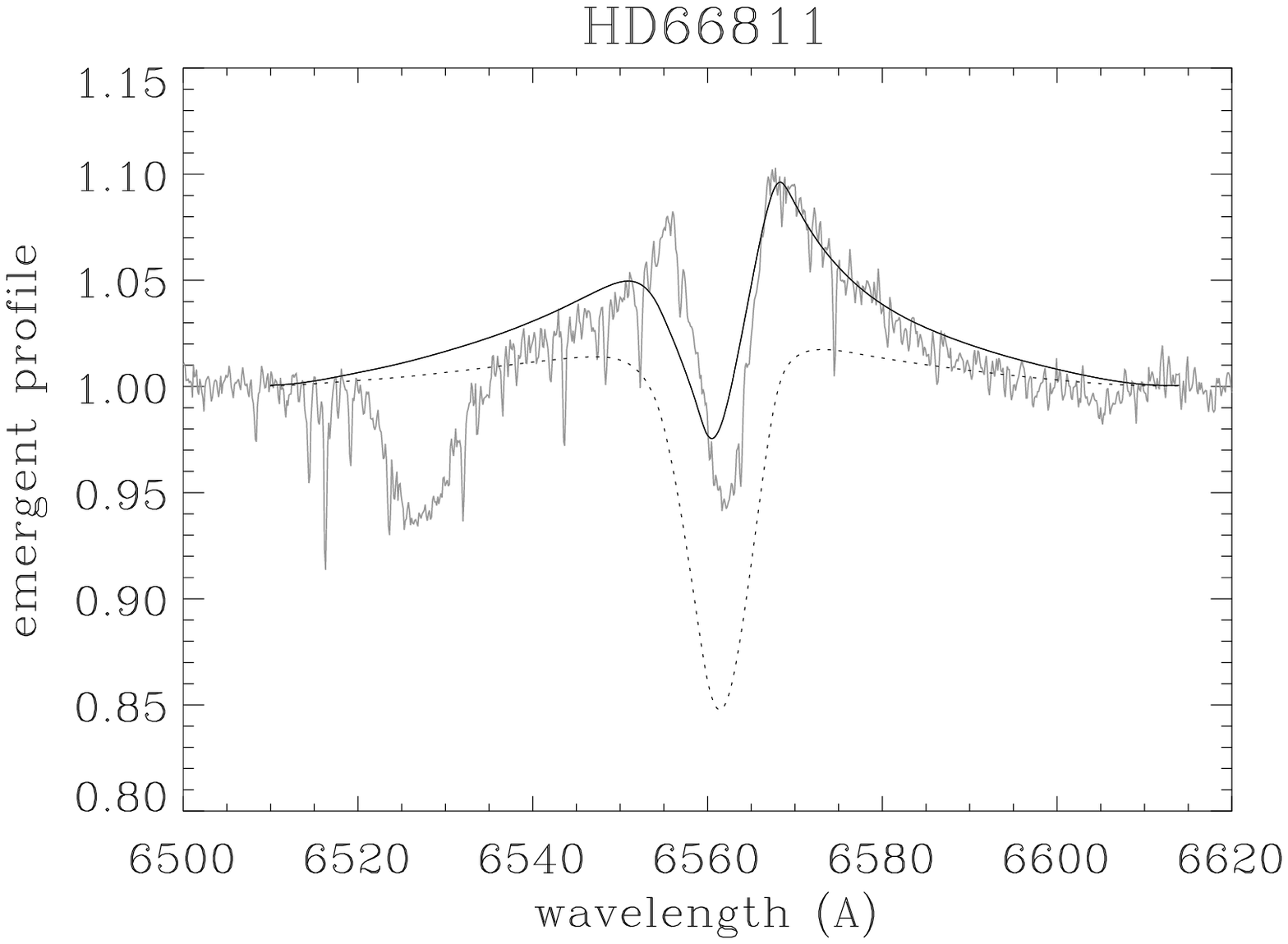}}
\end{minipage}
\hfill
\begin{minipage}{8.8cm}
   \resizebox{\hsize}{!}
   {\includegraphics{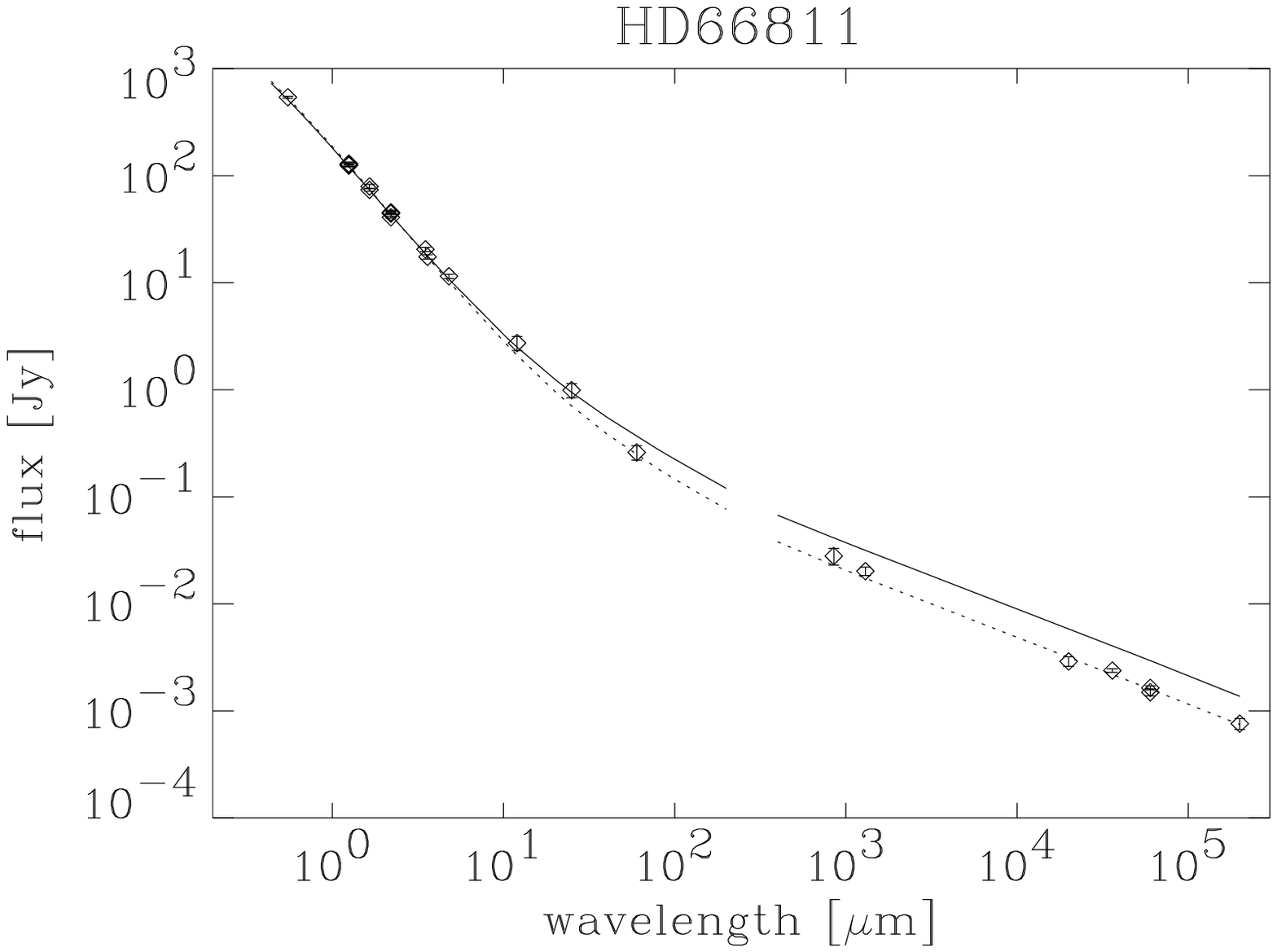}}
\end{minipage}
\caption{{\it Homogeneous} models for $\zeta$~Pup which {\it either} fit \Ha\
(\Mdot\ = 13.5 \Mdu, solid; cf. Table~\ref{haconsist_table}) {\it or} the radio
range (\Mdot\ = 8.5 \Mdu, dotted). A simultaneous fit cannot be achieved.
(Regarding the ``gap'' between 0.2 and 0.4 mm in the theoretical predictions,
see Sect.~\ref{radiofluxes}).}
\label{hd66811_1}
\end{figure*}

With respect to our models now, the inclusion of clumping effects in the
spirit as described above becomes very simple. Since all opacities entering
the calculations (bound-free, free-free and the \Ha\ line opacity) are
dependent on $\rho^2$, they are multiplied with a pre-described
clumping factor, whereas the corresponding source functions remain free from
such a manipulation, which is also true for the electron scattering
component, being proportional to $\rho$. Despite our caveats, we assume 
($i$) the clumps to be optically thin in \Ha\ and the IR/radio continuum, and
($ii$) the inter-clump matter to be void, since, anticipating our following
results, there is no need to require the inhomogeneity to start already from
the wind base on. 

In summary, our procedure is equivalent to other approaches used in the
literature (e.g., any analysis performed with {\sc cmfgen}), so that the
results can be easily compared.

Since we want to obtain constraints on the radial stratification of the
clumping factor, it would be dangerous to use a pre-prescribed law, and to
adapt only the parameters of such a law. An optimum solution would leave the
run of the clumping factor completely unconstrained, and would derive this
quantity at {\it all} depth points from a maximum likelihood method (or
other optimization algorithms) by fitting to the observed data. In view of
our interactive procedure, and particularly because of our desire to also
elaborate on the allowed range of the various possibilities\footnote{Note
that, e.g., the velocity-law-index, $\beta$, and the run of the clumping
factor are interrelated, and that for most of our objects observational data
in the far-IR are missing.}, we follow a simplified philosophy, by defining 
five different regions of the stellar wind with corresponding {\it average}
clumping factors, denoted by
\begin{center}
\tabcolsep1.45mm
\begin{tabular}{c|ccccc}
region & 1   & 2   & 3   & 4  & 5\\
\hline
r/$\Rstare$ & 1 \ldots $\rin$ & $\rin$ \ldots $\rmid$  & $\rmid$ \ldots $\rout$ & 
$\rout$ \ldots $\rfar$  & $ >\rfar$ \\
$\fcl$ & 1 & $\fin$ & $\fmid$ & $\fout$ & $\ffar$   
\end{tabular}
\end{center}
The boundaries of these regions and the clumping factors can be adapted
within our procedure. The first region with fixed clumping factor, $\fcl=1$,
has been designed to allow for a lower, unclumped wind region, in accordance
with theoretical predictions and our argument from above (namely that any
instability needs some time to grow before significant structure is formed).
But note also that by choosing $\rin = 1$ we are alternatively able to 
simulate a wind where the medium is clumped from the wind base on. 

Typical values for $\rin, \rmid$, $\rout$ and $\rfar$ are 1.05, 2, 15 and
50, respectively. For not too thin winds, this corresponds to the major
formation zones of \Ha\ (region 1 and 2), the mid-/far-IR (region 3), the mm
range (region 4) and the radio-flux (region 5). Note that for a number of
test cases we have used different borders, and sometimes combined region 4
and 5 into one outer region. All clumping factors derived in the following
are average values regarding the different regions, which admittedly are
rather extended. In almost all cases, however, with such a low number of
regions consistent fits could be obtained, with rather tight constraints on
the {\it global} behaviour of the clumping factor.

As a final comment, we like to stress a fact which has been mentioned
already in Sect.~\ref{intro}. Since (except for electron scattering) all
diagnostics used in this investigation have the same dependence on the
clumping properties, we are not able to derive {\it absolute} values for the
clumping factors, but only {\it relative} numbers. Note at first that in the
case $\rin =1$ all results derived for $\fcl(r)$ could be multiplied with an
arbitrary factor, if in parallel the mass-loss rate were reduced by the
square root of this value, without any loss in fit quality. The only
{\it physical} constraint is the requirement that the minimum value
(regarding all five regions) of the derived clumping factor must not be
lower than unity. The corresponding mass-loss rate is then the {\it largest
possible} one.

If, on the other side, $\rin \ne 1$, this scaling property is no longer
exactly preserved, because of the presence of an unclumped region not
affected by such a scaling. Since particularly the innermost core of \Ha,
but also the optical/near-IR fluxes (cf. Sect~\ref{dered}), are formed in
this region, they consequently deviate from this scaling. As it turned out
from the analysis performed in the next section, these deviations remain
fairly small, so that, unfortunately, the derivation of {\it absolute} 
values for $\fcl$ and \Mdot\ will require the use of different diagnostics.

\section{Constraints on the clumping factor: a combined \Ha, IR and radio
analysis}
\label{analysis}

\subsection{Two prototypical test cases: $\zeta$ Pup and HD\,209975}
\label{proto}

In this section, we will discuss two prototypical cases in some detail 
before presenting the results for our complete sample. We will consider
$\zeta$~Pup as a representative for a high-density wind, with \Ha\ in
emission (this star has the best wavelength coverage available within our
sample, including fluxes at 25, 60~$\mu{\rm m}$, 0.85, 1.3~mm and 20~cm), 
and HD\,209975 as a representative for a moderate-density wind (\Ha\ in
absorption).

\paragraph{$\zeta$~Pup.} In the following, we will usually display the
results of our simulations as done in Fig.~\ref{hd66811_1}, namely comparing
the observations and simulations for \Ha\ in parallel with the IR/radio
range. Fig.~\ref{hd66811_1} immediately shows the dilemma typical for all our
objects with \Ha\ in emission: the best fit for \Ha\ requires a mass-loss
rate typically twice as large as for the radio domain, if homogeneous models are
used. The far-IR fluxes are also closer to the low-\Mdot\ solution than
to the \Ha-fitting one. Let us point out already here that this finding is
in agreement with a recent comparison of consistent\footnote{i.e., using
identical stellar parameters and distances.} \Ha\ and radio mass-loss rates
performed by \citet{fulli06}, who found the same factor-of-two discrepancy
for a large number of objects. 

The derived radio mass-loss rate is considerably larger than the
corresponding result from \citet{blomme03} (using the same data set), due to
different parameters (larger distance and larger helium abundance adopted
here). With identical parameters, on the other hand, we obtain similar
results, \Mdot\ = 3.7 \Mdu, compared to 3.5 \Mdu. Note also the (small) flux
excess in the mm-range (with respect to the radio fluxes from a smooth
model, dotted line), in agreement with the findings by Blomme et al.

\begin{figure*}
\begin{minipage}{8.8cm}
\resizebox{\hsize}{!}
   {\includegraphics{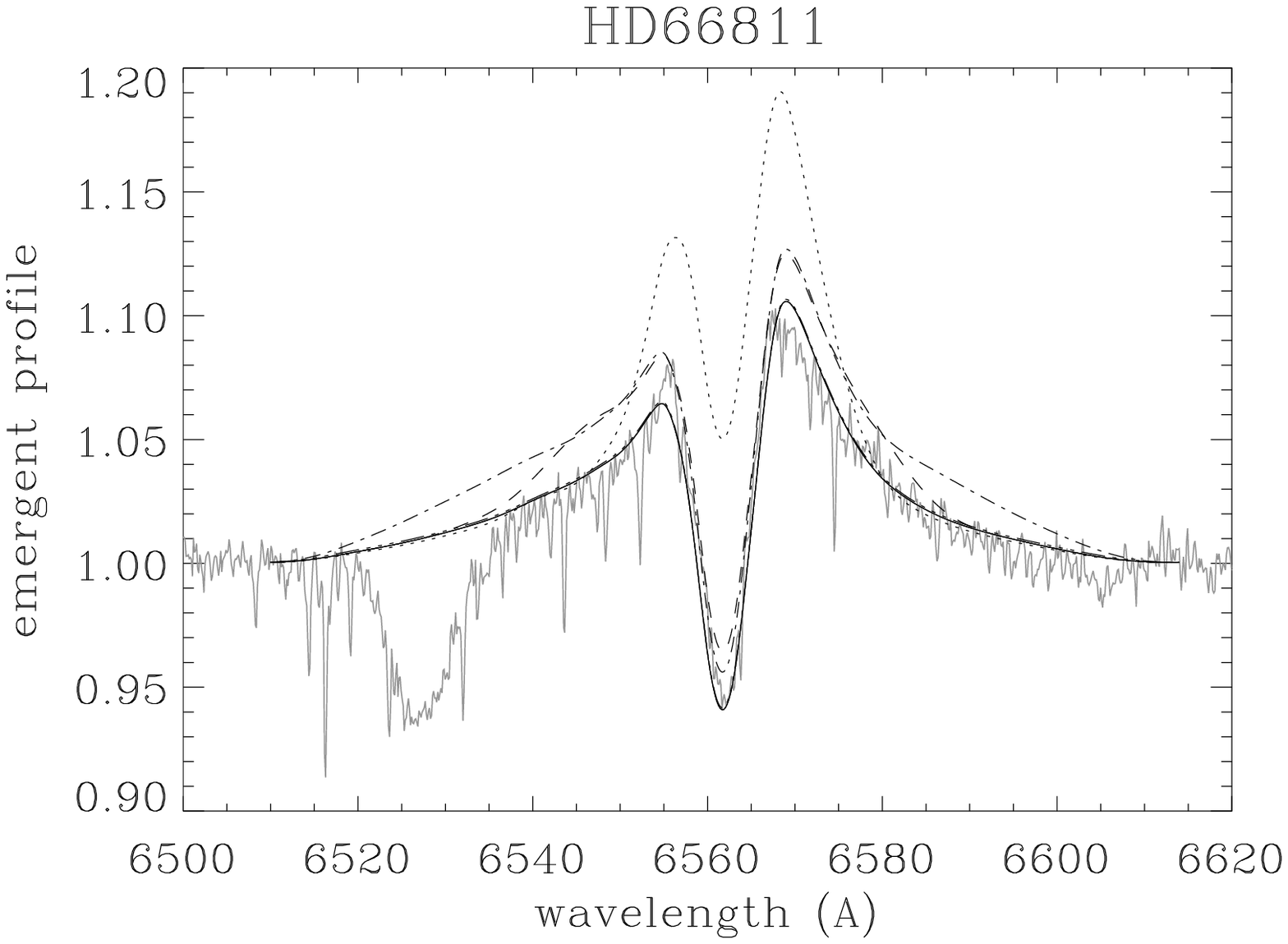}}
\end{minipage}
\hfill
\begin{minipage}{8.8cm}
   \resizebox{\hsize}{!}
   {\includegraphics{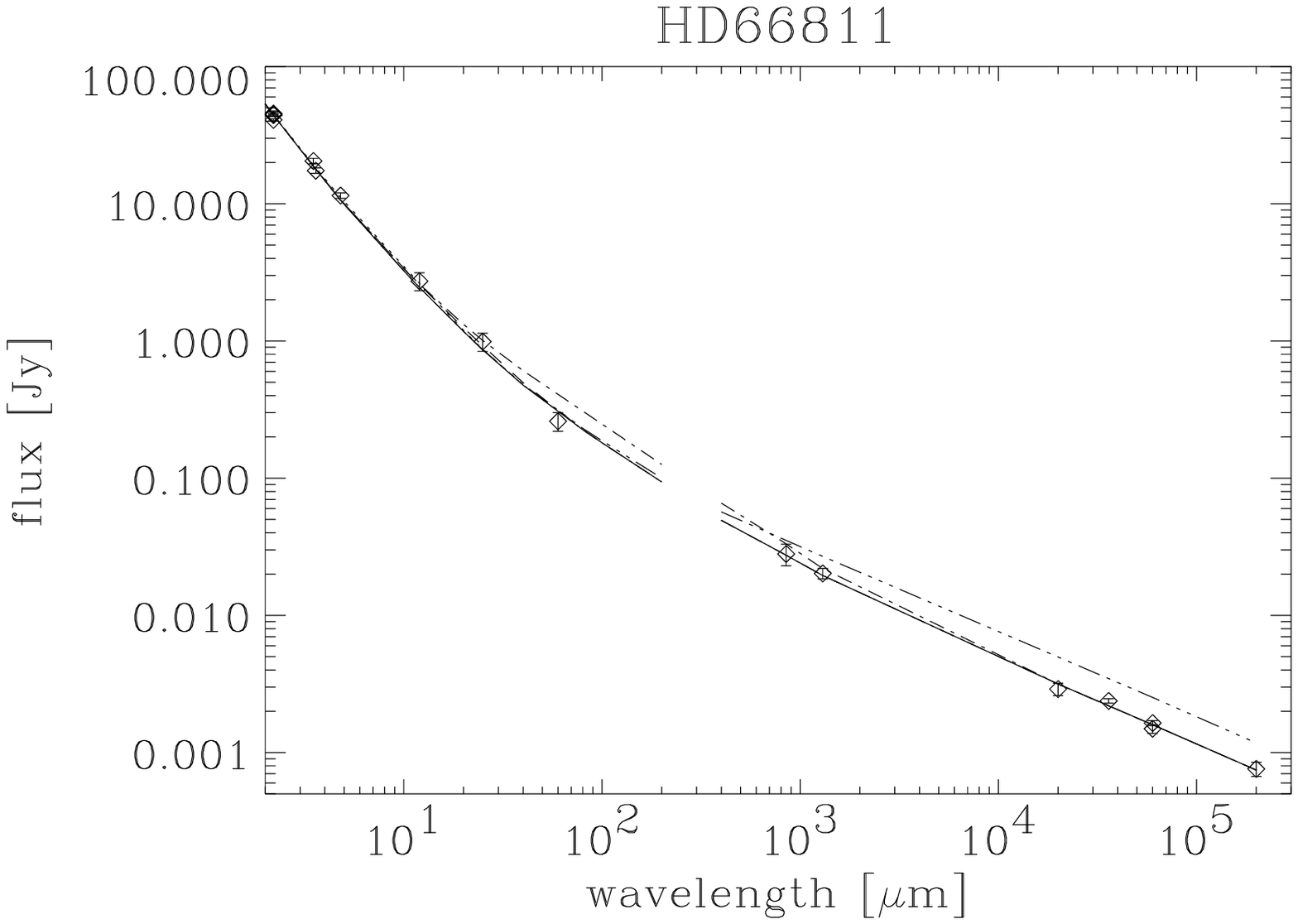}}
\end{minipage}
\caption{Clumped models for $\zeta$~Pup, compared to \Ha (left) and the
IR/radio continuum (right). The best-fitting model (cf.
Table~\ref{tab_zp_clumped}, first entry) is displayed in bold. Other curves:
variation of the clumping factor in individual regions, by a factor of two;
dotted: $\fcl(1.12{\ldots}1.5$ \Rstar) 5.5\rarrow11; dashed:
$\fcl(1.5{\ldots}2$ \Rstar) 3.1\rarrow6.2; dashed-dotted:
$\fcl(2{\ldots}15$~\Rstar) 2\rarrow4; dashed-dotted-dotted: $\fcl(>15$
\Rstar) 1\rarrow2. Note that \Ha\ remains sensitive to {\it all variations}
except for the last one. The mid-/far-IR, on the other hand, is sensitive
``only'' to variations in the range $2{\ldots}15$ \Rstar.} 
\label{hd66811_2}
\end{figure*}

Table~\ref{tab_zp_clumped} and Fig.~\ref{hd66811_2} (bold), on the other
hand, display our best solution for a clumped model which {\it consistently}
reproduces \Ha\ and the complete IR/radio band in parallel. In the spirit as
outlined above, the mass-loss rate has been chosen from the region with
lowest clumping, which in this case is the radio domain. By {\it setting}
$\ffar$ to unity then, the adopted mass-loss rate is the {\it largest
possible} one and corresponds to the ``homogeneous'' radio mass-loss rate,
\Mdot = 8.5 \Mdu, cf. Fig.~\ref{hd66811_1}, right panel. In this case,
the \Ha-forming region displays a typical clumping factor of 5.5 (from $r$ =
1.12 to 1.5) to 3.1 (from $r$ = 1.5 to 2), and $\beta$ has been adapted to
0.7 to provide a perfect \Ha\ fit.

\begin{figure}
\resizebox{\hsize}{!}
   {\includegraphics{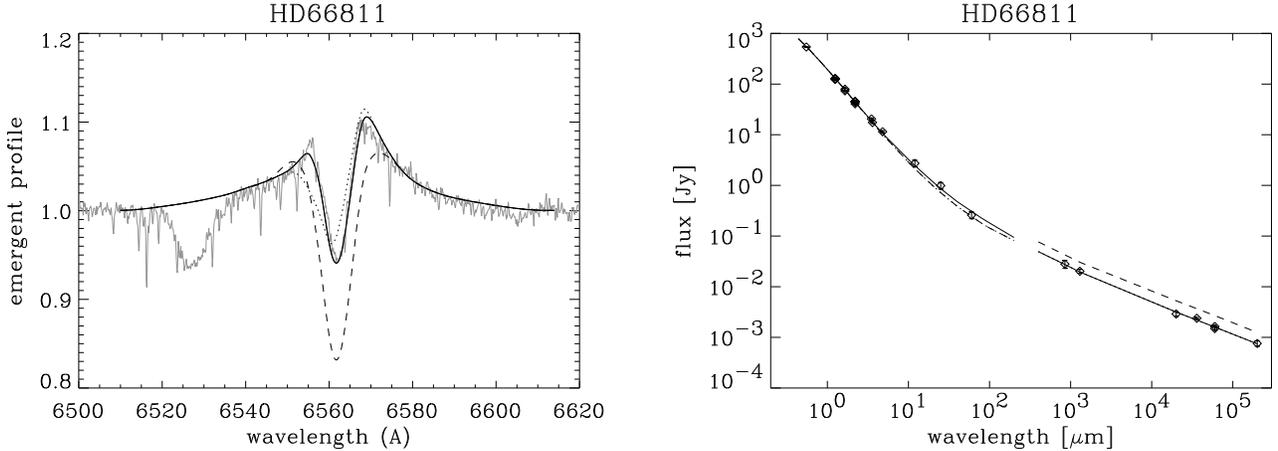}}
\caption{Clumped models for $\zeta$~Pup: influence of a different onset of
clumping on \Ha. Solid: best-fitting model, $\rin = 1.12$ \Rstar; dotted:
$\rin$ = \Rstar, i.e., clumping starting at the wind base; dashed: $\rin =
1.3$ \Rstar.}
\label{hd66811_3}
\end{figure}

Fig.~\ref{hd66811_3} displays the change in \Ha\ when a different onset of 
clumping was chosen. If $\rin$ were 1.3 (dashed profile), the central
emission would be missing, whereas for $\rin$ = 1.0 (dotted profile,
corresponding to a model which is clumped from the wind base on), the
absorption trough is not perfectly reproduced: the position
of maximum depth is located at too high velocities, and the trough becomes 
too broad, resembling our best solution for the homogeneous model. 

\begin{table}
\caption{Clumping factors, boundaries of different regions and mass-loss
rates (in units of $10^{-6}{\rm M_{\odot}/yr}$) for three equally well
fitting models of $\zeta$~Pup (strong wind, \Ha\ in emission, $\beta=0.7$)
and for our best fitting models of HD\,209975 (moderate wind, \Ha\ in
absorption, $\beta=0.9$). The first solution for $\zeta$~Pup (which
optimizes \Ha) is displayed in the following figures, whereas the second one
is almost indistinguishable from the first, though slightly worse in \Ha,
and slightly better in the mid-/far-IR and mm range. In the 3rd model it is
assumed that helium remains doubly ionized everywhere. Note the large
difference between the clumping properties of the two stars.}
\tabcolsep1.3mm
\begin{tabular}{c|c|c|c|c|c|c|c|l}
reg. & 1   & \multicolumn{2}{c|}{2}   & 3   & \multicolumn{2}{c|}{4/5} &
\Mdot & comment\\
\hline
\multicolumn{9}{c}{HD\,66811} \\
\hline
$r/\Rstare$  & $<$ 1.12 & $<$ 1.5 & $<$ 2 & $<$ 15 & \multicolumn{2}{c|}{$>$ 15} &          & best fit \\
$\fcl$       & 1        & 5.5       & 3.1     & 2    & \multicolumn{2}{c|}{1}      & \rb{8.5} & for \Ha \\   
\hline
$r/\Rstare$  & $<$ 1.12 & \multicolumn{2}{c|}{$<$ 2} & $<$ 15  & $<$ 50 & $>$ 50 &          & best fit for \\
$\fcl$       & 1        & \multicolumn{2}{c|}{5}   & 1.5       & 1.4    & 1   & \rb{8.5} & far-IR/mm \\   
\hline
$r/\Rstare$  & $<$ 1.12 & $<$ 1.5 & $<$ 2 & $<$ 15 & \multicolumn{2}{c|}{$>$ 15} &          & \HeIII \\
$\fcl$       & 1        & 11.8    & 10   & 2      & \multicolumn{2}{c|}{1}    & \rb{5.8} & everywhere \\   
\hline
\hline
\multicolumn{9}{c}{HD\,209975} \\
\hline
$r/\Rstare$  & $<$ 1.05 & $<$ 1.5 & $<$ 2 & $<$ 15 & \multicolumn{2}{c|}{$>$ 15} &          &  \\
$\fcl$       & 1        & 1       & 1-2  & 1-1.5 & \multicolumn{2}{c|}{1.3}    & \rb{1.2} & identical  \\   
\cline{1-8}
$r/\Rstare$  & 1        & \multicolumn{2}{c|}{$<$ 2} & $<$ 10  & $<$ 50 &$>$ 50  &          & fit quality \\
$\fcl$       & 1        & \multicolumn{2}{c|}{1}     & 1-1.5  & 1-10  & 1.3    & \rb{1.2} & \\   
\hline
\end{tabular}
\label{tab_zp_clumped}
\end{table}

From our arguments given at the end of Sect.~\ref{clumping}, it should be
clear that in particular the latter solution is not unique,
since an alternative model with {\it all} clumping factors multiplied by an
arbitrary factor $f$, in parallel with a mass-loss rate reduced by a factor
of $1/\sqrt f$, would result in an identical fit. If, on the other hand, the
perfectly matched absorption trough for our model with $\rin$ = 1.12 were
actually due to a clumping-free lower wind base (and not coincidentally
matched due to somewhat erroneous departure coefficients and/or the specific
observational snapshot\footnote{Concerning the temporal variability of \Ha\
in $\zeta$~Pup, see \citet{reid96} and references therein,
\citet{pulsetal93} and \citet{bergetal96}. From these data-sets, a moderate
variability of the absorption trough is visible indeed.}), such a scaling
would no longer work (because of the presence of an unclumped region), and
our solution would become ``almost'' unique, at least regarding the clumping
properties of the inner wind.

The ``almost'' refers to the fact that a different distribution of the
individual regions, combined with somewhat different clumping factors, gives
fits of similar quality. The 2nd entry of Table~\ref{tab_zp_clumped} is such
an example. In this case, we have combined the region between $r$ = 1.12 to
2 into one region, whereas we have split the outer region, beyond $r$ = 15,
into two regions, with a border at $r$ = 50. To fit \Ha\ (with a slightly
worse quality than displayed in Fig.~\ref{hd66811_2}), the innermost
clumping factors had to be reduced (from 5.5 and 3.1 to an average factor of
5.0), whereas, by adapting the clumping factors in the middle and outer
part, the fit quality at 60~$\mu{\rm m}$ becomes perfect and the quality at
0.85/1.3~mm remains preserved Note, however, that the overall stratification
of the clumping factors is rather similar.

Fig.~\ref{hd66811_2} displays the advantage of fitting \Ha\ and the IR/radio
range in parallel. Although the primary formation region of \Ha\ is below 2
\Rstar, it also remains sensitive to variations of the clumping factors in
the intermediate wind, $r \la 15$, as can be seen from the reaction in the
line wings if $\fcl$ is doubled from 2 to 4 (dashed-dotted profile). Of
course, a variation of the clumping factors in the inner regions (dotted and
dashed) has even more impact. On the other hand, as displayed in the right
panel of this figure, the IR/radio band reacts complementarily to variations
beyond $r$ = 2, although only from the mid-IR on ($\lambda \ga 10$~$\mu{\rm
m}$). Thus, a combined analysis is able to provide tight constraints on the
largest possible mass-loss rate and to scan the complete stratification of
$\fcl(r)$ (at least differentially, i.e., modulo a constant factor) if the
far-IR is well observed. Concerning the possible degeneracy of clumping factors
and $\beta$, we refer the reader to Sect.~\ref{errors}.

\begin{figure}
\resizebox{\hsize}{!}
   {\includegraphics{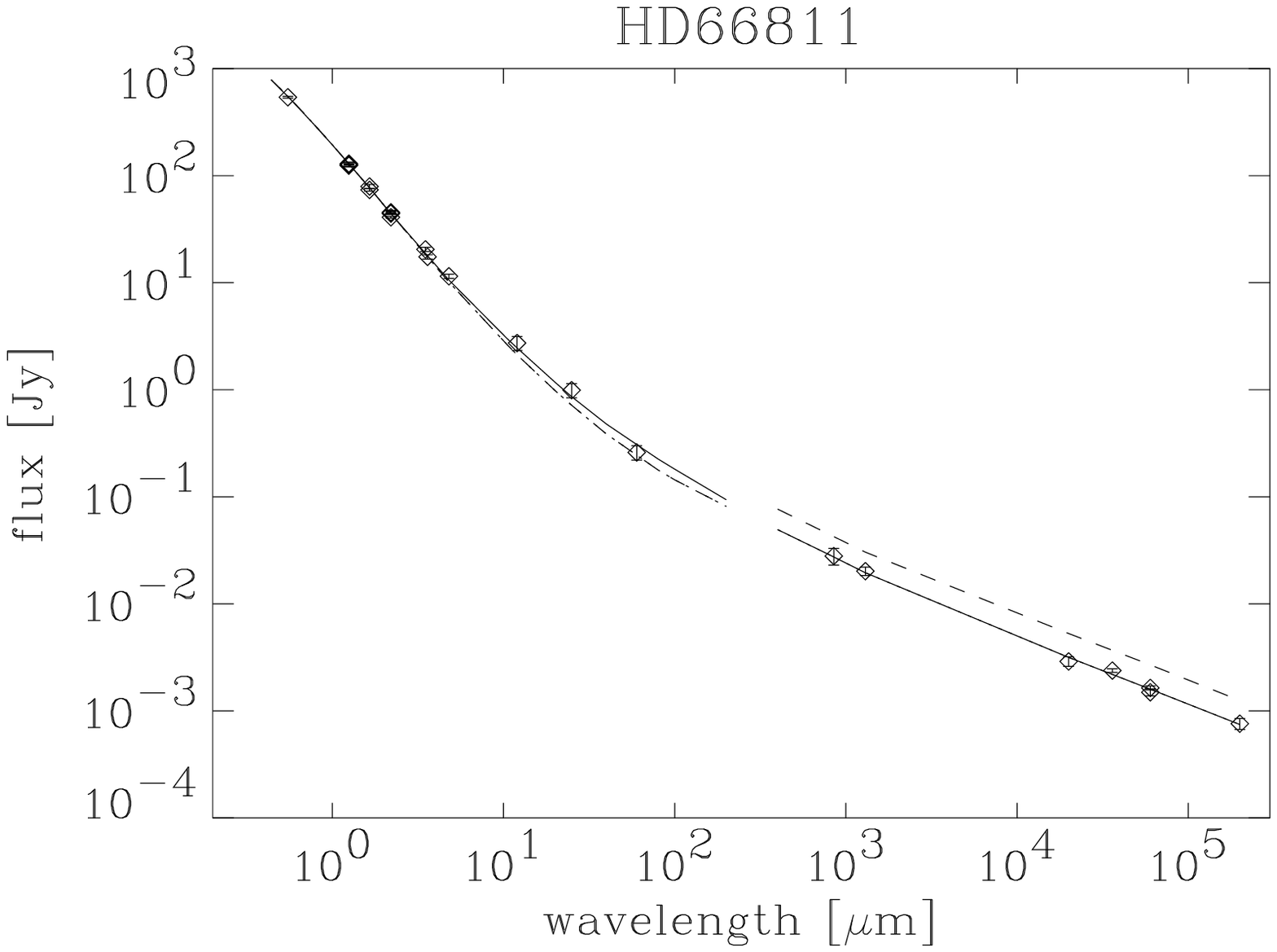}}
\caption{Clumped models for $\zeta$~Pup in the IR/radio band: influence of
helium ionization. Solid: best fitting model, with \HeIII\ as the major ion
for $v < v_{\rm rec} = 0.86$ (5.3~\Rstar), and \HeII\ as the dominant
ionization stage outside $v_{\rm rec}$; dashed-dotted: \HeII\ as the major ion
everywhere; dashed: \HeIII\ as the major ion in the radio emitting domain.}
\label{hd66811_4}
\end{figure}

Fig.~\ref{hd66811_4}, finally, displays the possible error if the helium 
ionization were different to that assumed here (cf. Sects.~\ref{irfluxes}
and \ref{radiofluxes}.) If helium were singly ionized throughout the
complete wind (instead of recombining only at $v_{\rm rec}$ = 0.86), the
synthetic 10 and 20~$\mu{\rm m}$ fluxes in particular would become too low;
compensating for this effect by increasing clumping factors is not possible,
because \Ha\ would then no longer be fit. If, on the other hand, helium were
to remain doubly ionized in the outermost region also, the radio/mm (and the
far-IR fluxes) would become larger than observed; in this case, a reasonable
fit is still possible, by lowering the mass-loss rate and increasing the
inner clumping factors (with a factor roughly corresponding to $(\Mdote_{\rm
old}/\Mdote_{\rm new})^2$). The parameters for such a model (which fits
both \Ha\ and the entire IR--radio range) is given in Table~\ref{tab_zp_clumped},
3rd entry. The rather large difference in the resulting (maximum) mass-loss
rate (factor 0.7) and clumping factors is due to the fact that our model of
$\zeta$~Pup has a helium content which is twice solar, \YHe\ = 0.2. For
solar helium abundance, as is typical for most of the other objects of our
sample, the corresponding factor would be 0.85, as outlined in
Sect.~\ref{radiofluxes}. Note again, however, that it is rather improbable
that helium is still doubly ionized in the radio-forming region. From the
consistency of the mm and radio fluxes, it is also clear then that 
the Helium ionization must be similar in the mm and the radio forming region, 
in agreement with our predictions for $v_{\rm rec}$.

\begin{figure*}
\begin{minipage}{8.8cm}
\resizebox{\hsize}{!}
   {\includegraphics{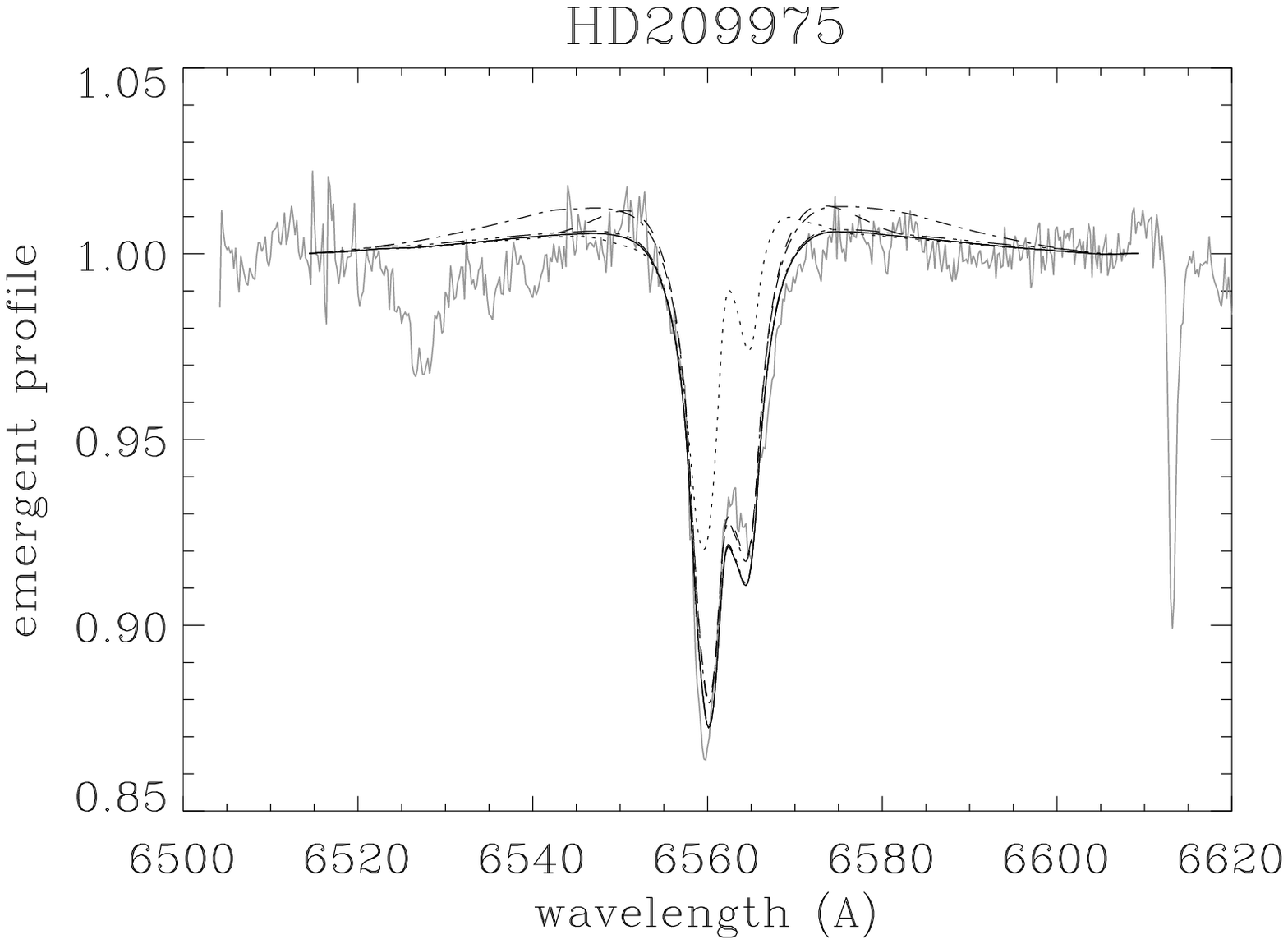}}
\end{minipage}
\hfill
\begin{minipage}{8.8cm}
   \resizebox{\hsize}{!}
   {\includegraphics{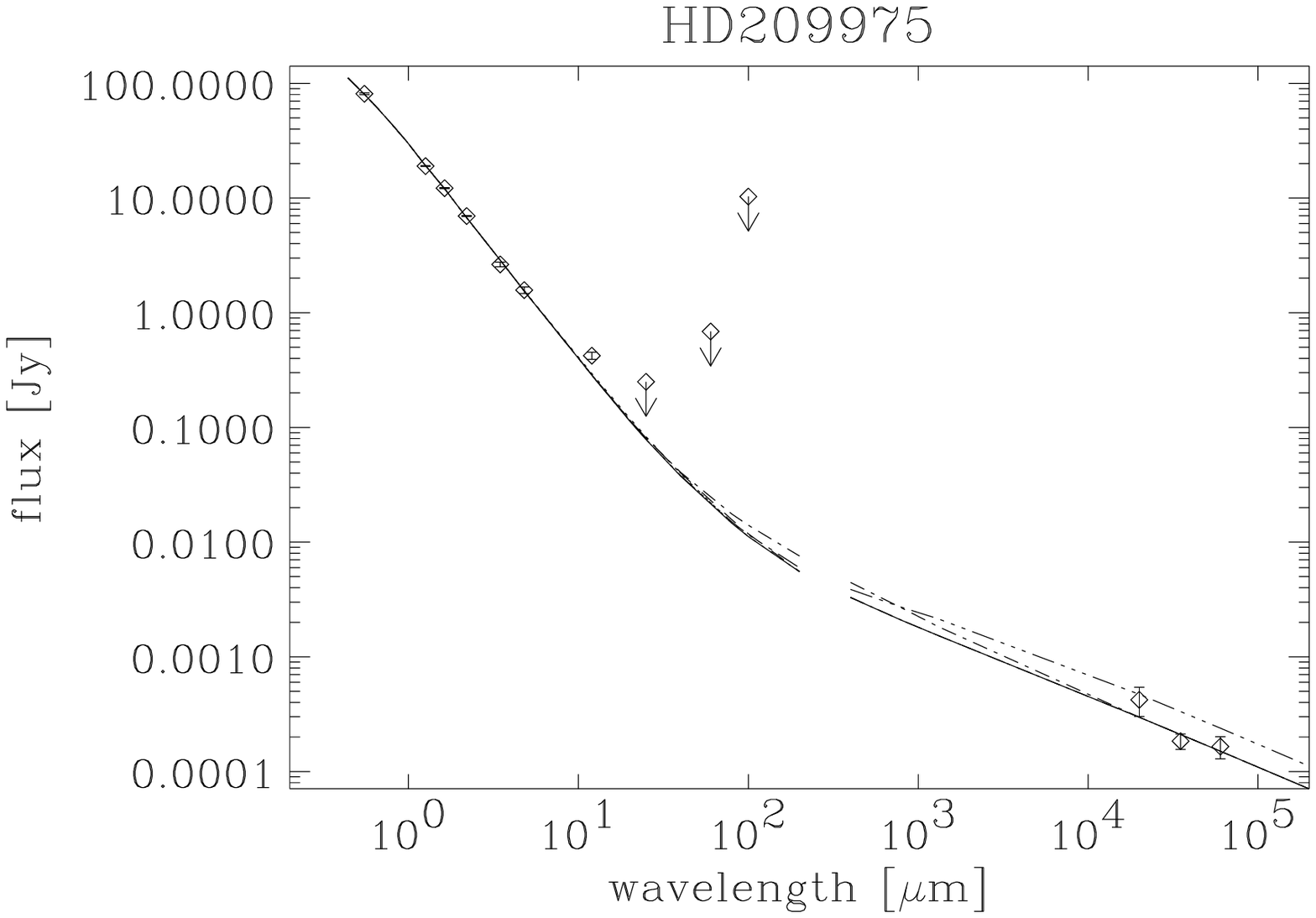}}
\end{minipage}
\caption{As Fig.~\ref{hd66811_2}, but for HD\,209975.  The best-fitting
model (with {\it all} clumping factors at or close to unity) is displayed in
bold. Other curves show the effects of varying, by a factor of 2, the
clumping factors in individual regions alone. Dotted: $\fcl(1.05{\ldots}1.5$
\Rstar) 1\rarrow2; dashed: $\fcl(1.5{\ldots}2$ \Rstar) 1\rarrow2;
dashed-dotted: $\fcl(2{\ldots}15$~\Rstar) 1\rarrow2; dashed-dotted-dotted:
$\fcl(>15$ \Rstar) 1.3\rarrow2.6. Again, \Ha\ remains sensitive to
variations below $r$ = 15 \Rstar\ (but see text), whereas the far-IR (not
constrained by observations) is mostly sensitive to variations in the range
2{\ldots}15 \Rstar. Note that the dashed solution is also consistent with
the observations.}
\label{hd209975_1}
\end{figure*}

\paragraph{HD\,209975.} Table~\ref{tab_zp_clumped} and Fig.~\ref{hd209975_1}
display the results of our combined fit procedure for this star, which has a
moderate wind density and \Ha\ in absorption. Again, we have indicated the
resulting profiles/fluxes when the derived clumping factors are varied by a
factor of two in specific regions, to check for their sensitivity. Most
interestingly, {\it this object can be fitted with almost constant clumping
factors throughout the wind}, in stark contrast to the above example. Indeed,
with slightly different \Mdot\ and $\beta$, an almost equally perfect fit is
possible with all clumping factors being unity. If at all, the (homogenous)
radio mass-loss rate is somewhat higher than the mass-loss rate derived from 
\Ha, so that in this case $\fin$ is set to unity.

Note that a moderate clumping factor of 2 for $1.5 < r < 2$ is still
consistent with the data, and that due to missing far-IR information (the
indicated data denote upper limits derived by IRAS), the clumping in the
intermediate wind remains somewhat unconstrained. After some
experimentation, it turned out that the data are also consistent with a
moderately clumped wind ($\fcl = 10$) in the region $10 < r < 50$, or a
weakly clumped wind ($\fcl =2$) in the region $3 < r < 50$ (not quoted in
Table~\ref{tab_zp_clumped}). Only for the outermost wind ($r > 50$), do the
clumping properties {\it have} to be similar to the inner wind conditions. 

Since the innermost wind has the lowest clumping, no statement concerning
its onset is possible within our approach. Thus, any scaled solution ($\fcl$
multiplied with $f$, \Mdot\ reduced by $1/\sqrt f$) provides an equally
perfect fit and cannot be excluded. 

The second entry for HD\,209975 in Table~\ref{tab_zp_clumped} refers to our
``standard'' division of the different regions used for winds with \Ha\ in
absorption, namely $\rmid=2$ and $\rout=10$. This scheme accounts for the
fact that in moderate/low density winds the IR and radio emission is formed
closer to the star. As can be seen from the best fitting clumping factors,
the results do (almost) not depend on details of the specific borders.

In summary, the inner and outer wind of this object have similar clumping
properties, whereas far-IR observations are required to constrain the
intermediate region.

\subsection{Clumping properties of the complete sample}

Before discussing the results of our analysis for the complete sample, let
us point out some general findings, and remind the reader that the derived 
clumping factors are independent of any uncertainty concerning radius and
distance, since all our diagnostics (\Ha/radio/IR) scale in an identical way
with respect to these quantities, cf. Eq.~\ref{radioflux_approx} and
corresponding discussion.

\paragraph{The core of \Ha\ as a tracer of wind clumping below $r \approx 2 
\Rstare$.} Our simulations show that the strength of the core of \Ha, 
whether in emission or in absorption, is quite sensitive to the value of the
clumping factor in the inner part of the wind, and thus can be used to
determine this parameter out to distances of about $r \approx 2$. If one
relies on the value of $\beta$ as derived by means of unclumped models, the
corresponding (average) clumping factors are very precise, with an accuracy
of roughly 10\% (but see Sect.~\ref{errors}). Note particularly that
clumping factors $\fin$ of order 2 or larger {\it are still visible for
objects with \Ha\ in absorption} (see Fig.~\ref{hd209975_1}, dotted
profile).

\paragraph{Constraints on the clumping factor beyond $r \approx 2 
\Rstare$.} In addition to constraining the clumping properties in the lower
wind, \Ha\ can even serve as an indicator of wind clumping in layers beyond
$r \approx 2$ (e.g, Fig.~ \ref{hd66811_2}, left panel). How much beyond? The
answer depends, of course, on the specific wind density, but some general
statements for stars with \Ha\ in emission are possible though. Usually, we
found that reducing the extent of the intermediate clumped region~3 from
$\rout=15$ to about $\rout=3$ has a noticeable effect on the strength of the
\Ha\ emission wings. The same is true if the boundary of region~2,
$\rmid=2$, is extended to a value of $\rmid = 3$. The effect becomes
visible when the outer boundary moves from $\rout=15$ to $\rout=5$ and is
insignificant if $\rout$ is set to 8 stellar radii instead of $\rout=15$. 

For those objects with \Ha\ in emission and missing far-IR/mm information, 
in Table~\ref{table_clf}, column 14, we have indicated the outermost radius,
$\rout'$, to which \Ha\ {\it alone} can provide information on the clumping
factor, on the assumption that the region, $\rout' < r < \rfar$, is
``unclumped'' (or, more precisely, has the same clumping properties as
region~5). In parallel, we also quote the corresponding value, $\fmid$,
which is somewhat larger than the original one (for $\rout=15$), due to the
reduced width of region 3. Indeed, for almost all objects, $\rout'$ is of
the order of 5 \Rstar, except for HD\,14947 and HD\,192639, where \Ha\
provides information only out to 3 \Rstar. Thus, it is safe to conclude that
\Ha\ constrains the clumping factor up to distances of $r=3{\ldots}5
\Rstare$ {\it if in emission}. Note, however, that in some cases,
significant clumping in region 4 (from $\rout$ to $\rfar$) has an effect on
\Ha, which leads to an additional constraint on the clumping in this region.

For objects with \Ha\ in absorption, on the other hand, the intermediate
region remains much less constrained (Fig.~\ref{hd209975_1}, left panel),
and we will comment below on the corresponding limits.

\begin{table*}
\caption{Clumping properties as derived from our combined \Ha/IR/radio
analysis. Stars are ordered according to \Ha\ profile type (``pt'') and
spectral type. Entries with {\bf name} in bold are objects with extremely
well-constrained clumping parameters.
\newline
\Teff\ is given in kK, and $\Mdote_{\rm cl}$ is the largest possible mass-loss
rate, in units of $10^{-6}{\rm M_{\odot}/yr}$. ``ratio'' gives the ratio of
``clumped'' mass-loss rate to optical results using unclumped models (cf.
Table~\ref{sample}). $\beta_{\rm cl}$ is the velocity field exponent as
derived or adopted here, $v_{\rm rec}$ and $r_{\rm rec}$ are the velocity (in
units of \vinf) and radius where He recombines (see Sect.~\ref{irfluxes}),
respectively, and $r(\tau_2)$ is the radius where the radio continuum becomes 
optically thick\protect\footnotemark\ at 2\,cm ($r_{\rm rec}$ and 
$r(\tau_2)$ in units of \Rstar).
\newline
Clumping factors and boundaries are defined as in Sect.~\ref{clumping}. For
all models, region~1 with $\fcl=1$ (not tabulated) extends from $r=1$ to
$\rin=1.05$, except for HD\,66811 where $\rin=1.12$, and $\rfar$ (defining
the border between region 4 and 5) has been set to 50 \Rstar\ always. For
objects with \Ha\ in emission or of intermediate type, and missing far-IR/mm
data, $\rout'$ (with corresponding clumping factor) indicates the maximum
radius to which \Ha\ {\it alone} can provide constraints on the clumping, on
the assumption that the outer wind is ``unclumped'' (see text). For objects
with \Ha\ in absorption, $\fmidm$ gives the maximum possible clumping factor
in region 3, which is still consistent with the data. $\foutm$ is defined
similar to $\fmidm$, but for region 4. For comments on individual objects
and corresponding fits, see Appendix~\ref{comments}.}

\tabcolsep1.6mm
\begin{tabular}{lcrcrrrrrr|rr|rrr|rr|r}
\hline
\multicolumn{10}{c|}{} & \multicolumn{2}{c|}{region 2} & 
\multicolumn{3}{c|}{region 3} & \multicolumn{2}{c|}{region 4} & reg. 5 \\
{\rule[-2mm]{0mm}{6mm}Star} & pt& \Teff& \Mdot$_{\rm cl}$ & ratio & 
$\beta_{\rm opt}$& $\beta_{\rm cl}$ & $v_{\rm rec}$ &$r_{\rm rec}$ & $r(\tau_2)$ 
& $\fin$ &  $\rmid$ & $\fmid$ &  $\fmid(\rout')$ & $\rout$ &  $\fout$ & $\foutm$ & $\ffar$ \\ 
\hline
  Cyg\,OB2\#7    & e & 45.8 &  $\le$4.0$^{\rm a,b}$ &  0.38 & 0.77 & 0.90 & 1.00 &    inf &   29.7 &   5.0 &   2.0 &  4.0-6.0 & 7.0(5)      &  15 &   1.0 &  10.0 &   1.0\\
    HD\,190429A  & e & 39.2 &  9.5                  &  0.59 & 0.95 & 0.95 & 0.85 &    6.2 &   49.6 &   3.0 &   2.0 &   3.0 & 3.5(5)      &  15 &   1.0 &   2.0 &   1.0\\
                 &   &      &  7.5                  &  0.46 &      &      &      &        &        &   5.0 &   2.0 &   5.0 & 5.8(5)      &  15 &   1.0 &   2.0 &   1.0\\
{\bf HD\,15570}  & e & 38.0 &  6.5                  &  0.38 & 1.05 & 1.05 & 0.84 &    6.3 &   45.0 &   5.5 &   2.0 &   4.0-6.0 &             &  15 &  13.0 &  20.0 &   1.0\\
{\bf HD\,66811}  & e & 39.0 &  8.5                  &  0.51 & 0.90 & 0.70 & 0.86 &    5.3 &   36.1 &   5.0 &   2.0 &   1.5 &             &  15 &   1.4 &   1.8 &   1.0\\
                 &   &      &  4.2                  &  0.51 & 0.90 & 0.70 & 0.86 &    5.0 &   36.5 &   5.0 &   2.0 &   1.5 &             &  15 &   1.4 &   1.8 &   1.0\\
{\bf HD\,14947}  & e & 37.5 & 10.0                  &  0.59 & 0.95 & 0.95 & 0.81 &    5.0 &   37.9 &   3.1 &   2.0 &   2.5 & 4.0(3)      &  15 &   1.0 &   5.0 &   1.0\\
{\bf Cyg\,OB2\#11}& e& 36.5 &  5.0                  &  0.62 & 1.03 & 1.10 & 0.81 &    5.6 &   30.7 &   3.0 &   2.0 &   5.0 & 6.0(5)      &  15 &   1.0 &  15.0 &   1.0\\
{\bf HD\,210839} & e & 36.0 &  3.0                  &  0.38 & 1.00 & 1.00 & 0.83 &    5.9 &   24.7 &   6.5 &   4.0 &  10.0 &             &  15 &   1.0 &   8.0 &   1.0\\
  HD\,192639     & e & 35.0 &  $\le$3.0$^{\rm a}$   &  0.48 & 0.90 & 1.14 & 0.82 &    6.3 &   27.7 &   3.5 &   2.0 &   3.5 & 6.0(3)      &  15 &   1.0 &  10.0 &   1.0\\
{\bf HD\,30614}  & e & 29.0 &  1.5                  &  0.49 & 1.15 & 1.15 & 0.16 &    1.2 &   25.7 &   2.6 &   2.0 &   3.0 & 3.5(5)      &  15 &   1.0 &   4.0 &   1.0\\
\hline
 Cyg\,OB2\#8A    & i & 38.2 &  $\le$8.0$^{\rm c}$   &  0.71 & 0.74 & 0.74 & 0.84 &    4.7 &   33.6 &   2.5 & 2.0 & 1.0-2.0 & 2.5(3)      &  10 &   1.0 &  10.0 &   1.0\\
{\bf Cyg\,OB2\#10}& i& 29.7 &  2.74                  &  1.00 & 1.05 & 1.05 & 0.17 &    1.2 &   23.2 &   1.4 &   2.0 &   1.8 & 2.0(3)      &  10 &   1.0 &   4.0 &   1.0\\
\hline
                 &   &      &                        &       &      &      &      &        &        &       &       &       & $\fmidm$    &     &       &       &     \\
 Cyg\,OB2\#8C    & a & 41.8 &  $\le$3.5$^{\rm d}$   &  0.82 & 0.85 & 1.00 & 0.94 &   17.3 &   33.0 &   1.0 &   2.0 &   1.0 &   -         &  10 &   1.0 &   5.0 &   1.0\\
   HD\,34656     & a & 34.7 &  3.0                  &  1.15 & 1.09 & 1.00 & 0.60 &    2.5 &   28.2 &   1.0 &   2.0 &   1.0 &   -         &  10 &   1.0 &   8.0 &   6.0\\
   HD\,24912     & a & 35.0 &  $\le$2.3$^{\rm a}$   &  0.94 & 0.80 & 0.90 & 0.85 &    6.1 &   16.4 &   2.1 &   2.0 &   5.0 &   7.0       &  10 &   1.0 &   2.0 &   1.0\\
                 &   &      &  $\le$1.2$^{\rm a}$   &  0.49 &      &      &      &        &        &   8.0 &   2.0 &  20.0 &  25.0       &  10 &   1.0 &   3.0 &   1.0\\
{\bf HD\,203064} & a & 34.5 &  1.1                  &  1.12 & 0.80 & 0.90 & 0.57 &    2.2 &   23.3 &   1.0 &   2.0 &   1.0 &   2.0       &  10 &   1.0 &   8.0 &   1.0\\
   HD\,36861     & a & 33.6 &  $\le$0.4$^{\rm a}$   &  0.54 & 0.80 & 0.90 & 0.51 &    1.9 &   10.2 &   2.0 &   2.0 &   1.0 &  20.0       &  10 &   1.0 &   2.0 &   1.0\\
{\bf HD\,207198} & a & 36.0 &  1.0                   &  0.95 & 0.80 & 0.90 & 0.82 &    5.2 &   22.5 &   1.0 &   2.0 &   1.0 &   2.0       &  10 &   1.0 &  15.0 &   1.0\\
   HD\,37043     & a & 31.4 &  0.8                  &  0.78 & 0.85 & 0.90 & 0.29 &    1.3 &   14.4 &   1.0 &   2.0 &   1.0 &   4.0       &  10 &   1.0 &   2.0 &   1.0\\
                 &   &      &  0.25                  &  0.24 &     &       &      &        &        &  12.0 &   1.3 &   1.0 &  20.0       &  10 &   1.0 &  10.0 &   1.0\\
{\bf HD\,209975} & a & 32.0 &  1.2                  &  1.08 & 0.80 & 0.90 & 0.42 &    1.6 &   27.1 &   1.0 &   2.0 &   1.0 &   1.5       &  10 &   1.0 &  10.0 &   1.3\\
\hline
\end{tabular}

$^{a)}$ only upper limits of radio fluxes available; \Mdot\ maximum radio mass-loss rate. \\
$^{b)}$ He assumed to be recombined in radio region (see Appendix~\ref{comments}). \\
$^{c)}$ upper limit, since non-thermal radio emitter; \Mdot\ from 2\,cm flux. \\
$^{d)}$ \Mdot\ from \Ha, since radio fluxes (upper limits only) give larger value. 
\label{table_clf}
\end{table*}

\smallskip
\noindent
Table~\ref{table_clf} summarizes the results of our simultaneous
\Ha/IR/radio analysis for the two objects already discussed in
Sect.~\ref{proto} and for the remaining ones. We have ordered the sample
according to \Ha\ profile type and spectral type. For almost
all objects, we have used identical boundaries, $\rin=1.05$, $\rmid=2.0$ and
$\rfar=50$, to obtain comparable results. The default values for $\rout$
correspond to 15 (\Ha\ in emission) and 10 (\Ha\ in absorption or of
intermediate type), but have been adapted where necessary. Detailed comments
regarding the individual objects are given in Appendix~\ref{comments}, where
all fits are displayed as well. 

Overall, our simulations show that for stars with \Ha\ in emission, a
simultaneous fit of the observed radio fluxes and the shape and strength of
\Ha, requires clumping factors which are always higher in the \Ha-forming
region than in the radio-forming one. For stars with \Ha\ in absorption, the
situation seems to be different: in most cases, the required clumping
factors are of similar order in the inner and outer regions, as already
discussed for the case of HD\,209975. Note, however, that this preliminary
impression is dependent on the actual value of $\beta$, a problem
which will be discussed in our error analysis further below.

For all objects quoted with a definite mass-loss rate (and not only an upper
limit), this value represents the {\it largest possible value} (for given
\Rstar), usually derived from adopting an outer, unclumped wind with
$\ffar=1$ or, for weaker winds, $\fout$ = 1. These mass-loss rates
correspond to the ``usual'' radio mass-loss rate. Only for one object,
HD\,34656, did the maximum mass-loss rate have to be derived from \Ha, since
the radio regime seems to be more strongly clumped than the lower wind, at
least if the radio emission is purely thermal. Remember that the radio and
\Ha\ mass-loss rates for HD\,209975 are consistent to within the error bars.
\footnotetext{more precisely, where the optical
depth $\tau =1$ is reached along the radial ray, not to be confused with the
so-called ``effective radius'' located at $\tau \approx 0.24$, e.g.
\citet{wright75} and \citet{lamerswaters84a}.} 

Because all our diagnostics depend on $\rho^2$, different solutions with
lower mass-loss rates and scaled clumping factors are consistent with the
observational data to a similar accuracy as obtained from our fits, except
for the innermost cores of \Ha\ (particularly if of P~Cygni shape), due
to our assumption of an unclumped innermost region. As already noted, these
deviations remain very small for the derived values of $\rin$ though.

For six objects, the maximum mass-loss rate could not be uniquely
constrained, and the quoted limits correspond to the largest value
consistent with the data. In five of theses cases (denoted by superscripts
``a'' and ``d''), all radio fluxes are upper limits only, and consequently
the derived mass-loss rates as well. One object (Cyg\,OB2\#8A) is a
confirmed non-thermal emitter \citep{bieging89}, and the adopted maximum
mass-loss rate relies on the 2 cm which gives the lowest (radio) \Mdot\
within the available data set (see Sect.~\ref{variab}).

Of course, all objects with only upper limits for the radio flux(es) might 
be non-thermal emitters, and our interpretation depends on the assumption
that the radio excess is due to thermal emission alone. In addition to these
objects, three more stars (HD\.190429A, HD\,34656 and HD\,37043 (SB2!))
have somewhat peculiar radio fluxes, and might also be non-thermal emitters.

Mostly because of these peculiarities (for more details, see
Appendix~\ref{comments}), we have given two possible solutions for
HD\,190429A, HD\,37043 and also HD\,24912 in Table~\ref{table_clf},
comprising a minimum and maximum solution with respect to the (relative)
clumping properties. For HD\,37043 and HD\,24912, the 2nd entries are the
more plausible ones (as discussed in the appendix), whereas for
HD\,190429A both solutions have similar problems (though the difference is
not as large as for the other two stars).

Indicated by their name appearing in bold face, the remaining objects (six
with \Ha\ in emission, one with intermediate type and three with \Ha\ in
absorption) have well-constrained clumping properties, i.e., the derived
results are robust {\it if $\beta$ is not too different from the values
derived or adopted here.}

The latter quantity has been specified as follows. For objects with \Ha\ in
emission and of intermediate type, we have used the values from our
unclumped analyses (see Tables~\ref{sample} and \ref{haconsist_table}) 
wherever possible, i.e., if satisfactory fits could be achieved. This turned
out to be true in almost all cases, with the notable exception of
$\zeta$~Pup, where our clumped analysis favours a much lower value
($\beta_{\rm cl}$ = 0.70) than previously found. For most objects with \Ha\
in absorption (except Cyg\,OB2\#8C and HD\,34656, see
Appendix~\ref{comments}), because of missing constraints we used the
``standard'' value (from hydrodynamical models) of $\beta$ = 0.9,
to obtain at least consistent results. Further consequences of this
uncertainty are discussed in the next section.

For those stars where \Ha\ is of P-Cygni shape or displays a well-refilled
absorption trough, conclusive limits could be derived regarding the maximum
value of $\rin$, i.e., the maximum extent of a potentially unclumped region.
In all cases, this region lies below 1.2 \Rstar. 

In addition to the derived clumping factors which represent the best-fitting
solution, we also provide maximum values for $\fmid$ and $\fout$ which are
still consistent with our data and can be restricted further only by
additional far-IR and sub-mm observations. For all objects with entries 
``above'' Cyg\,OB2\#8C in Table~\ref{table_clf}, $\fmid$ could be
constrained from the wings of \Ha, either for the entirety of region 3 or,
if indicated, at least out to $\rout'$. For the other objects, the wind
density is too low to induce significant reactions in either \Ha\ or the IR
when the clumping properties in region~3 are changed, such that more
definite statements are not possible.

\begin{table*}
\caption{Upper and lower limits for the clumping factors in regions~2 and 3,
corresponding to a variation of $\beta_{\rm cl}$ as indicated (``used''
refers to the best fitting values tabulated in Table~\ref{table_clf}. For
Cyg\,OB2\#7, HD\,15570 and Cyg\,OB2\#8A we display the solutions for the
larger values of $\fmid$, which fit \Ha\ but somewhat overestimate the
10~$\mu$m fluxes, see Appendix~\ref{comments}).
No entries are given for Cyg\,OB2\#11 and HD\,34656 due to the very unclear
situation encountered for these objects (see Appendix~\ref{comments}).
Usually, the minimum value of $\fcl$ refers to the maximum of $\beta_{\rm
cl}$, and vice versa. For objects with an uncertainty in \Mdot\ being 
larger than typical, column~3 indicates the corresponding range (in units of
$10^{-6}{\rm M_{\odot}/yr}$). For entries with purely negative $\Delta
\Mdote$, the correction refers to the maximum value of $\beta_{\rm cl}$; in
these cases, the outer wind must also be clumped, with values as indicated
by $\ffar$. For HD\,209975, the positive correction refers to $\beta_{\rm
cl}$ = 0.7 with $\ffar=1$, no correction but $\ffar=1.3$ refers to
$\beta_{\rm cl}$ = 0.9, and the negative correction and $\ffar=3.5$ refers
to $\beta_{\rm cl}$ = 1.1.}
\begin{center}
\begin{tabular}{lrc|rrr|rrr|rrr|l}
\hline
\multicolumn{3}{c|}{} & \multicolumn{3}{c|}{$\beta_{\rm cl}$} &
\multicolumn{3}{c|}{$\fin$} & \multicolumn{3}{c|}{$\fmid$} & \\
\rb{Star} & \rb{\Mdot$_{\rm cl}$} & \rb{$\Delta \Mdote_{\rm cl}$} & min & used & max & min &
used & max & min & used & max & \rb{$\ffar(\beta_{\rm max})$} \\
\hline
  Cyg\,OB2\#7&  $\le$ 4.0 &            & 0.80 & 0.90 & 1.10 &  3.1 &  5.0 &  7.0 &  5.5 &  6.0 &  7.0 &           \\
 HD\,190429A &  9.5       &            & 0.85 & 0.95 & 1.10 &  2.0 &  3.0 &  3.8 &  2.5 &  3.0 &  3.5 &           \\
             &  7.5       &            & 0.85 & 0.95 & 1.10 &  3.2 &  5.0 &  6.5 &  4.5 &  5.0 &  6.5 &           \\
   HD\,15570 &  6.5       &            & 0.85 & 1.05 & 1.15 &  3.8 &  5.5 &  7.5 &  4.5 &  6.0 &  7.5 &           \\
   HD\,66811 &  8.5/4.2  &            & 0.60 & 0.70 & 0.90 &  3.0 &  5.0 &  6.0 &  1.5 &  1.5 &  2.0 &           \\
   HD\,14947 & 10.0       &     +/-2.0 & 0.85 & 0.95 & 1.15 &  1.7 &  3.1 &  3.8 &  2.0 &  2.5 &  3.0 &           \\
 Cyg\,OB2\#11&  5.0       &     +/-0.5 & 1.00 & 1.10 & 1.40 &  1.8 &  3.0 &  4.0 &  3.5 &  5.0 &  5.3 &           \\
  HD\,210839 &  3.0       &            & 0.80 & 1.00 & 1.10 &  5.0 &  6.5 &  8.0 &  5.0 & 10.0 & 12.0 &           \\
  HD\,192639 &  $\le$ 3.0 &            & 1.00 & 1.14 & 1.25 &  2.8 &  3.5 &  5.0 &  2.5 &  3.5 &  4.5 &           \\
   HD\,30614 &  1.5       &            & 1.00 & 1.15 & 1.25 &  2.5 &  2.6 &  3.5 &  2.0 &  3.0 &  4.0 &           \\
\hline
 Cyg\,OB2\#8A&  $\le$8.0  &            & 0.65 & 0.74 & 1.10 &  1.2 &  2.5 &  3.0 &  1.5 &  2.0 &  3.0 &           \\
 Cyg\,OB2\#10&  2.74       &            & 0.80 & 1.05 & 1.15 &  1.1 &  1.4 &  2.2 &  1.5 &  1.8 &  2.3 &           \\
\hline
   HD\,24912 & $\le$ 2.3  &            & 0.70 & 0.90 & 1.10 &  1.0 &  2.1 &  6.0 &  1.0 &  5.0 &  7.0 &           \\
             & $\le$ 1.2  &            & 0.70 & 0.90 & 1.10 &  3.0 &  8.0 & 20.0 &  1.0 & 20.0 & 25.0 &           \\
  HD\,203064 &  1.1       &       -0.4 & 0.70 & 0.90 & 1.10 &  1.0 &  1.0 &  2.0 &  1.0 &  1.0 &  2.0 &        2.5\\
   HD\,36861 & $\le$ 0.4  &            & 0.70 & 0.90 & 1.10 &  1.0 &  2.0 & 15.0 &  1.0 &  1.0 & 20.0 &           \\
   HD\,207198&  1.0       &      -0.35 & 0.70 & 0.90 & 1.10 &  1.0 &  1.0 &  2.5 &  1.0 &  1.0 &  2.0 &        3.0\\
   HD\,37043 &  0.8       &       -0.3 & 0.70 & 0.90 & 1.10 &  1.0 &  1.0 &  3.0 &  1.0 &  1.0 &  4.0 &        2.6\\
             &  0.25       &            & 0.70 & 0.90 & 1.10 &  3.0 & 12.0 & 30.0 &  1.0 &  1.0 & 20.0 &           \\
  HD\,209975 &  1.2       &  +0.1/-0.4 & 0.70 & 0.90 & 1.10 &  1.0 &  1.0 &  1.3 &  1.0 &  1.0 &  1.5 &    1.3/3.5\\
\hline
\end{tabular}
\end{center}
\label{table_errors}
\end{table*}

\subsection{Errors in the derived clumping factors}
\label{errors}
In the following, we will concentrate on the errors introduced into the
derived clumping factors; errors in the mass-loss and modified wind-momentum
rates are dominated by errors in the angular diameter and radius, but do not
affect the major outcome of our investigation.

Let us first mention that during our detailed fits we found no {\it
systematic} problem concerning an underestimation of the $N$- and $Q$-band
fluxes, so that at least our absolute flux calibration seems to be
appropriate (see Sect.~\ref{absolute_fluxes}). On the contrary, for some
objects ($Q$-band: HD\,15570; $N$-band: Cyg\,OB2\#11, \#10 and HD\,207198),
these fluxes lie above our predictions for the best-fitting model. To
investigate this point in more detail, however, additional fluxes in the
mid- and far-IR are required.

\paragraph{Uncertainties introduced by the radio continuum.} To determine
the uncertainty in the derived clumping factors due to uncertainties in the
observations (e.g., intrinsic errors and/or temporal variability of the
observed radio fluxes), we have varied $\fin$ and $\fmid$ by identical
factors and adapted \Mdot\ accordingly, until the observed radio fluxes
could no longer be matched. From these experiments, it turned out that the
clumping factors in the regions traced by \Ha\ (i.e., below $r=3{\ldots}5$)
are accurate (on an absolute scale) to within 20 to 50\%, whereas the {\it
ratio} of the clumping factors in the various regions remains preserved.
Remember that the derived clumping factors scale inversely with $\Mdote^2~({\rm radio})
\propto F_\nu^{1.5}$, i.e., $\delta \fcl/\fcl \approx -1.5\, \delta
F_\nu/F_\nu$. Extreme cases regarding this uncertainty in the radio fluxes
are HD\,190429A, HD\,14947 and Cyg\,OB2\#11 (cf. Table~\ref{table_errors},
3rd column).

\paragraph{The degeneracy of $\beta$ and clumping factors in the inner wind.}

As noted above, the strength of the core of \Ha\ is highly sensitive to the
value of the clumping factor in the inner part of the wind, below $r \approx
2 \Rstare$. It is also sensitive to the value of the velocity exponent,
$\beta$, and in a similar way: larger values of both $\beta$ and clumping
factors lead to more emission in the line core, giving rise to an
unfortunate degeneracy. Note, however, that the well-known $\beta$ vs.
\Mdot\ degeneracy (e.g., \citealt{puls96}) has ``vanished'', since the
(maximum) mass-loss rate is determined from the radio regime. Except for the
weakest winds (which cannot be observed in the radio anyway), 
the radio fluxes remain unaffected by the shape of the velocity
field (cf. Table~\ref{table_clf}, column 10).

This new degeneracy requires an investigation into the question of how far
any uncertainty in $\beta$ will propagate into the errors of $\fcl$. To this
end, we have varied $\beta$ and determined the appropriate values of $\fin$
and $\fmid$ such that the quality of our \Ha\ fit remained preserved. For
profiles with \Ha\ in emission and of intermediate type, the minimum and
maximum values of $\beta$ were taken from those solutions which were still
compliant with the observed profile shape. For objects with \Ha\ in
absorption, we used reasonable limits, at $\beta=0.7$ and $\beta=1.1$,
respectively. Larger values could usually be excluded from the profile
shape, whereas in certain cases a lower value (though being larger than the
physical limit, $\beta \ge 0.5$) might still be possible. This procedure is
somewhat similar to our approach to resolving the alternative $\beta$ vs.
\Mdot\ degeneracy in homogeneous winds, when \Mdot\ is derived from \Ha\
alone (cf. Paper~I).

In Table~\ref{table_errors} we have summarized the results of our
simulations. As expected, for stars with \Ha\ in emission, the uncertainty
in $\beta$ is not dramatic. This uncertainty leads to an average uncertainty
in $\fin$ of about $\pm$\,30\%, whereas for objects with \Ha\ in absorption,
much larger uncertainties are possible (factors of between 2 and 7), if
$\beta$ were 0.7 instead of 0.9.

For most of the objects with \Ha\ in absorption, a larger value of $\beta$
(1.1 instead of 0.9) would have some interesting consequences. Since for
these objects the inner clumping factors are of the order of unity for
$\beta=0.9$, an increase of $\beta$ cannot be compensated for by diminished
clumping. Consequently, the mass-loss rate must be decreased in this case,
to reduce the wind emission. Table~\ref{table_errors}, 3rd column, shows
that the required amount is of the order of 30\%. To still obtain a
consistent fit in the radio domain, $\ffar$ has to become larger than unity,
of the order of 2. Thus, {\it if} low-density winds were to have a velocity
field exponent larger than the standard one, the differences to the objects
with emission profiles would become even more pronounced: in this case, the
outer region would be even more clumped than the inner one. Only if $\beta$
were close to its lower limit, would the clumping properties of some of the
thin winds become similar to those of high-density winds.

Concerning the resulting uncertainties for $\fmid$ (region~3), the situation
for \Ha\ emission type objects is similar as for $\fin$. The average minima
and maxima lie $\sim \pm\,20\%$ below and above the best-fitting value of
$\beta$.  For the objects with weaker winds, on the other hand, $\fmid$
still remains unconstrained, and in all cases the upper limits as already
quoted in Table~\ref{table_clf} remain valid.

One last comment. Concerning our model(s) for $\xi$~Per (HD\,24912), we note
in Appendix~\ref{comments} that large values for the clumping factors in
region~3 ($\fmid$) are required if the small emission humps bluewards and
redwards of the \Ha\ absorption trough are to be explained by clumping. If,
on the other hand, $\beta$ were 0.7 for this object, these humps can be
created from region~2 {\it alone}.

\begin{figure}
\resizebox{\hsize}{!}
   {\includegraphics{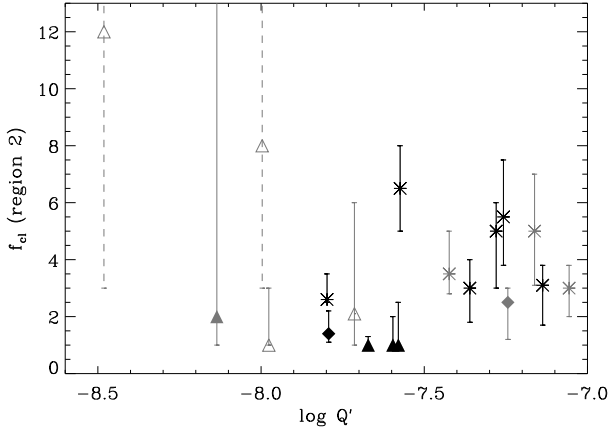}}
\caption{Clumping factors, $\fin$ (region 2), for our sample (cf.
Table~\ref{table_errors}), as a function of the distance-invariant quantity,
$\log Q'$ ($Q' = \Mdote/\Rstare^{1.5}$, with \Mdot\ the largest possible
mass-loss rate, in units of ${\rm M_{\odot}/yr}$ and \Rstar\ in units of
\Rsun). Remember that most clumping factors refer to outermost clumping
factors set to unity. Asterisks: objects with \Ha\ in emission; diamonds:
objects with intermediate \Ha\ profile type; triangles: objects with \Ha\ in
absorption. Black colors: objects with definite maximum mass-loss rates
(corresponding to bold-face entries in Table~\ref{table_clf}). Grey colors:
objects with upper limits for \Mdot\ and corresponding lower limits for
$\fin$. Maximum values of $\fin$ correspond to minimum values of $\beta_{\rm
cl}$, and vice versa for the minimum values. The open triangles with solid
error bars display the high-\Mdot--weak-clumping solution for HD\,24912 and
HD\,37043, and the open triangles with dashed error bars the alternative
low-\Mdot--strong-clumping solution for these objects.}
\label{clfin}
\end{figure}

\begin{figure}
\resizebox{\hsize}{!}
   {\includegraphics{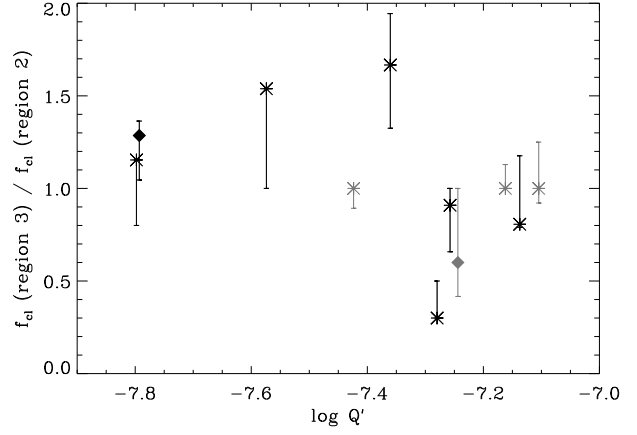}}
\caption
{As Fig.~\ref{clfin}, but for the ratio $\fmid/\fin$, and objects with \Ha\
in emission or of intermediate profile type only. The star with the lowest
ratio (0.3) is $\zeta$~Pup. For the three objects with a given
interval for $\fmid$ (Cyg\,OB2\#7, HD\,15570 and Cyg\,OB2\#8A, see
Table~\ref{table_clf}), we have used the mean value regarding this
interval.}
\label{clfratio}
\end{figure}

\section{Discussion}
\label{discussion}

\subsection{Clumping properties as a function of wind density}

Fig.~\ref{clfin} displays the derived clumping factors for region 2 (i.e.,
the first clumped region) as a function of $\log Q' = \log
\Mdote/\Rstare^{1.5}$, i.e., a quantity which is closely related to the mean
wind density, but is additionally distance invariant. Remember that in the
present context \Mdot\ is the largest possible mass-loss rate, and that most
of the derived factors refer to outermost clumping factors set to unity. In
other words, they have to be regarded as a measurement of the clumping
properties of the inner wind {\it relative to the outermost one}. Details of
the figure are given in the corresponding caption. 

The most important conclusions which can be drawn from this figure are the
following. For thinner winds with $\log Q' \la -7.5$ (a regime which is
populated by objects with \Ha\ in absorption or of intermediate type, but
also by the supergiant $\alpha$ Cam), {\it the inner wind seems to be
clumped by a similar degree as the outermost one}, at least if we discard
the alternative low-\Mdot--strong-clumping solutions for HD\,24912 and
HD\,37043 (open triangles with dashed error bars). Note that if the
latter solutions were the actual ones (and we have indicated that this is
rather possible), then both stars are behaving completely different to the
other absorption-type stars.

On the other hand, for stronger winds (almost all stars with emission
profiles, plus Cyg\,OB2\#8A), {\it the inner wind seems to be more
strongly clumped than the outermost one}, with an average ratio of
4.1\,$\pm$\,1.4. Of course, for this class of objects there is also the
possibility that we encounter moderately ($\fin \approx 3$) and stronger
($\fin \approx 5$) clumped lower wind regions, or that the degree of
clumping decreases again towards the largest wind densities. However, due to
the restricted number of objects, the influence of temporal variations 
(Sect.~\ref{variab}) and
the error introduced by the uncertainty of the continuum flux level, such
statements cannot be verified at the present time.

Fig.~\ref{clfratio} displays the ratio of clumping factors in the
intermediate and inner part of the wind, for objects with \Ha\ in emission
or of intermediate type; for those objects, this ratio could be constrained
in a rather robust way. In most cases, the clumping properties in both
regions are either similar, or the (average) clumping factors increase
moderately from region 2 towards region 3, at most by a factor of 2. Let us
reiterate, however, that region 3 is rather extended (i.e., local values
might deviate from their average ones), and that we cannot derive definite
values for radii larger $\rout' \approx 5 \Rstare$, except for few cases,
because \Ha\ becomes insensitive in this region, and strong constraints from
the IR continuum are missing. Future observations will help to clarify this
situation.

For objects with \Ha\ in absorption, at least upper limits for the clumping
factors in region 3, $\fmidm$, could be derived (see Tables~\ref{table_clf}
and \ref{table_errors}). For three well-constrained objects, HD\,203064,
HD\,207198 and HD\,209975, these upper limits lie between 1.5 and 2, i.e.,
they might be twice as large as the corresponding values for $\fin$, but are
still rather low. For the remaining stars, the maximum values for $\fmid$
lie in between 4 and 25, but only for HD\,24912 is a large value actually
{\it needed}, if the observed emission humps are to be interpreted in terms
of clumping and $\beta$ were of order 0.9 or larger (see above).

Concerning the clumping properties in region 4 ($15 \le r \le 50$), finally,
definite statements are only possible for those 3 stars observed in the
mm region (see below). For the rest, solutions with $\fout=1$ are consistent
with the observations, but larger values ($\foutm$ = 2{\ldots}20, cf.
Table~\ref{table_clf}) are possible as well. For HD\,190429A, HD\,14947, 
HD\,30614 and Cyg\,OB2\#10, \Ha\ still reacts to variations of the clumping
factor in region 4, and $\fout$ could be restricted to values from 2 to 4.
Since for weaker winds the radio-forming region can extend into region 4,
for a number of objects with \Ha\ in absorption, $\fout$ is better defined
than for the rest, particularly for HD\,24912 and HD\,36861, with $\fout \la
2{\ldots}3$.

\begin{figure*}
\begin{minipage}{9cm}
\resizebox{\hsize}{!}
   {\includegraphics{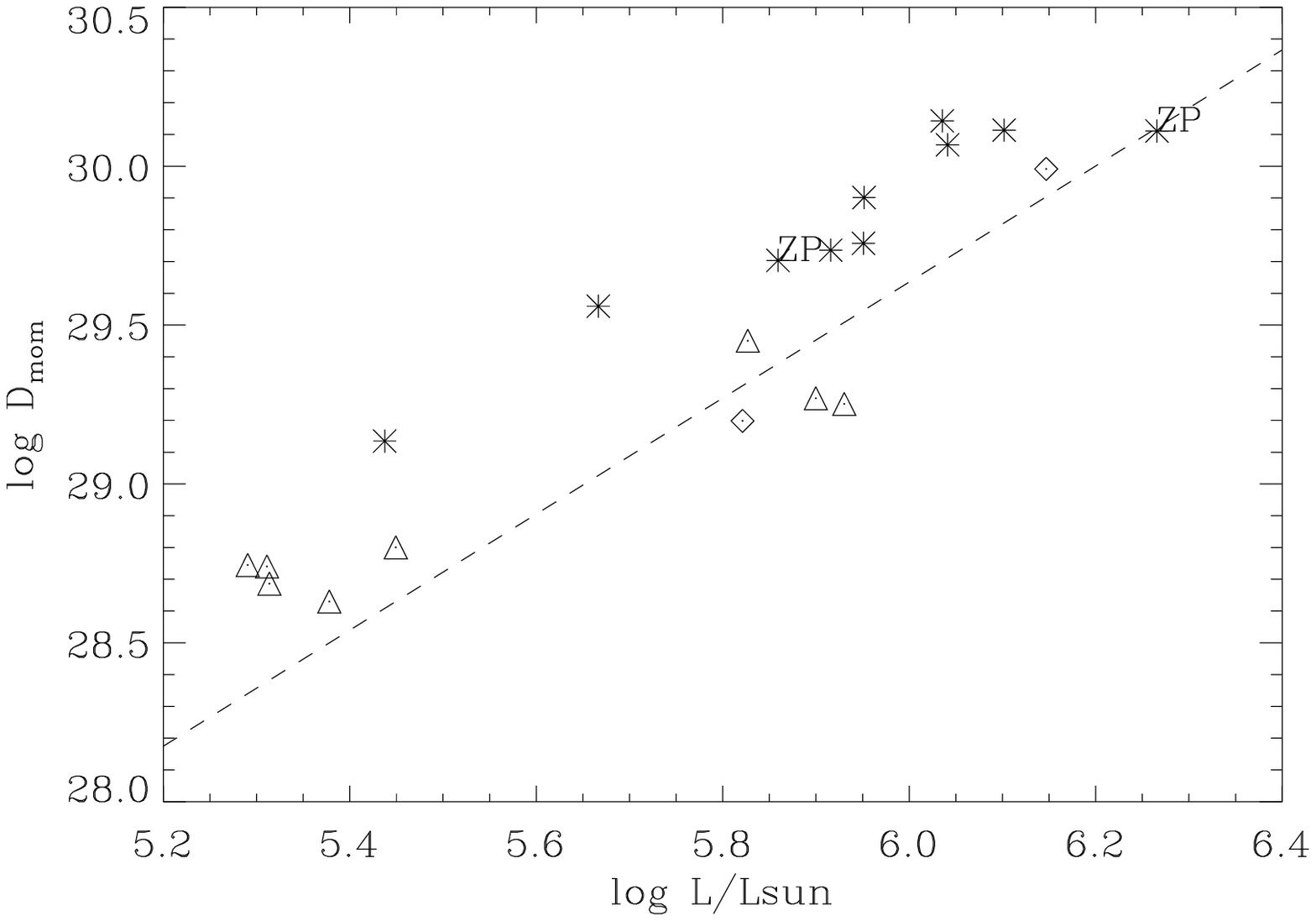}}
\end{minipage}
\hfill
\begin{minipage}{9cm}
   \resizebox{\hsize}{!}
   {\includegraphics{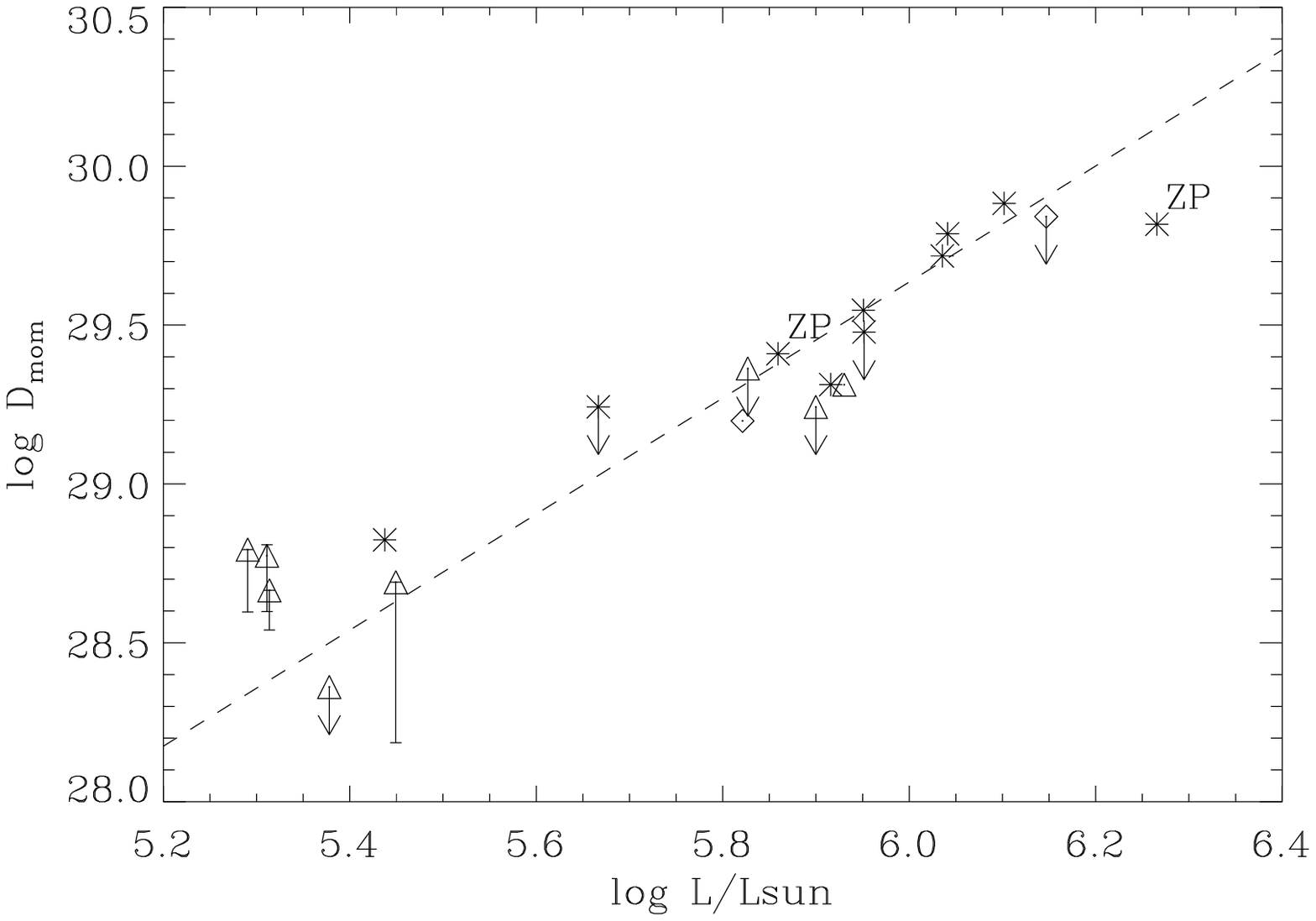}}
\end{minipage}
\caption{Wind-momentum--luminosity relation for our sample. Modified 
wind-momentum rate, $D_{\rm mom} = \Mdote \vinfe (\Rstare/\Rsune)^{0.5}$, 
in cgs units. Left panel: mass-loss rates derived from \Ha, using {\it
homogeneous} models, cf. Table~\ref{sample}. Right panel: largest possible
mass-loss rates, from this investigation. Upper limits indicate those cases
where radio fluxes are upper limits and/or non-thermal emission cannot be
excluded. Asterisks: objects with \Ha\ in emission; diamonds: objects with
intermediate profile type; triangles: objects with \Ha\ in absorption.
Dashed line indicates theoretical prediction by \citet{vink00}. ``ZP''
indicates the large and low distance solution for $\zeta$~Pup (see text).
For the three objects at $\log L/\Lsune = 5.3$ (HD\,203064, HD\,207198 and
HD\,209975), the lowermost solution indicates their position if the velocity
exponent was larger than expected ($\beta = 1.1$ instead of $\beta = 0.9$).
In this case, the (unclumped) \Ha\, mass-loss rate would be lower than the
radio mass-loss rate, and the wind would have to be more strongly clumped in
the radio regime than in the innermost region. For HD\,37043, at $\log
L/\Lsune = 5.45$, the lowermost solution corresponds to the 2nd entry in
Table~\ref{table_clf}.} 
\label{wlr}
\end{figure*}

The best-constrained objects within our sample are $\zeta$~Pup, HD\,15570
and HD\,210839, due to IRAS (for $\zeta$~Pup) and mm observations.  The
first of these objects, $\zeta$~Pup, displays the only notable exception
concerning the ratio of $\fmid$ and $\fin$, namely that region 3 is much less 
clumped than region 2. In other words, maximum clumping must be close
to 2~\Rstar, or even lower (cf. Fig.~\ref{hd66811_2} and
Table~\ref{tab_zp_clumped}). For this star, the {\it derived} clumping
factor for region 4 (extending from 15 to 50 \Rstar) is even lower than for
region 3: at most, $\fout \la 1.8$. 

For HD\,15570, on the other hand, regions 2 and 3 are similarly clumped, and
the derived clumping factor might increase even further towards region 4,
with $\fout$ being 5 to 20 times larger than the average clumping in the
radio-emitting region. In the unlikely case that the wind is not recombined
at 1.3\,mm, even $\fout=1$ is possible. For this object, the mm measurements
from {\sc scuba} are extremely valuable, though the rather large error bars
leave the situation not as clear as desirable. 

For $\lambda$~Cep (HD\,210839), finally, the intermediate region is more
heavily clumped than the inner one, whereas region~4 could be constrained
(again via {\sc scuba} observations) to display clumping factors between 1
and 8. It remains to be clarified whether the two different observed flux
levels (Tab.~\ref{scubatable}, Fig~\ref{fits_haem_2}, 2nd row) are a sign of
significant temporal variability of the outer wind (indicating a temporal
variation of clumping or a non-negligible effect of macro-structures) or the
``truth'' lies in between both measurements, which are still consistent
within the claimed error bars.

In summary, at least one of these three objects is rather weakly clumped in
region 4. Although the same might be true for the other two stars
(accounting for the lowest possible fluxes), a significantly clumped outer
region is more probable.

\subsection{Wind-momentum--luminosity relation}

Before discussing some further implications of our findings, let us consider
the wind-momentum--luminosity relation for our stellar sample, accounting
for the results derived in the present paper. Fig.~\ref{wlr} displays two
such relations, in comparison with the theoretical predictions by 
\citet{vink00}.\footnote{which are consistent with independent
investigations by our group, see \citet{puls03}, and also \citet{kud02}.} In
the left panel, we show the results using \Ha\ mass-loss rates derived by
unclumped models, updated for a re-determined stellar radius
(Sect.~\ref{obs}). As already noted in Paper~I and outlined in the
introduction, objects with \Ha\ in absorption and of intermediate type are
perfectly consistent with the predictions (except for a few objects at $\log
L/\Lsune < 5.35$; see below), whereas objects with \Ha\ in emission populate
a strip parallel to, but above, the predictions.  Only the large-distance
solution for $\zeta$~Pup lies {\it on} the relation, whereas the
low-distance solution displays the same discrepancy as the other stars (both
solutions indicated by ``ZP'').

In the right panel, we display our new results, with mass-loss rates from
Table~\ref{table_clf}. These mass-loss rates are the largest possible ones,
and are essentially the radio mass-loss rates if the winds were unclumped in
the radio-forming regime. Except for this assumption, the largest errors
present in this figure are due to errors in the distance estimate. We have
deferred from a rigorous error analysis concerning this problem, as this is
beyond the scope of the present investigation.

What is obvious from this plot, however, is that the agreement between 
observations and theoretical predictions has significantly improved. Almost
all objects now lie very close to the theoretical relation, {\it independent
of profile type}.

The reason, of course, is that the newly derived (radio) mass-loss rates for
emission-profile objects are smaller than the \Ha\ mass-loss rates (see
Table~\ref{table_clf}, column ``ratio''), by an average factor of
$0.49\pm0.10$. Most interestingly, this is almost exactly the same factor
which has been claimed in Paper~I (0.48, drawn from a much larger sample), and
which has been used a priori in our de-reddening procedure (see
Sect.~\ref{dered}). A factor of the same order (0.42) has also been found by
\citet{fulli06}, for a sample comprising objects similar to those considered
here. For objects with \Ha\ in absorption and of intermediate type, \Ha\ and
radio mass-loss rates agree well, and they remain at their ``old'' position.
Note that for the only absorption-type object in the sample of Fullerton et
al. with \Ha\ and radio data available in parallel (HD\,149757), a
comparable agreement was found, supporting our results.

Due to the shift in wind-momentum rate, the new position of $\zeta$~Pup
(larger distance) is completely inconsistent with the rest, whereas the
conventional, lower-distance solution matches the predictions perfectly. The
same problem was found in Paper~I (after applying an average down-scaling of
wind momenta, in anticipation of clumping effects), and our present result
(which confirms this expectation) seems to favour a lower radius. 

In accordance with our reasoning in Paper~I, however, we like to point out
that $\zeta$ Pup is a ``bona fide'' runaway star, (i.e., its parent
association, Vela R2, has been identified by \citealt{Sahu93}). Based on
Hipparcos data, \citet{VLR98} argued that $\zeta$~Pup could have become a
runaway as a result of a supernova explosion in a massive close binary,
which might explain its peculiar characteristics, such as enhanced He and N
abundances at the stellar surface, high peculiar and rotational velocities,
and its overluminosity. The reason why the wind-momentum rate should be
lower than for other objects remains to be clarified though.  

Whereas the ``new'' WLR agrees extremely well with the theoretical
predictions for objects with $\log L/\Lsune > 5.35$, the three best-defined
absorption-type stars at the lower luminosity end of our sample (HD\,203064,
HD\,207198 and HD\,209975) lie too high, by a factor of $\approx 2.5$. A
similar effect was found in Paper~I, though at the time it was not clear
whether or not their winds are clumped in the \Ha-forming region to a
similar degree as emission type objects. In addition to wind momenta based
on nominal radio mass-loss rates, we have also indicated (by the lower end
of the displayed bars) their position if the velocity exponent were to be
larger than expected ($\beta = 1.1$ instead of $\beta = 0.9$). In this case,
the (unclumped) \Ha\, mass-loss rate would be lower than the radio mass-loss
rate, and the wind would have to be more strongly clumped in the radio
regime than in the innermost region (cf. Table~\ref{table_errors}). Even in
this case, a discrepancy of a factor of $\approx 1.7$ would still be
present. To unify these objects with the others by clumping arguments {\it
alone} would require that they have to be much more clumped in the radio
regime (on an absolute level). 

Of course, one might argue that this problem is not related to (unknown)
physics but to wrong distances and radii. Though this might be possible
accounting for the mean errors in modified wind-momentum rate (0.13 dex) 
and luminosity (0.19 dex) derived for Galactic objects in
Paper~I, it is more plausible to invoke physical reasons, since we have to
explain an identical problem for three different stars (with different
\Teff) at identical positions in the diagram.

Again, we stress that all displayed positions rely on the derived, {\it
largest possible} mass-loss rates. If the radio regime were clumped,
downward corrections become necessary. In this case, however, the displayed
agreement would be pure coincidence.

\subsection{Implications and conclusions}

The results from the previous section have confirmed our earlier hypothesis 
that the ``old'' \Ha\ mass-loss rates for objects with \Ha\ in emission are
contaminated by clumping, and that, compared to theoretical predictions,
these mass-loss rates are overestimated by a factor of {\it at least} 2{\ldots}3.
Regarding the different behaviour of objects with emission and those with 
absorption profiles, however, we seem to have invoked a somewhat erroneous
explanation (see our arguments recapitulated in Sect.~\ref{intro}).

Indeed, if objects with \Ha\ in absorption were clumped in the lower-wind
region, in a similar way to emission-type objects, we would have seen this:
note that the presence of clumping with factors as low as $\fin = 2$ is
clearly visible (cf. Fig.~\ref{hd209975_1}). It must be stressed, however,
that our present sample consists of supergiants and giants only, and that
dwarfs (with a very low wind emission inside the core of \Ha) are missing.
At least for the latter luminosity class, our old arguments might still be
valid. For example, if the wind base was actually unclumped (as allowed for by our
analysis, but in contrast to the findings by \citealt{bouret05}), and \Ha\
predominantly forms in this region, we would not see the clumping effects,
though they would be present in, e.g., the mid-/far-IR.

Assuming for the moment that, on an absolute scale, the {\it outer} clumping
properties are independent of wind density, our results imply that the
different degree of consistency between the theoretical and observed WLR is
likely related to a physical effect: inside the \Ha-forming region, denser
winds {\it are} more strongly clumped than (most of) the weaker winds, at
least if $\beta$ is of the order of 0.9 or larger in the weaker winds (see
Fig.~\ref{clfin}).

What might be the origin of such a difference? Objects with \Ha\ in emission
have a large wind density and are usually supergiants with low gravity and a
considerable Eddington-$\Gamma$. Thus, it is rather possible that they are
subject to photospheric instabilities and/or pulsations, triggering a
somewhat larger structure formation in the lower wind, compared to lower
density winds from higher gravity objects. Indications of such a dependence
are consistent with investigations regarding photospheric line-profile
variability (increasing with stellar radius and luminosity), as outlined by
\citet{fulli96}.

On the other hand, our findings are in some contrast to hydrodynamical
simulations, at least regarding {\it self-excited} structure formation. If
there was any dependence on wind density predicted at all, thin winds should
be more strongly clumped than thick winds, because of the missing
stabilization due to the continuum \citep{op99}, which induces a more
heavily structured wind in the lower part. Note also, that in thin winds,
(transonic) velocity curvature terms become important, leading to gradient
terms in the source functions and modification of the line acceleration
\citep{puls98}.  Simulations by \citet{op99} accounting for this effect
resulted in a highly structured wind, with a moderately reduced mass-loss
rate and a rather steep velocity law in the lower region. Thus, even the
possibility that $\beta$ {\it is} low (which would increase the derived,
lower clumping factors, cf. Table~\ref{table_errors}) cannot be completely
excluded, although in this case, $\fin \la 2$ (for the three well-defined
objects) is still rather low.

Let us now compare our results with the predicted radial stratification of
$\fcl$ itself \citep{ro02, ro05}. As is true for our results, theory also
depends on a number of assumptions. Most important in this context are:-
\begin{itemize}
\item 
the dimensionality of the hydrodynamical treatment, which is mostly 1-D
(because of the complexity of calculating the radiative line force). First
results from a 2-D approach \citep{desowo03} might indicate somewhat lower
(factor of 2) clumping factors than those resulting from a 1-D treatment.
\item 
the excitation of the line-driven instability. Almost all models 
investigated with respect to the clumping factor refer to self-excited
perturbations. Unfortunately, externally triggered perturbations, such as
sound waves and photospheric turbulence (see \citealt{feld97}), and
photospheric pulsations, have not been examined with regard to this
quantity.
\item
the so-called line-strength cut-off, \kmax. In order to keep the problem
numerically treatable (i.e., to avoid too fine a grid resolution), \citet{ocr}
introduced an opacity cut-off regarding the driving lines, which is
typically three dex below the actual value. Experiments performed by
\citet{ro02} showed that the clumping factor in the outer wind (around 50
\Rstar), in particular, can increase if more realistic values are used. The
inner and outermost part seem to remain rather insensitive, at least if 
very low values for \kmax\ are avoided.
\end{itemize}

Thus, the numbers which will be quoted in the following might be considered
in a qualitative sense, especially since, for our comparison, we have to
estimate appropriate spatial averages over the different regions. In
our approach, we have used clumping factors assumed to be spatially constant
within certain regions, whereas \citet{ro02} display the clumping factor as
a function of $r$. The most decisive quantity regarding radiative transfer 
is the optical depth, being proportional to the spatial integral over
$\fcl(r) \rho(r)^2$ (assuming the source function to be unaffected by
clumping), so that a meaningful comparison requires the predicted
clumping factors, $\fcl(r)$, to be averaged over $\rho^2$ inside the
regions considered.\footnote{By adopting this approach, we discard certain
details, such as the fact that \Ha\ reacts to averages over constant
velocity surfaces (and not along the radial direction), as well as
optical depth effects.} To this end, we have used the results
displayed in the various figures provided by Runacres \& Owocki.

\begin{table}
\caption{Clumping factors as predicted by hydrodynamical simulations from
\citet{ro02}, for the different regions as used in this investigation. The
first average is a straight one, the 2nd is weighted with $\rho^2$ (see
text). Note that these numbers are only approximate ones, since they have
been derived from figures and not from tables.}
\begin{tabular}{l|lll}
\hline
region    & $\fcl$      &  $\langle \fcl \rangle_1$ & $\langle \fcl \rangle_2$ \\
\hline
1         & 1           &                 1         &  1     \\
2         &1{\ldots}4   &                 2.5       &  $\ga$2.1 \\
3         &4{\ldots}13&                   8.5       &  $\ga$4.7 \\   
4$^a$     &13{\ldots}5&                   9         &  $\la$11.6\\
5$^a$     &5 {\ldots}4 &                  4.5       &  $\la$4.7 \\ 
4$^b$     &13{\ldots}20&                  16        &  $\ga$14 \\
5$^b$     &20{\ldots}4 &                  12        &  $\la$15 \\ 
\hline                
\end{tabular}

$^a$ \kmax\ from \cite{ocr}; $^b$ \kmax\ larger by a factor of 10. 
\label{clf_theo}
\end{table}

Table~\ref{clf_theo} summarizes the predictions. Region 1 (the inner,
unclumped region) typically extends to 1.3 \Rstar\ (for thin winds, it might
be narrower; see above), which is fairly consistent with the {\it derived}
maximum extent of such a potentially unclumped domain ($\rin \la$
1.1{\ldots}1.2 \Rstar).

Regarding the other regions, we have to discriminate between absolute
numbers and numbers referring to the average clumping factor in region 5,
which is of the order of 4{\ldots}5 or even larger, if \kmax\ is increased
beyond its ``standard'' value. Such large averages depend on results indicating
that the outermost wind (beyond 1000 \Rstar) is also considerably clumped;
cf. \citet{ro05}. Only for rather low values of \kmax\ is a smooth radio
regime predicted. 

For region 2, we find {\it average} values $\fin \approx 2{\ldots}3$ (lower
than in region 5!), for region 3 values around 4{\ldots}5, and for region 4
values around 11, which again might be even larger for large \kmax. Note
that for different wind densities and wavelengths, the calculated averages
for regions 3 and 4 might be higher and lower, respectively, than the
indicated ones, depending on the radial position at which $\tau=1$ is
reached. Finally, the predicted maximum is located at the border between
regions 3 and 4 (around 15 \Rstar), but might be shifted towards larger
radii for larger \kmax.

Compared to our results, these predictions are significantly different, at
least {\it if} the average clumping factor in the radio domain is of the
order of 4 or larger. In this case, all \Ha\ mass-loss rates should be lower
than the radio mass-loss rates, which is definitely not true. Thus, either
the clumping factors in region 2 are predicted as too low, or those in
region 5 as too large!

Disregarding this problem, the average clumping factor should increase
monotonically from region 2 to 4 according to theory, and at least {\it
some} of our emission type objects (e.g., HD\,15570) are compatible with
this result (though for others $\fout$ is of the same order or even lower
than $\fmid$). Only concerning the differential behaviour of region 2 to
region 3, do {\it most} objects behave as predicted (Fig.~\ref{clfratio}).
As outlined already above, the notable exception to this rule is
$\zeta$~Pup, where the complete run of $\fcl(r)$ and the position of its
maximum definitely deviate from the predictions (and from the other objects
investigated). Such a deviation was already found by \citet{pulsetal93}, who
tried to simulate the observed \Ha\ profile and IR continuum for
$\zeta$~Pup, based on hydrodynamical models from S.~Owocki. Though they were
quite successful in fitting \Ha\ with a mass-loss rate just a factor of 2
lower than when using homogeneous models (and consistent with present
estimates), the IR continuum was too strong at this \Mdot,
indicating lower clumping factors than predicted in region~3.  

The real question, of course, concerns the absolute value of the clumping
factors, and their dependence on stellar parameters. What has been derived
in this investigation is the behaviour of the inner clumping
properties {\it relative} to the outermost ones. To reiterate, if the outer
clumping properties were independent of wind density and/or stellar
parameters, thinner winds would be less clumped in the inner region than
stronger winds, and we have indicated above a possible reason for this. If,
on the other hand, the (absolute) clumping factors in the inner part were to
be equal or even larger in thinner winds than in denser ones, we would meet
a number of other problems requiring explanation. In this case:-
\begin{itemize}
\item 
the outer region of thinner winds {\it has} to be more clumped than in
thicker winds.
\item
the consistency with the theoretical WLR would completely vanish.
\item
the WLR would again show a strong dependence on luminosity class
and/or \Ha\ profile type (even if the theoretically predicted off-set was
wrong). Such a dependence is presently not understandable, since
the major prediction of radiation-driven wind theory is that the modified
wind-momentum rate should be dependent on luminosity alone (at least if
the slope of the corresponding line-strength distribution function 
is not too different from its presently derived value).
\end{itemize}
The only way to clarify this situation is the inclusion of processes which
do not depend on $\rho^2$. One such diagnostic is \Pv\ \citep{massa03,
fulli04, fulli06} which under favourable circumstances scales $\propto \rho$
alone. The major problem here arises from the uncertainties regarding the
ionization fraction of this ion, which might be additionally contaminated by
the UV-tail of the X-ray emission.  Assuming that \Pv\ is a major ion
between O4 and O7, \citet{fulli06} derived a median reduction in \Mdot\
(compared to homogenous \Ha\ and radio diagnostics) by a factor of 20, where
thin winds seemed to be more affected than thicker ones. Note that this
would imply clumping factors of the order of 100 in the radio regime!

Detailed NLTE investigations accounting for clumping, on the other hand, are
only in their infancy, and again, the inclusion of X-ray effects is a
difficult task. The only object within our sample which can be compared with
such an investigation is HD\,190429A, analyzed by \citet{bouret05}. In their
conclusions, they quote a reduction of a factor of three in \Mdot, compared
to a homogeneous mass-loss rate of 6~\Mdu\ {\it derived from the far-UV},
exploiting $\rho$- and $\rho^2$-dependent processes in parallel, and
accounting for a consistent ionization equilibrium.

The derived homogeneous UV mass-loss rate is much lower than our homogeneous
\Ha\ value (radius and distance are comparable), and they speculate on
strong variations in \Ha, referring also to \citet{scu98}, who report an
increase of the \Ha\ equivalent width between 1988 and 1991, by a factor of
2 (but see also \citealt{markova05}, who found no indications of such large
changes in \Ha, at least over an interval of one year between 1997 and
1998). Though the implied clumping factor (from a comparison of homogeneous
and clumped UV mass-loss rates) would be not too different from ``our''
value, on an absolute scale there are much larger differences. Comparing
their final mass-loss rate (1.8~\Mdu, with \Rstar\ = 19.5~\Rsun\ and \vinf\
= 2300\,\kms) with our radio mass-loss rate (7.5{\ldots}9.5~\Mdu, with
\Rstar\ = 22.7~\Rsun\ and \vinf\ = 2400\,\kms), this would suggest a
strongly clumped radio regime, with $\ffar \approx 10{\ldots}16$, at least
if there have been no major changes in the average wind properties between
their UV and our radio observations. Additionally, \citet{bouret05} point to
the fact that the predictions by \citet{len04} concerning Br$_{\alpha}$
indicate that the outer winds ``would be less affected by clumping'',
compared to the regions they could access. Thus far, the situation remains
unclear. 

Notably, the other object investigated by \citet{bouret05} is an object with
\Ha\ in absorption, and for this object they find a reduction in \Mdot\ by a
factor of 7 (again with respect to UV observations alone). This result would
agree with our statement from above that thin winds are expected to be more
strongly structured than thick winds, at least if the latter are not
externally triggered by photospheric disturbances.

Accounting for these findings and other investigations with similar results
(e.g., \citealt{hil03, bouret03}), there seems to be increasing evidence
that the agreement between the theoretical and observed WLR (which, if real,
would imply a smooth wind in the radio regime) is indeed just coincidence,
and that the radio regime must be strongly clumped, maybe even more strongly
than presently described by hydrodynamics. 

Aside from the major implications such a reduction of mass-loss rates would
have, e.g., regarding stellar evolution in the upper HRD and feedback from
massive stars, such a result would also lead to the following problem: since
the present theoretical WLR originates from consistent calculations of the
radiative line force, lower wind momenta would imply that too much radiative
pressure is available. A reduction of this quantity, however, is rather
difficult (but see below).

Finally, let us note that a significant down-scaling of mass-loss rates would
unfortunately also affect stellar parameters (again!). For the
$\rho^2$-dependent results derived here, such scaling is easily possible,
without modifying any result. {\it Photospheric} lines, on the other hand,
might be differently affected by a strongly clumped, but weaker wind, since
they do not always scale with $Q$, but depend on other combinations of
\Mdot, \Rstar\ and \vinf\ as well.

\section{Summary and future work}
\label{summary}

In this investigation, we have performed a simultaneous analysis of \Ha, IR,
mm (if present) and radio data to constrain the radial stratification of the
clumping factor in a sample of 19 O-type supergiants/giants, with dense and
moderate winds (\Ha\ in emission and absorption). All analysis tools used
involve certain approximations, but we have ensured that the derived results
comply with state-of-the art NLTE model atmospheres, by comparing and
calibrating to a large grid of such models. Clumping has been included in
the conventional approach, by manipulating all $\rho^2$-dependent opacities
and assuming the inter-clump matter to be void. Caveats have been given to
this assumption and other problems inherent to this approach, namely the
neglect of disturbances of the velocity field due to the clumps, and the
assumption of small length scales, related to the problem of porosity.

Instead of adapting the clumping-factor at {\it each} radial grid point
(which is possible only if using optimization methods, requiring a
well-sampled observed wavelength grid), we have introduced 5 different
regions, with constant clumping factors inside each region. Because all our
diagnostics depends on $\rho^2$ (except for the small contribution by
electron scattering), the most severe restriction within our approach is
given by the fact that we cannot derive absolute clumping factors, but only
factors normalized to a certain minimum. Since in all but one case
(HD\,34656) this minimum was found to be located in region 5 (or, in other
words, since in all those cases the radio mass-loss rate is the lowest), our
normalization refers to the radio regime, and the corresponding (radio)
mass-loss rate as derived here is the largest possible one. Other solutions
are possible as well, with all clumping factors multiplied by a constant
factor, $f$, and a mass-loss rate reduced by $\sqrt f$. 

Our analysis is based on \Ha\ line profiles, near-/mid-/far-IR fluxes taken from our
own observations and the literature (de-reddened as detailed in
Sect.~\ref{dered}), mm fluxes observed by {\sc scuba/sest} (own and
literature data), and radio data taken from our own VLA observations and the
literature. We have discussed the issue of non-simultaneous
observations: based on present-day observational facts, the
\Ha, IR and radio variability of thermal emitters is low enough so as not to pose any
problems for our study, at least if the derived results are considered in a
statistical sense. Within our sample, there is only one confirmed
non-thermal emitter (Cyg\,OB2\#8A), and three more objects display somewhat
peculiar radio fluxes (HD\,190429A, HD\,34656, see above, and HD\,37043).
These objects might be non-thermal emitters as well, but this has to be
confirmed by future observations. In any case, the derived mass-loss rates
(from the minimum radio flux) can be considered as an upper limit. 

As it turns out, the core of \Ha\ provides very useful diagnostics for the
clumping properties in the inner wind ($r \la 2 \Rstare$), and, if in
emission, the wings can be used to constrain the clumping inside the first
five stellar radii, with an additional check provided by IR data. If mm fluxes were
available, the outer wind ($15 \Rstare \la r \la 50 \Rstare$) could be
constrained as well. Only the region between $5 \Rstare \la r \la 15 \Rstare$ 
remains ``terra incognita'' in most cases, due to missing far-IR fluxes.

For ten stars in our sample (six with \Ha\ in emission, one of intermediate
type and three with \Ha\ in absorption), the derived clumping factors are
robust and lie within well-constrained error bars. For six stars (including
Cyg\,OB2\#8A), only upper limits for the radio mass-loss rate are available,
and the derived clumping factors have to be considered as lower limits.
Obvious differences to the best-constrained objects were not found though,
except for HD\,24912, which behaves atypically. The three remaining objects
constitute HD\,34656, which is the only object in our sample with an \Ha\
mass-loss rate lower than the radio mass-loss rate (and as such has been
discarded from our further analysis), HD\,37043, which exhibits similar
problems to HD\,24912 (but has a better-constrained radio mass-loss rate),
and HD\,190429A, which displays a certain degree of radio-variability.
Taking the various results together, we can summarize our findings as
follows:-
\begin{itemize}
\item for almost all objects (except for 3 stars with \Ha\ in absorption 
and $\log L < 5.35\, \Lsune$), the derived (radio) mass-loss rates are in
very good agreement with the predicted wind-momentum--luminosity relation
\citep{vink00}, in contrast to previous results relying on unclumped \Ha\
data alone. If $\zeta$~Pup is located at the ``close'' distance, then it
behaves as the rest. If, on the other hand, it is located further away, its
(radio) wind-momentum rate would lie considerably below the predictions. 
\item the mean ratio of radio mass-loss rates to unclumped \Ha\ mass-loss rates
for stars with \Ha\ in emission is 0.49~$\pm$~0.10. This is almost exactly
the same factor as found in Paper~I, by shifting the {\it
observed} WLR (using unclumped models) for these objects onto the {\it
predicted} one. It also agrees well with recent findings from
\citet{fulli06}.  
\item the average, normalized  clumping factor in the innermost region 
($r \la 2 \Rstare$) of stars with \Ha\ in emission is $\sim 4.1 \pm 1.4$. 
\item thinner winds with \Ha\ in absorption have lower normalized
clumping factors in this region. For all three stars with robust
constraints, these factors are similar to those in the radio region, at
least if the velocity exponent is not too different from the hydrodynamical
prediction, $\beta \approx 0.9$. Factors of the order of $\fin \ga 2$ can
be excluded, due to the sensitive reaction of \Ha.
\item for all objects where \Ha\ is of P~Cygni shape, or displays a
well-refilled absorption trough, the maximum extent of a potentially unclumped
region can be limited to lie inside $r \la 1.2 \Rstare$. 
\item in most cases, the clumping factors in the inner and adjacent region
($2 \Rstare \la r \la 5{\ldots}15 \Rstare$)  are comparable or increase 
moderately from inside to outside. Only for $\zeta$~Pup, does our analysis
restrict the maximum clumping at $r \la 2 \Rstare$.
\item the presence of clumping introduces a new degeneracy in the results,
namely between the velocity field exponent, $\beta$, and the
clumping factors. If $\beta$ is lower than assumed or derived from the fits,
the clumping factors are larger, and vice versa. Extreme deviations of
$\beta$ from values obtained from an unclumped analysis can be excluded though.
Interestingly, a perfect fit for $\zeta$~Pup requires $\beta = 0.7$,
contrasted with $\beta =0.9$ from unclumped diagnostics \citep{repo04}.
\item two of the three stars with mm-observations (HD\,15570 and HD\,210839) 
indicate a certain probability that the outer region~4 ($15 \Rstare \la r
\la 50 \Rstare$) is considerably more clumped than the radio domain (but
remember the rather large error bars on the mm data), whereas the third
star, $\zeta$~Pup (with negligible observational errors), displays similar
clumping properties in both regions.
\item Our results differ from hydrodynamical predictions (incorporating 
the intrinsic, self-excited line-driven instability, \citealt{ro02,
ro05}) at least in one respect: the latter imply a larger radio than
\Ha\ mass-loss rate (or, alternatively, lower clumping in the inner than the
outer wind), which is definitely not true for our sample.
\end{itemize}
In addition to the conclusion that one of the best-observed massive stars,
$\zeta$~Pup, might be a rather atypical representative of its kind
(maybe due to its possible expulsion from a close binary system), the major
implications of these findings can be stated within three different
assumptions concerning the clumping properties of the outermost regions:-
\begin{list}{}{}

\item[assump. {\bf (a)}:] {\it The radio region is not, or only weakly,
clumped.} \\
In this case, our ``old'' hypothesis (concerning a shift of mass-loss rates 
for objects with \Ha\ in emission, due to clumping) would be confirmed, but
there would be a physical difference between denser and thinner winds, in
the sense that thinner winds would be less clumped than
thicker winds in the inner region. This difference might then be related to different excitation
mechanisms of structure formation. If assumption (a) were true, the
theoretical WLR would be perfectly matched. On the
other hand, the absolute numbers for clumping factors and mass-loss rates 
would be in severe contrast to results from other investigations that
have used alternative diagnostics, not directly affected by clumping
(e.g., the P{\sc v} resonance lines).
\item[assump. {\bf (b)}:] {\it The radio region is strongly clumped, but the 
outermost clumping factors are independent of wind density.}\\ 
In this case, a unification with results from other diagnostics is
possible, and the present mass-loss rates would have to be significantly
revised, with serious implications for the evolution of, and feedback
from, massive stars. Again, weaker winds would be less clumped in the inner
region, and the theoretical WLR would no longer be matched. One of the most
robust predictions from radiation-driven wind theory, namely that the
modified wind-momentum rate should depend almost exclusively on luminosity
(and not on mass or gravity), would still be consistent with our data, even
if there were an offset between the theoretical and observed WLR.  
\item[assump. {\bf (c)}:] {\it The radio region is strongly clumped, but 
the degree of clumping is different for different wind densities.} \\
This case is also consistent with present data, but would again imply, in
addition to different offsets between the theoretical and observed WLR, that the
observed WLR is dependent on a second parameter.
\end{list}
Obviously, the implications of all three assumptions pose their individual
problems, and would have different consequences regarding the urgent
question about the ``true'' mass-loss rates of massive stars. Since there is
no direct way to measure the clumping in the radio regime, for further
progress we suggest the following steps.

On the observational side, we have to: ($i$) re-observe some problematic 
objects at radio frequencies, to check their variability and to obtain
further clues as to whether their emission is of thermal or non-thermal
origin; and ($ii$) most importantly, accumulate far-IR and mm observations,
to constrain the (normalized) clumping factor in the intermediate wind.

Once a reliable, normalized stratification has been obtained, it
can be used as an {\it input} into state-of-the-art model
atmosphere codes allowing for the inclusion of clumping and X-ray emission,
with the mass-loss rate/velocity field adapted until all diagnostics (including
the FUV/UV) are reproduced. This would also clarify the question concerning the
ionization fraction of P{\sc v}. After having analyzed a significant number
of objects, covering a large parameter space, we should be able to
determine the importance of clumping, how it varies with spectral type and
wind density, and what the actual mass-loss rates are.

Additionally, the derived clumping factors have to be incorporated into {\it
stationary} wind-dynamics models. Using such models, we can investigate how
far the corresponding wind properties differ from models without clumping,
and check whether they are consistent with those derived from our
observational diagnostics. Remember that if assumption (b) or (c) were true,
the presently predicted line acceleration is much too large. It has to be
clarified whether strong clumping is able to induce such a large shift in
the ionization balance (see Sect.~\ref{clumping}) that the bulk of the
accelerating lines are shifted away from the flux maximum, such that a
reduction in the acceleration is possible.

Finally, time-dependent hydrodynamic simulations must also continue.
In particular, differences between self-excited and triggered structure
formation have to be investigated, and conditions found 
which might allow for a much more strongly clumped radio domain than presently
predicted (implied if assumptions (b) or (c) were true).

In this context, the following, concluding remark is relevant.
Though the usual interpretation of clumping relies on a relation to the 
intrinsic instability of radiative line-driving, the issue of whether the 
redistribution of wind material occurs predominantly on small ($\sim$0.01 
R$_\star$) or large ($\sim$1 R$_\star$) spatial scales has not 
yet been resolved. Small-scale clumping is suggested by observations of 
emission-line micro-variability in one of our targets (HD\,66811; see 
\citealt{evers98}). However, structuring of hot-star winds on 
large scales is indicated by the ubiquitous presence of recurrent wind 
profile variability in the form of discrete absorption components (DACs;
see, e.g., \citealt{prinhow86, kaper96}). Since there is no 
consensus on the physical origin of DACs, the structure responsible for
them is not included in the present generation of models. 

Future studies will help to address this issue by determining whether
objects with particularly well-studied DACs (e.g., HD\,24912, HD\,203064,
HD\,210839) can be modeled successfully without including large-scale
structure. The presence of unexplained residuals from our self-consistent
models (which cannot be discounted, due to missing
far-IR observations, and which might already have been identified in the mid-IR
fluxes of HD\,24912, or in the somewhat discordant mm-observations of
HD\,210839, cf. Appendix~\ref{comments}) with small-scale clumping would
imply that large-scale structures also play a role in the redistribution of
wind material, and would help to address the issue of whether
DACs represent localized enhancements in the mass flux.

\acknowledgements{Part of this investigation was supported by NATO
Collaborative Linkage Grant No. PST/CLG 980007, and by the National
Scientific Fundation of the Bulgarian Ministry of Education and Science,
under grant F-1407/2004. AWB acknowledges PPARC support. The very
constructive suggestions and comments of our anonymous referee are
gratefully acknowledged.}

\appendix 
\section{The journal of the VLA observations}
is given in Table~\ref{radioobs} (see Sect.~\ref{radio}).

\label{radiolog}
\begin{table}
\caption{Journal of the VLA observations, including observation dates, 
observing frequencies, time on targets, calibrators for flux-density
bootstrapping and VLA configuration.}
\tabcolsep1.5mm
\begin{tabular}{llcccc}
\hline
\multicolumn{1}{l}{Star}
&\multicolumn{1}{c}{date}
&\multicolumn{1}{c}{freq.}
&\multicolumn{1}{c}{time}
&\multicolumn{1}{c}{cal}
&\multicolumn{1}{c}{conf.}\\
\multicolumn{1}{c}{}
&\multicolumn{1}{c}{}
&\multicolumn{1}{c}{(GHz)}
&\multicolumn{1}{c}{(min)}
&\multicolumn{1}{c}{} 
&\multicolumn{1}{c}{}\\
\hline                
Cyg\,OB2\#7  & Feb 15, 2004 & 4.86 & 60 & 3C286,3C48  &CnB \\
Cyg\,OB2\#7  & Feb 15, 2004 & 8.46 & 45 & 3C286,3C48  &CnB \\
Cyg\,OB2\#10 & Feb 15, 2004 & 4.86 & 20 & 3C286,3C48  &CnB \\
Cyg\,OB2\#10 & Feb 20, 2004 & 4.86 & 40 & 3C286,3C48  &C \\
Cyg\,OB2\#10 & Feb 15, 2004 & 8.46 & 20 & 3C286,3C48  &CnB \\
Cyg\,OB2\#10 & Feb 20, 2004 & 8.46 & 25 & 3C286,3C48  &C \\
Cyg\,OB2\#10 & Feb 15, 2004 &14.94 & 20 & 3C286,3C48  &CnB \\
Cyg\,OB2\#10 & Feb 20, 2004 &14.94 & 40 & 3C286,3C48  &C \\
Cyg\,OB2\#11 & Feb 15, 2004 & 4.86 & 30 & 3C286,3C48  &CnB \\
Cyg\,OB2\#11 & Feb 15, 2004 & 8.46 & 30 & 3C286,3C48  &CnB \\
Cyg\,OB2\#11 & Feb 15, 2004 &14.94 & 30 & 3C286,3C48  &CnB \\
HD\,14947   & Apr 04, 2004 & 4.86 & 30 & 3C48        &C \\
HD\,14947   & Apr 04, 2004 & 8.46 & 30 & 3C48        &C \\
HD\,14947   & Apr 04, 2004 &14.94 & 30 & 3C48        &C \\
HD\,24912   & Mar 09, 2004 & 4.86 & 20 & 3C286       &C \\
HD\,24912   & Mar 09, 2004 & 8.46 & 15 & 3C286       &C \\
HD\,24912   & Mar 09, 2004 &14.94 & 20 & 3C286       &C \\
HD\,24912   & Mar 09, 2004 &43.34 & 20 & 3C286       &C \\
HD\,34656   & Apr 04, 2004 & 4.86 & 40 & 3C48        &C   \\
HD\,34656   & Feb 09, 2004 & 8.46 & 45 & 3C147,3C48  &CnB \\
HD\,34656   & Mar 09, 2004 &14.94 & 20 & 3C286       &C   \\
HD\,34656   & Apr 04, 2004 &14.94 & 20 & 3C147,3C48  &C   \\
HD\,36861   & Feb 09, 2004 & 8.46 & 40 & 3C147,3C48  &CnB \\
HD\,36861   & Feb 09, 2004 & 4.86 & 40 & 3C147,3C48  &CnB \\
HD\,37043   & Mar 09, 2004 & 4.86 & 30 & 3C286       &C \\
HD\,37043   & Mar 09, 2004 & 8.46 & 30 & 3C286       &C \\
HD\,37043   & Mar 09, 2004 &14.94 & 30 & 3C286       &C \\
HD\,190429A & Mar 01, 2004 & 4.86 & 20 & 3C48        &C \\
HD\,190429A & Mar 01, 2004 & 8.46 & 20 & 3C48        &C \\
HD\,190429A & Mar 01, 2004 &14.94 & 20 & 3C48        &C \\
HD\,190429A & Feb 26, 2004 &43.34 & 20 & 3C48        &C \\
HD\,203064  & Mar 01, 2004 & 4.86 & 60 & 3C48        &C \\
HD\,203064  & Mar 01, 2004 & 8.46 & 60 & 3C48        &C \\
HD\,203064  & Apr 04, 2004 &14.94 & 40 & 3C48        &C \\
HD\,207198  & Feb 20, 2004 & 4.86 & 60 & 3C286,3C48  &C \\
HD\,207198  & Feb 20, 2004 & 8.46 & 60 & 3C286,3C48  &C \\
HD\,207198  & Feb 20, 2004 &14.94 & 60 & 3C286,3C48  &C \\
HD\,209975  & Feb 20, 2004 & 4.86 & 30 & 3C286,3C48  &C \\
HD\,209975  & Feb 20, 2004 & 8.46 & 30 & 3C286,3C48  &C \\
HD\,209975  & Feb 20, 2004 &14.94 & 30 & 3C286,3C48  &C \\
HD\,210839  & Feb 26, 2004 & 4.86 & 20 & 3C48        &C \\
HD\,210839  & Feb 26, 2004 & 8.46 & 20 & 3C48        &C \\
HD\,210839  & Feb 26, 2004 &14.94 & 20 & 3C48        &C \\
HD\,210839  & Feb 26, 2004 &43.34 & 20 & 3C48        &C \\
\hline
\end{tabular}
\label{radioobs}
\end{table}

\section{Comments on individual objects} 
\label{comments} 

\begin{figure*}
\begin{center}
\begin{minipage}{7.8cm}
\resizebox{\hsize}{!}
   {\includegraphics{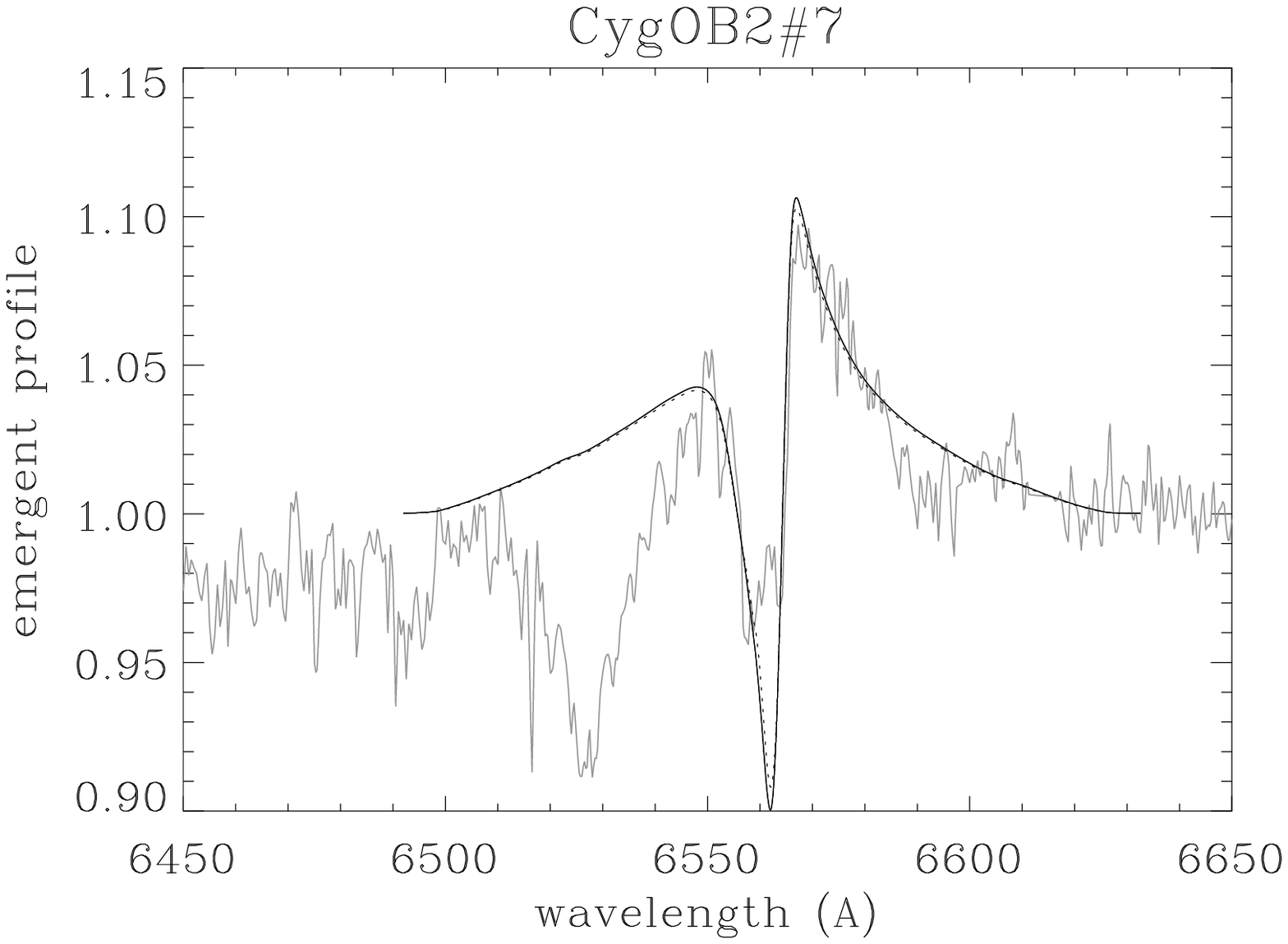}}
\end{minipage}
\begin{minipage}{7.8cm}
   \resizebox{\hsize}{!}
   {\includegraphics{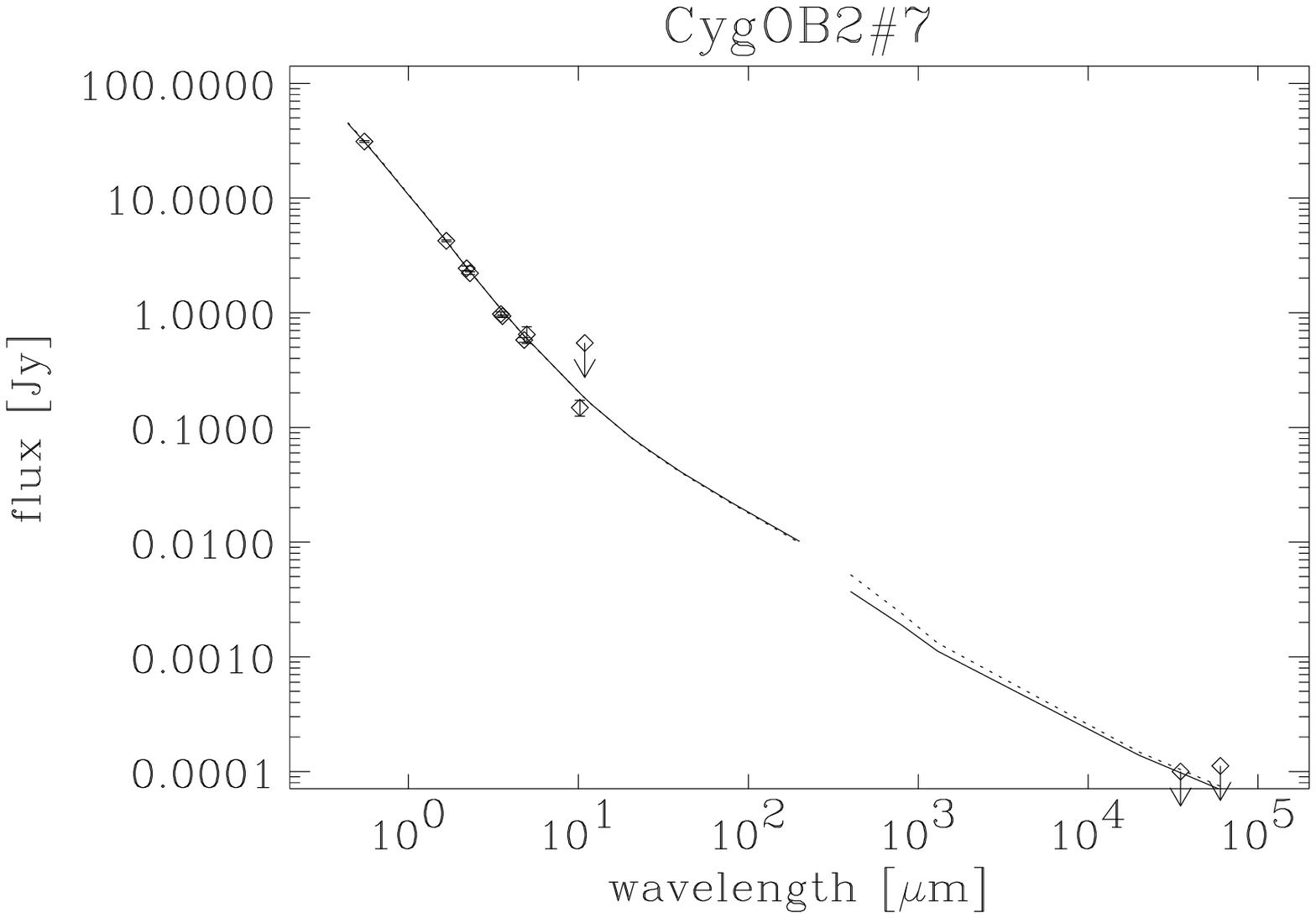}}
\end{minipage}

\begin{minipage}{7.8cm}
   \resizebox{\hsize}{!}
   {\includegraphics{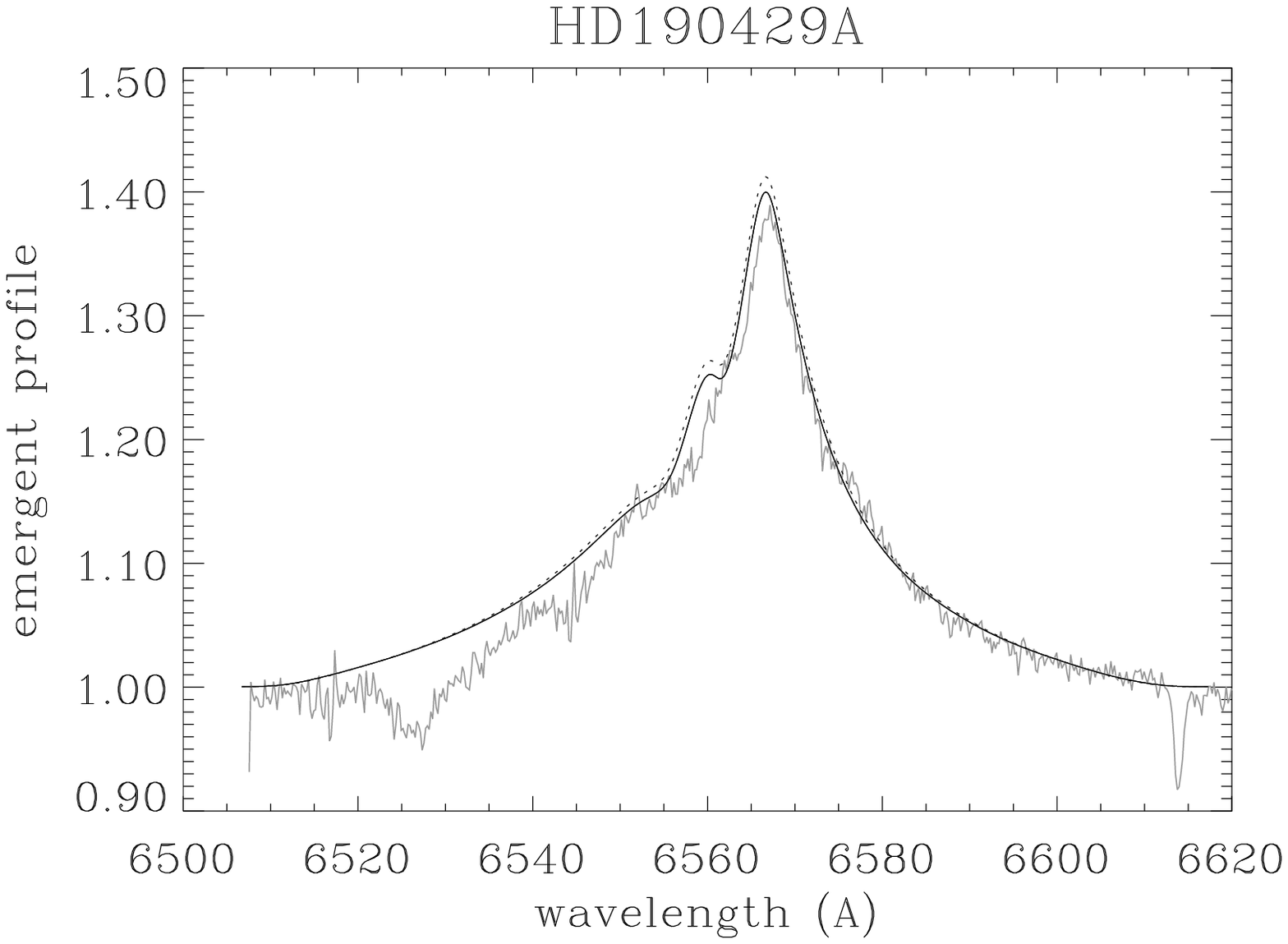}}
\end{minipage}
\begin{minipage}{7.8cm}
   \resizebox{\hsize}{!}
   {\includegraphics{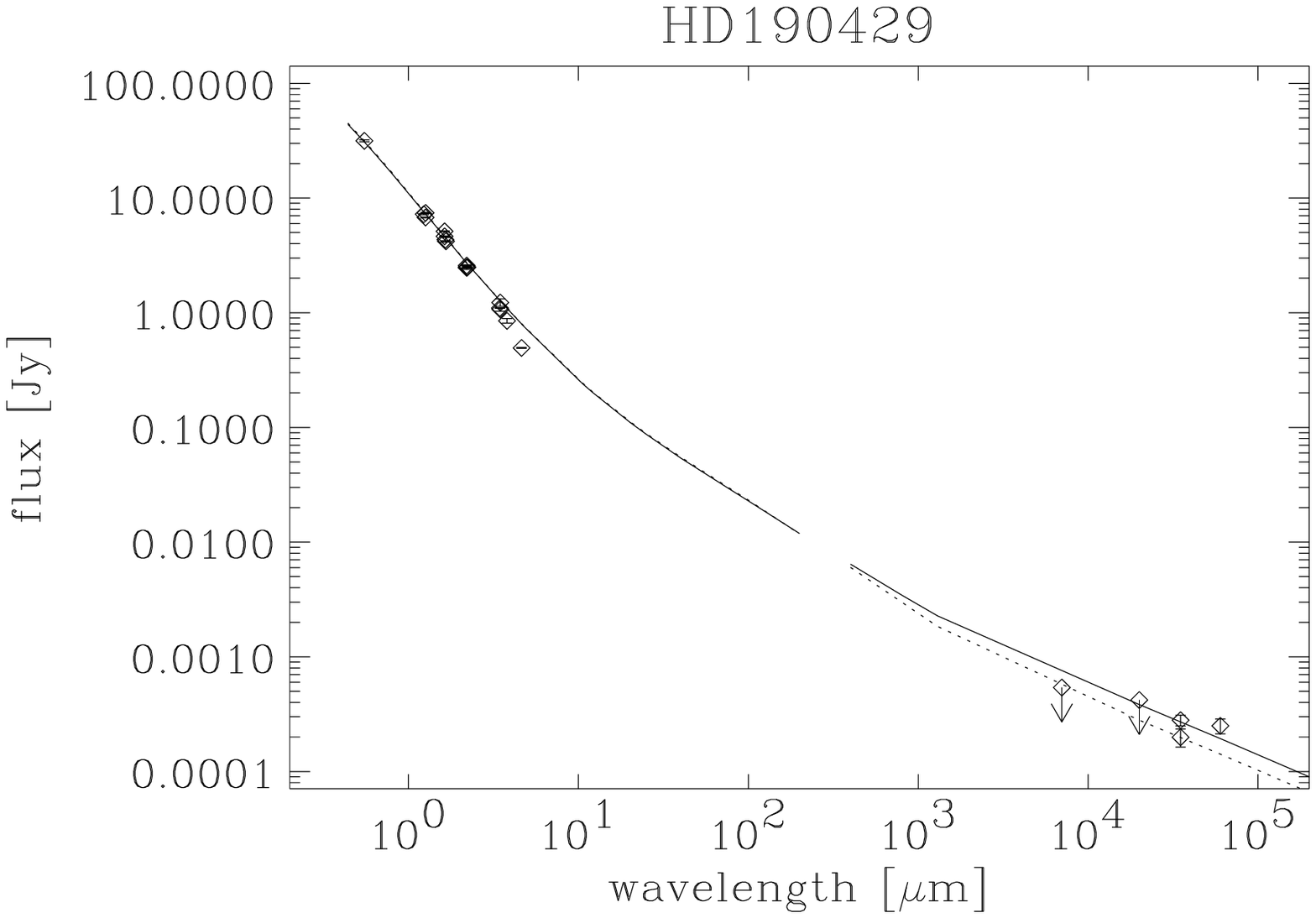}}
\end{minipage}

\begin{minipage}{7.8cm}
   \resizebox{\hsize}{!}
   {\includegraphics{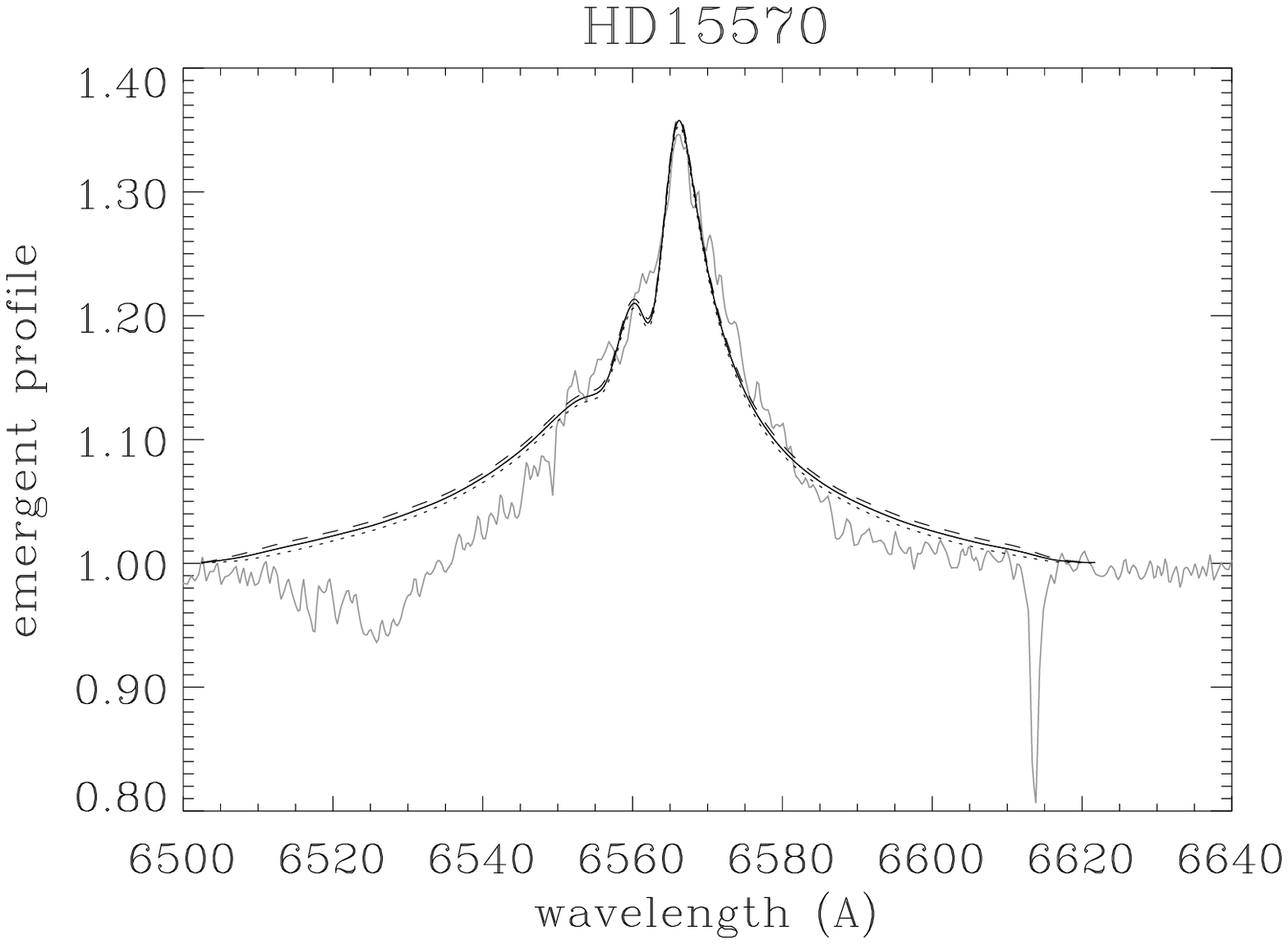}}
\end{minipage}
\begin{minipage}{7.8cm}
   \resizebox{\hsize}{!}
   {\includegraphics{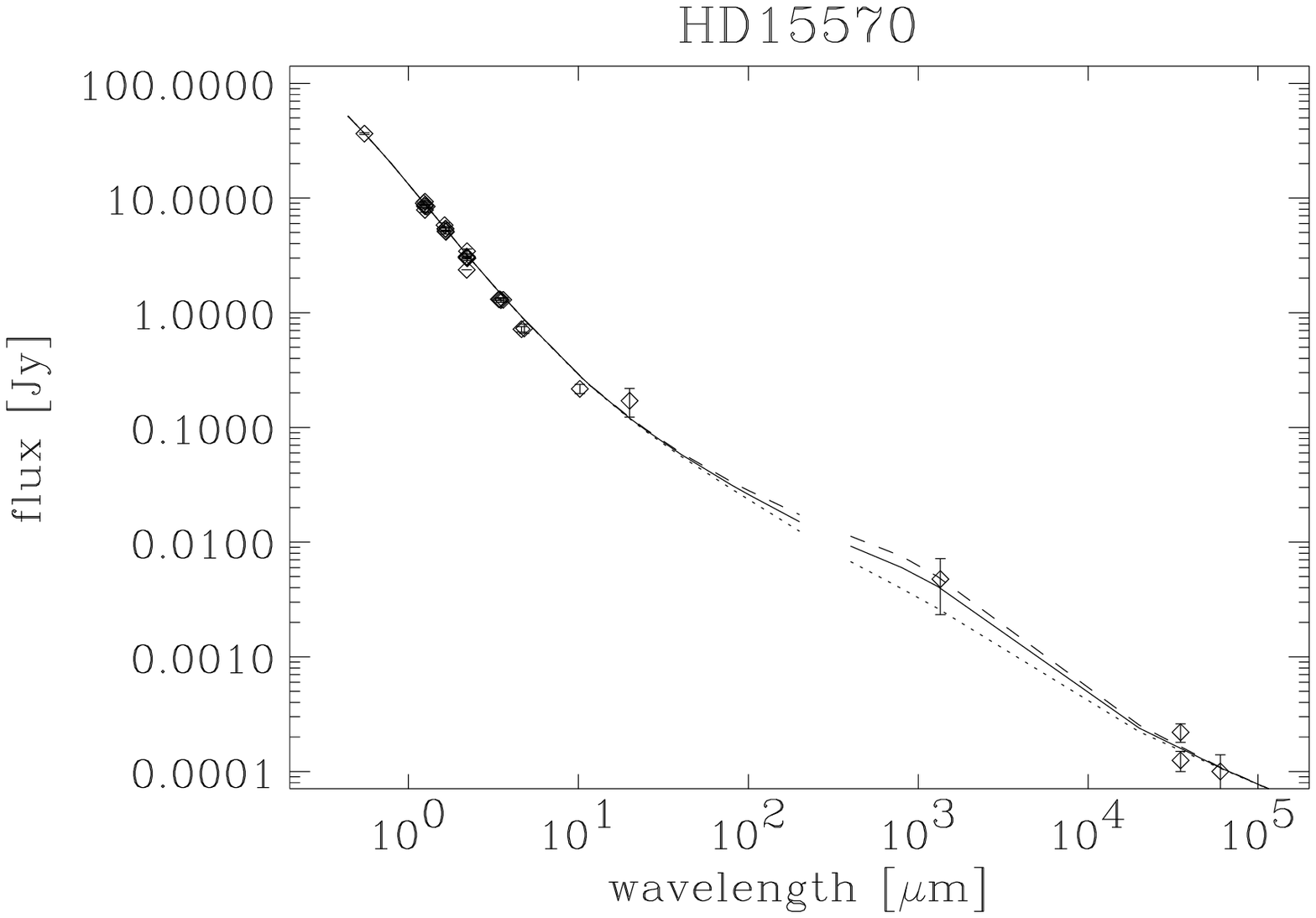}}
\end{minipage}

\begin{minipage}{7.8cm}
   \resizebox{\hsize}{!}
   {\includegraphics{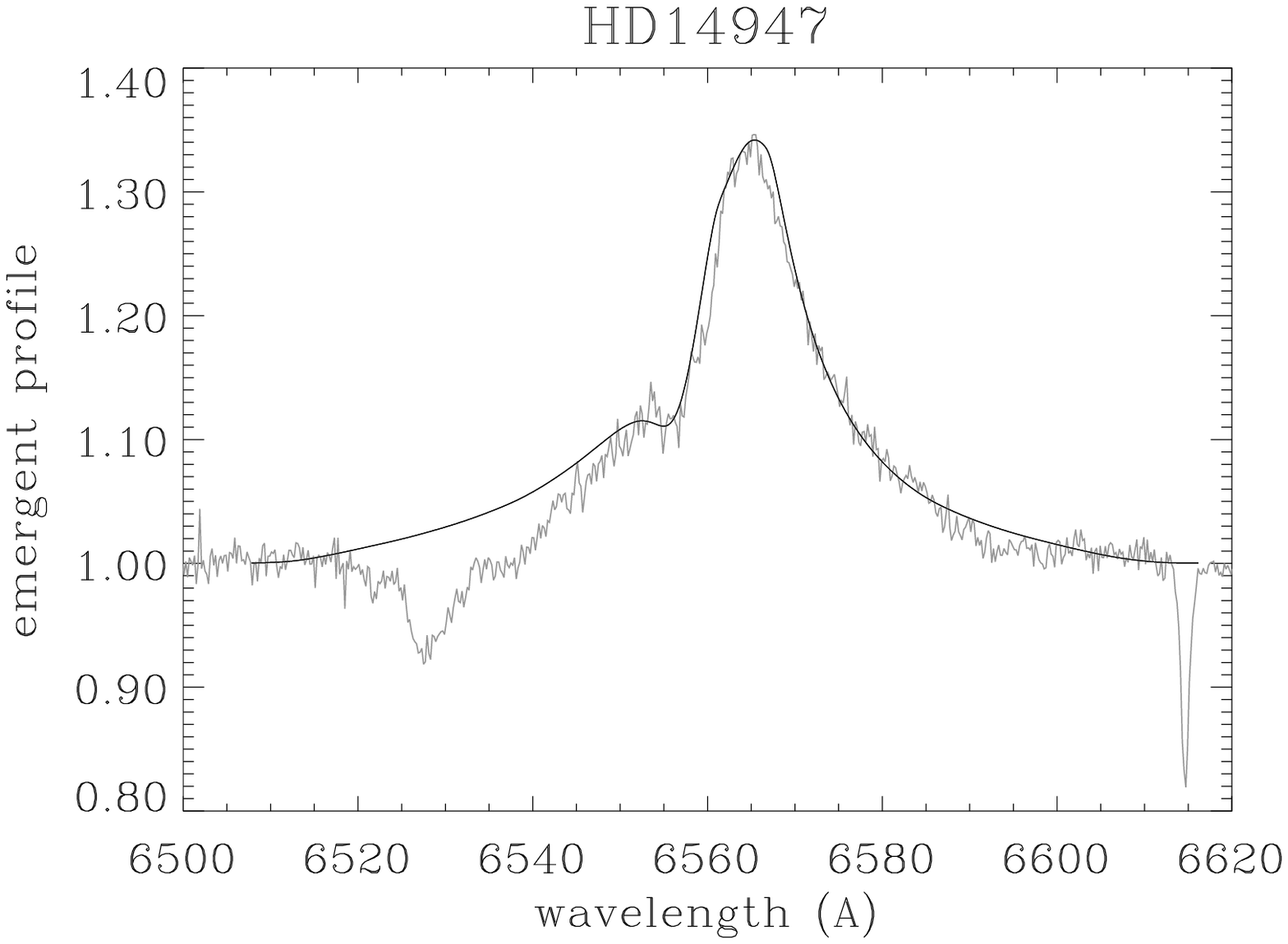}}
\end{minipage}
\begin{minipage}{7.8cm}
   \resizebox{\hsize}{!}
   {\includegraphics{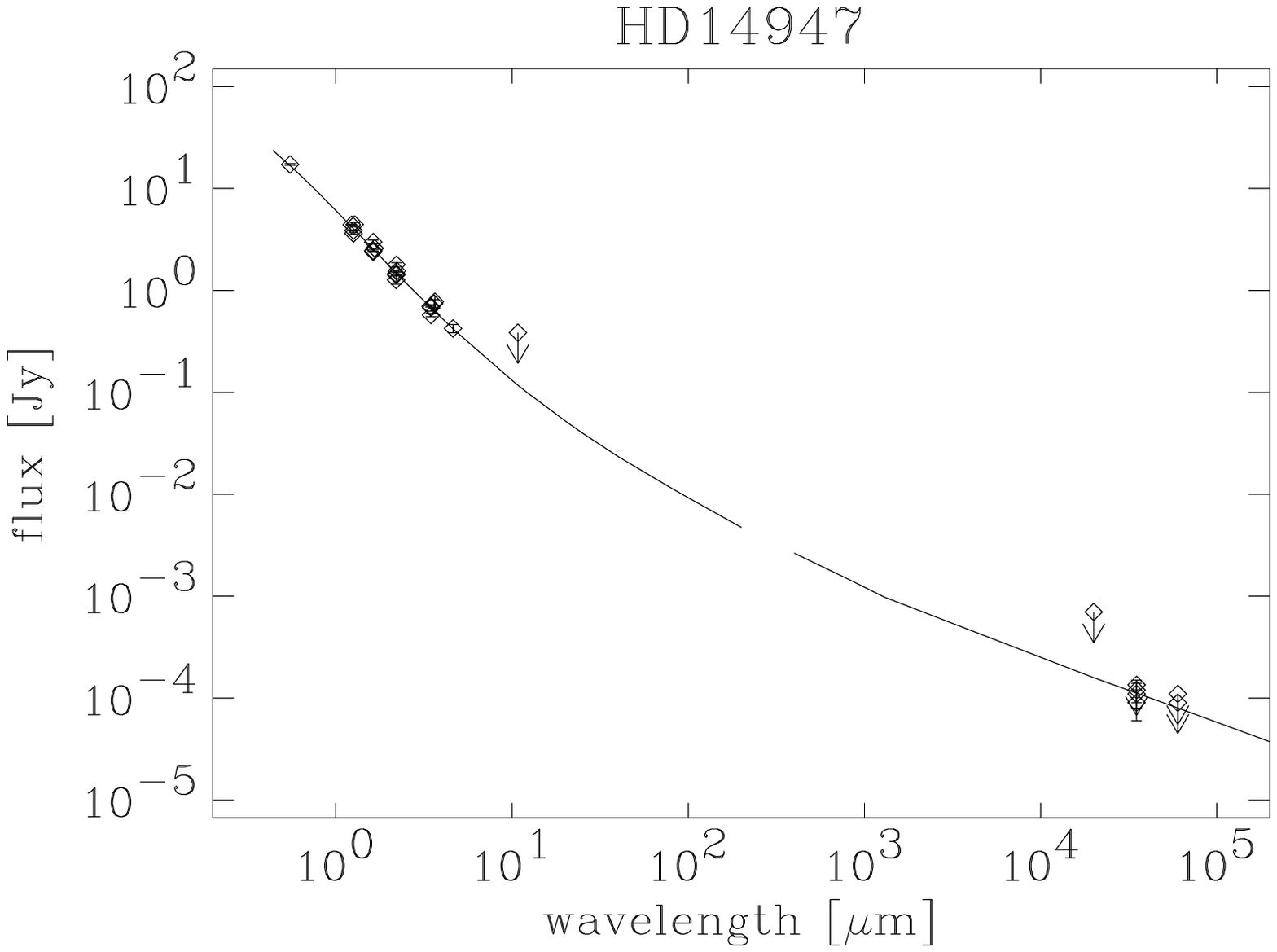}}
\end{minipage}
\end{center}
\caption{Fit diagrams (left: \Ha\ profile; right: IR/radio continua) for 
objects with \Ha\ in emission. Arrows indicate upper limits. For parameters,
see Table~\ref{table_clf}. Alternative solutions (dotted, dashed) are
discussed in the comments on individual objects in Appendix~\ref{comments}.}
\label{fits_haem_1}
\end{figure*}

\begin{figure*}
\begin{center}
\begin{minipage}{7.8cm}
\resizebox{\hsize}{!}
   {\includegraphics{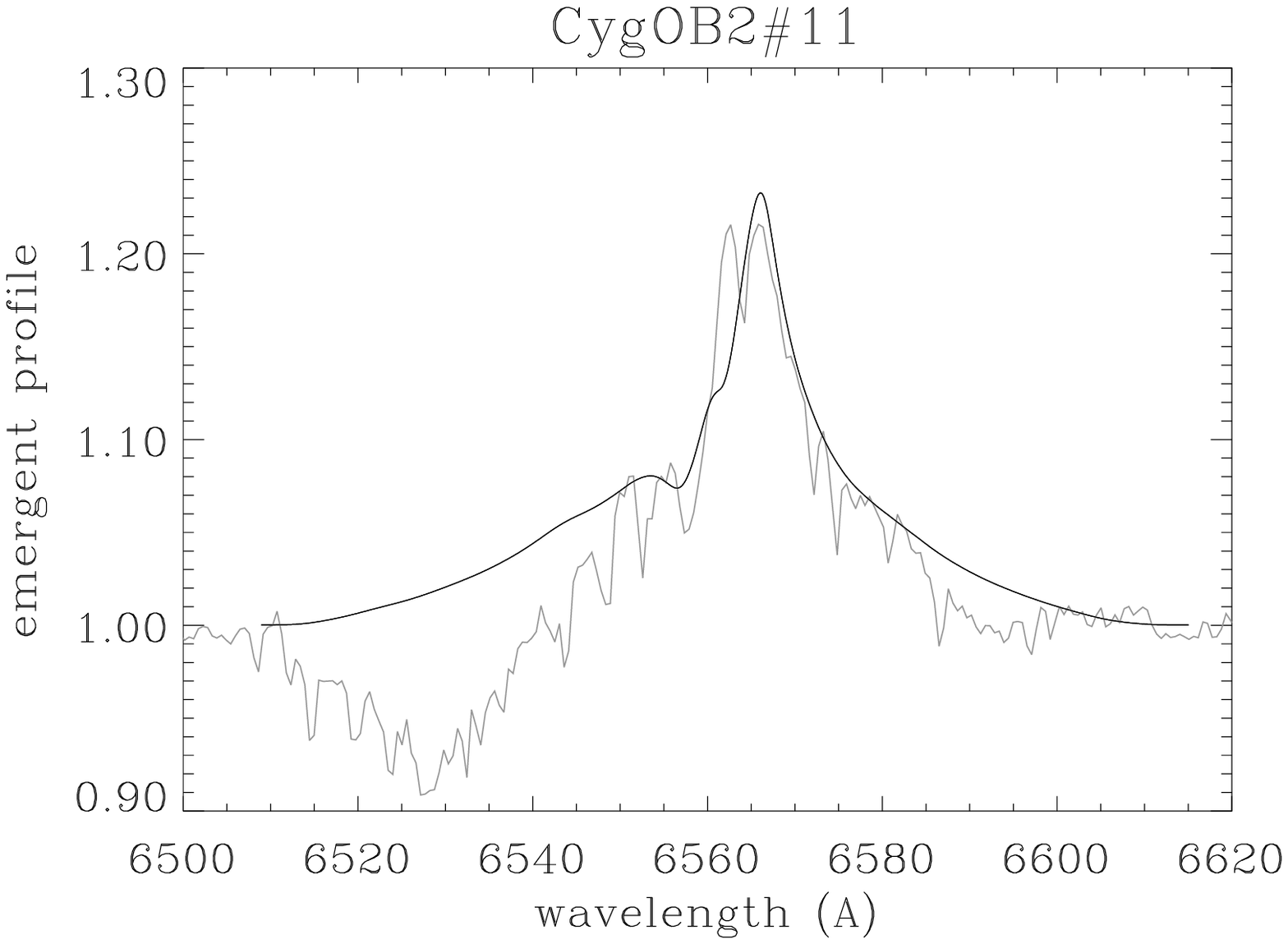}}
\end{minipage}
\begin{minipage}{7.8cm}
   \resizebox{\hsize}{!}
   {\includegraphics{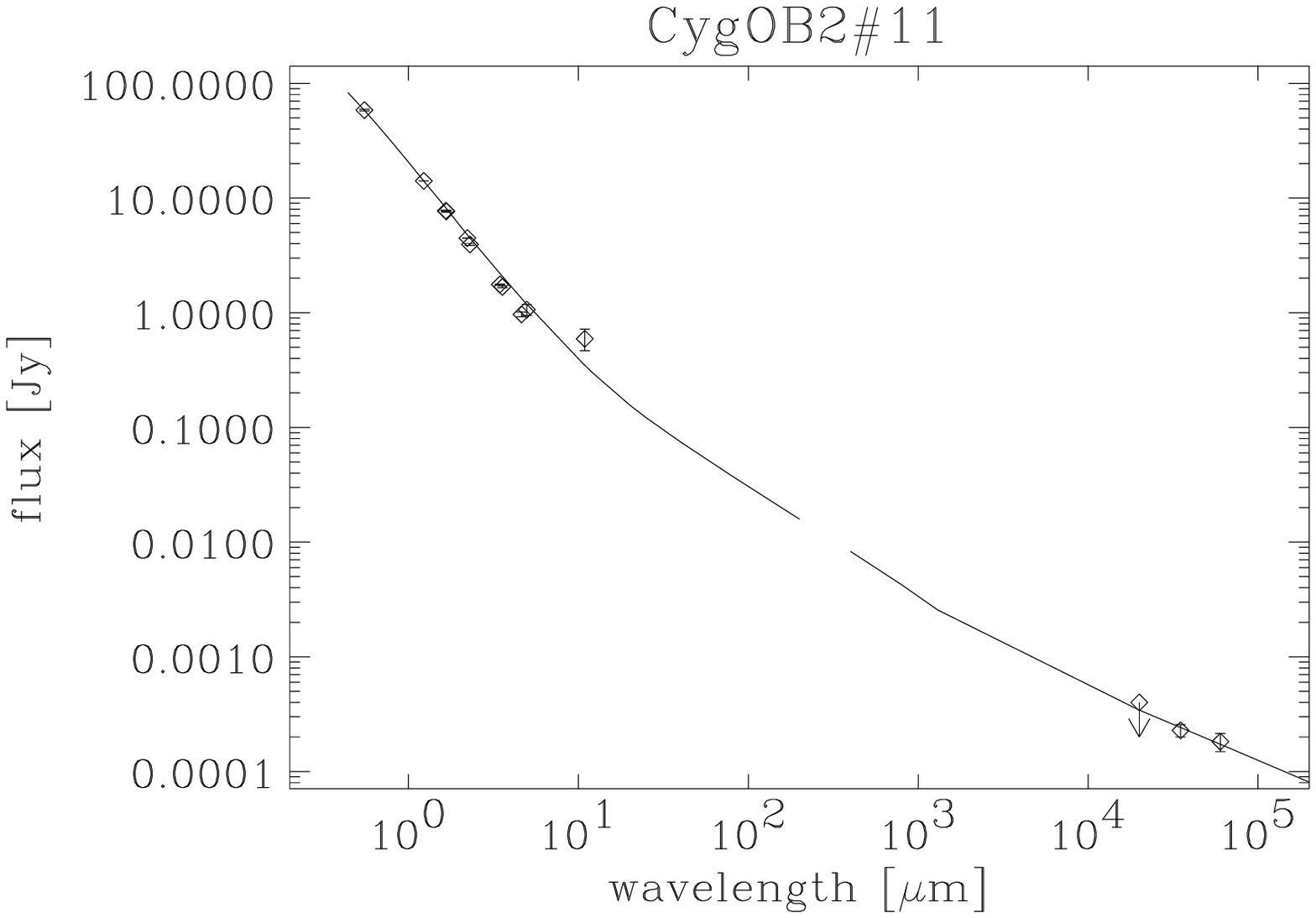}}
\end{minipage}

\begin{minipage}{7.8cm}
   \resizebox{\hsize}{!}
   {\includegraphics{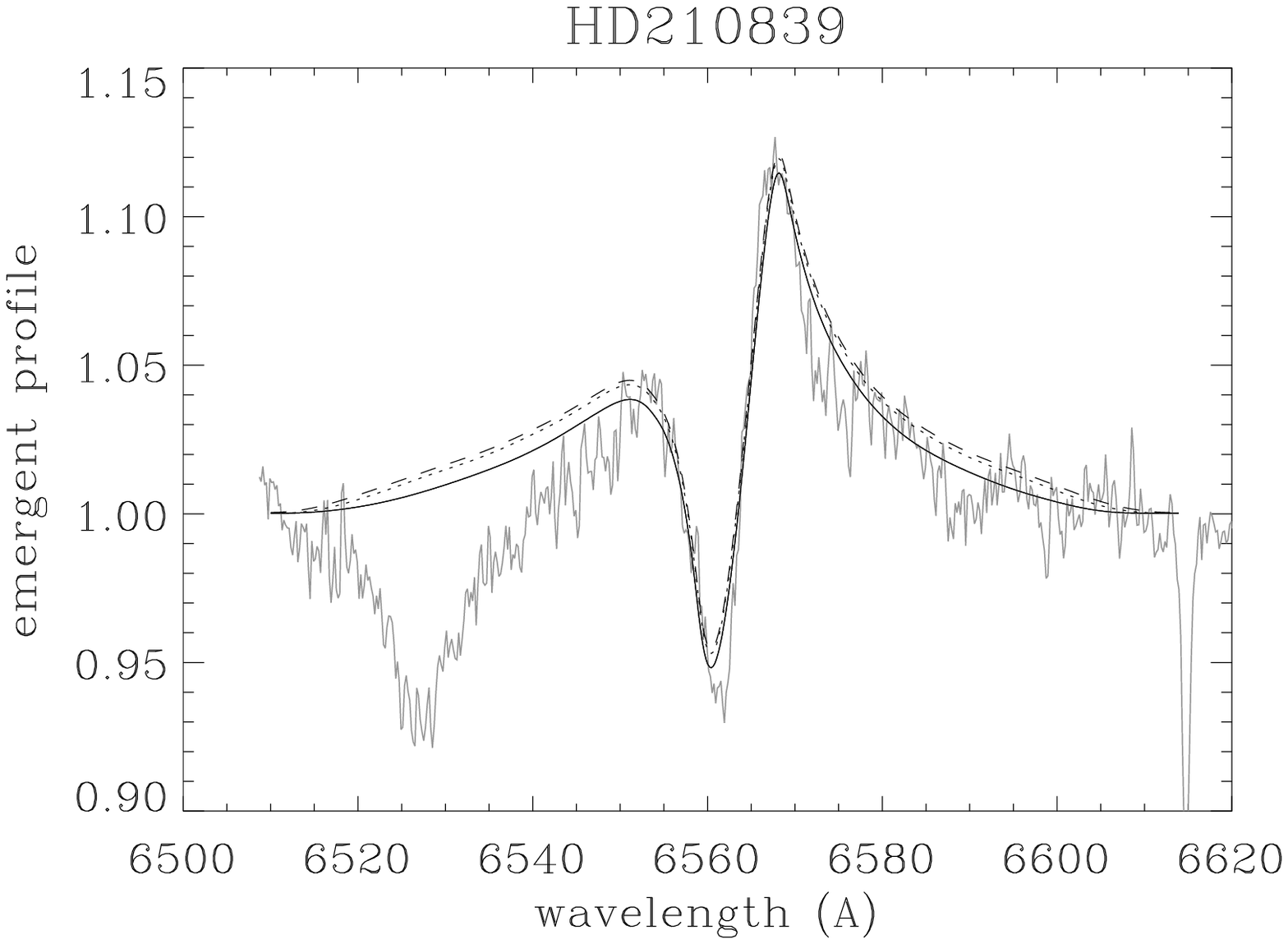}}
\end{minipage}
\begin{minipage}{7.8cm}
   \resizebox{\hsize}{!}
   {\includegraphics{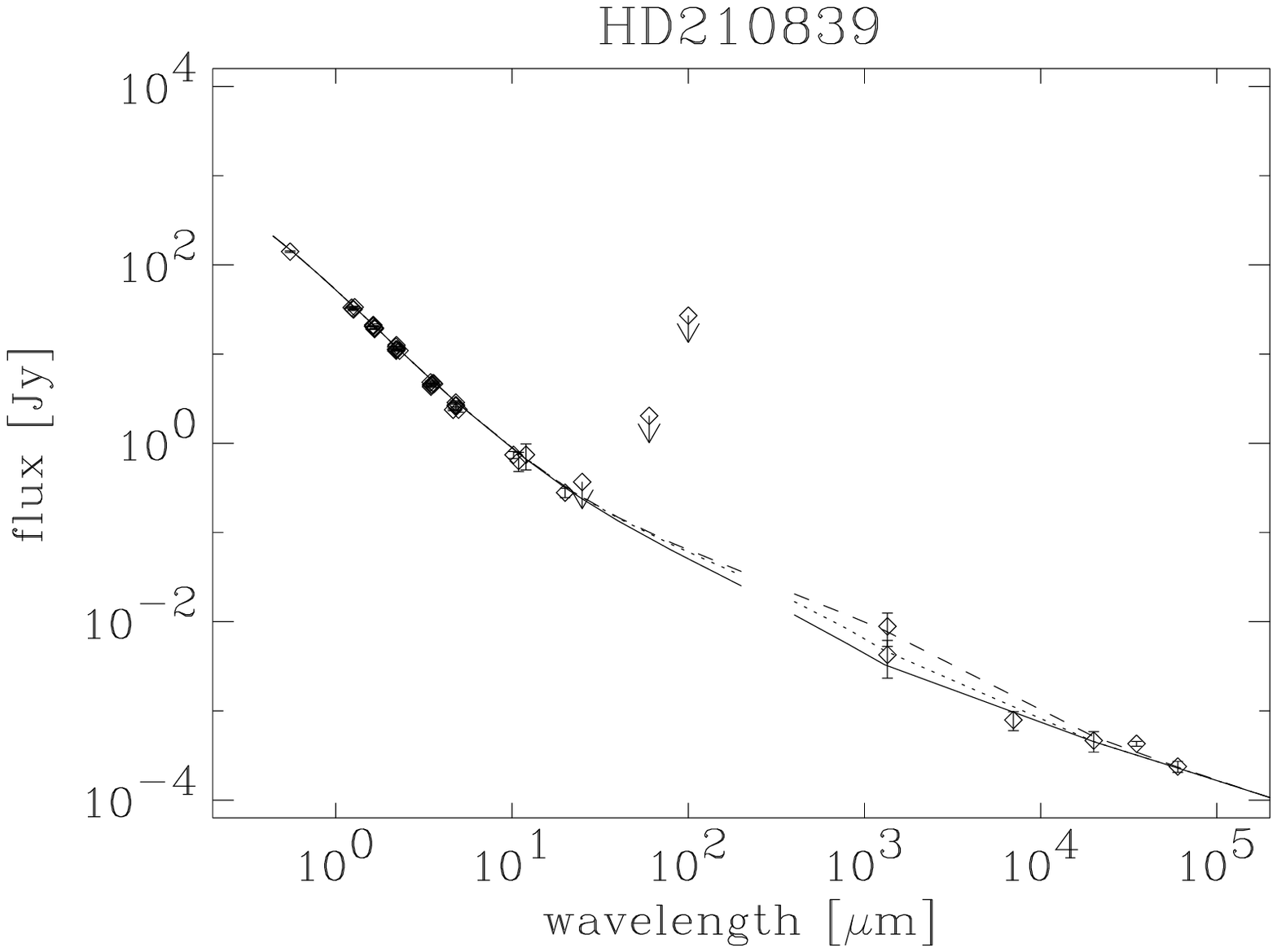}}
\end{minipage}

\begin{minipage}{7.8cm}
   \resizebox{\hsize}{!}
   {\includegraphics{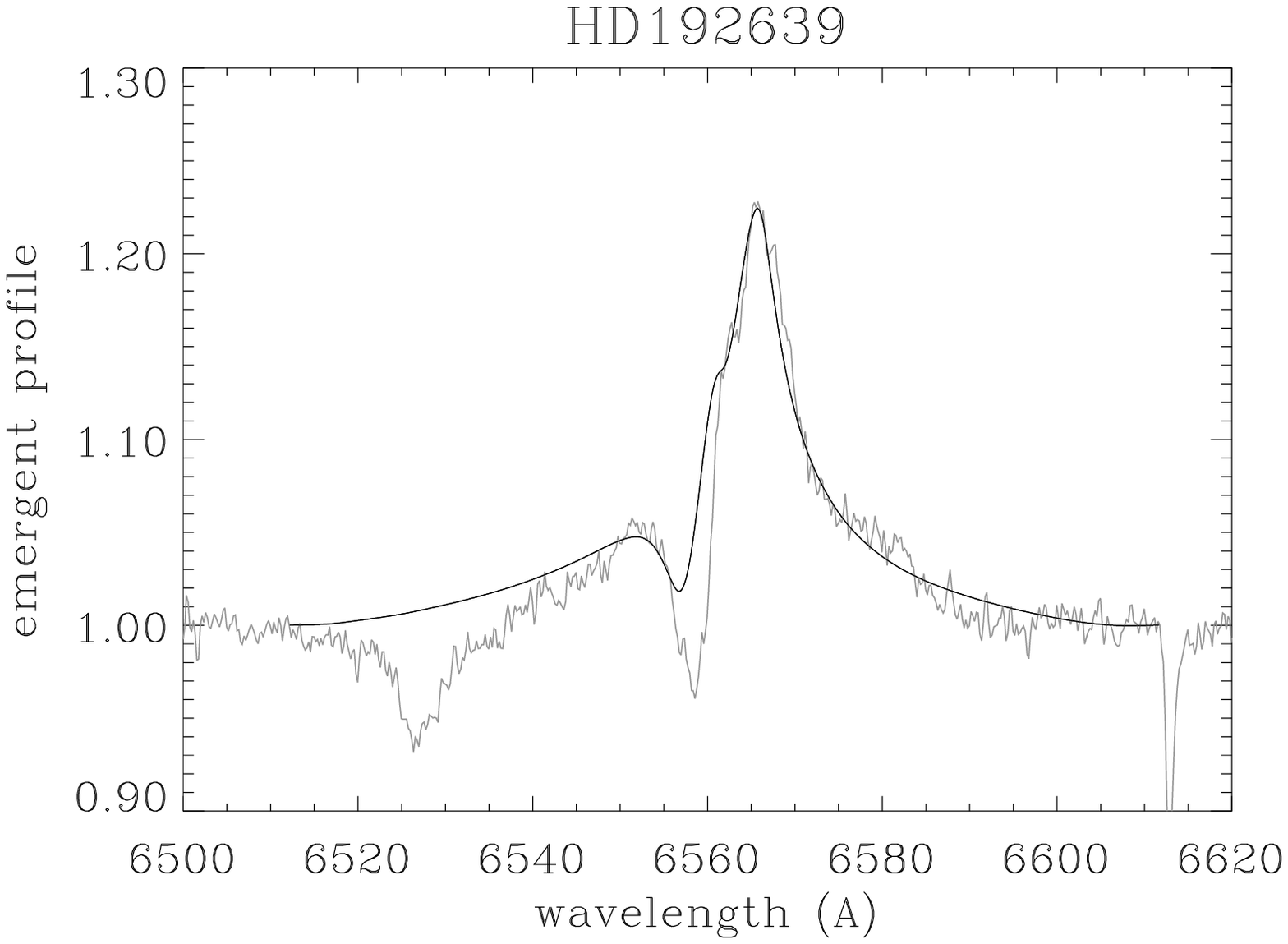}}
\end{minipage}
\begin{minipage}{7.8cm}
   \resizebox{\hsize}{!}
   {\includegraphics{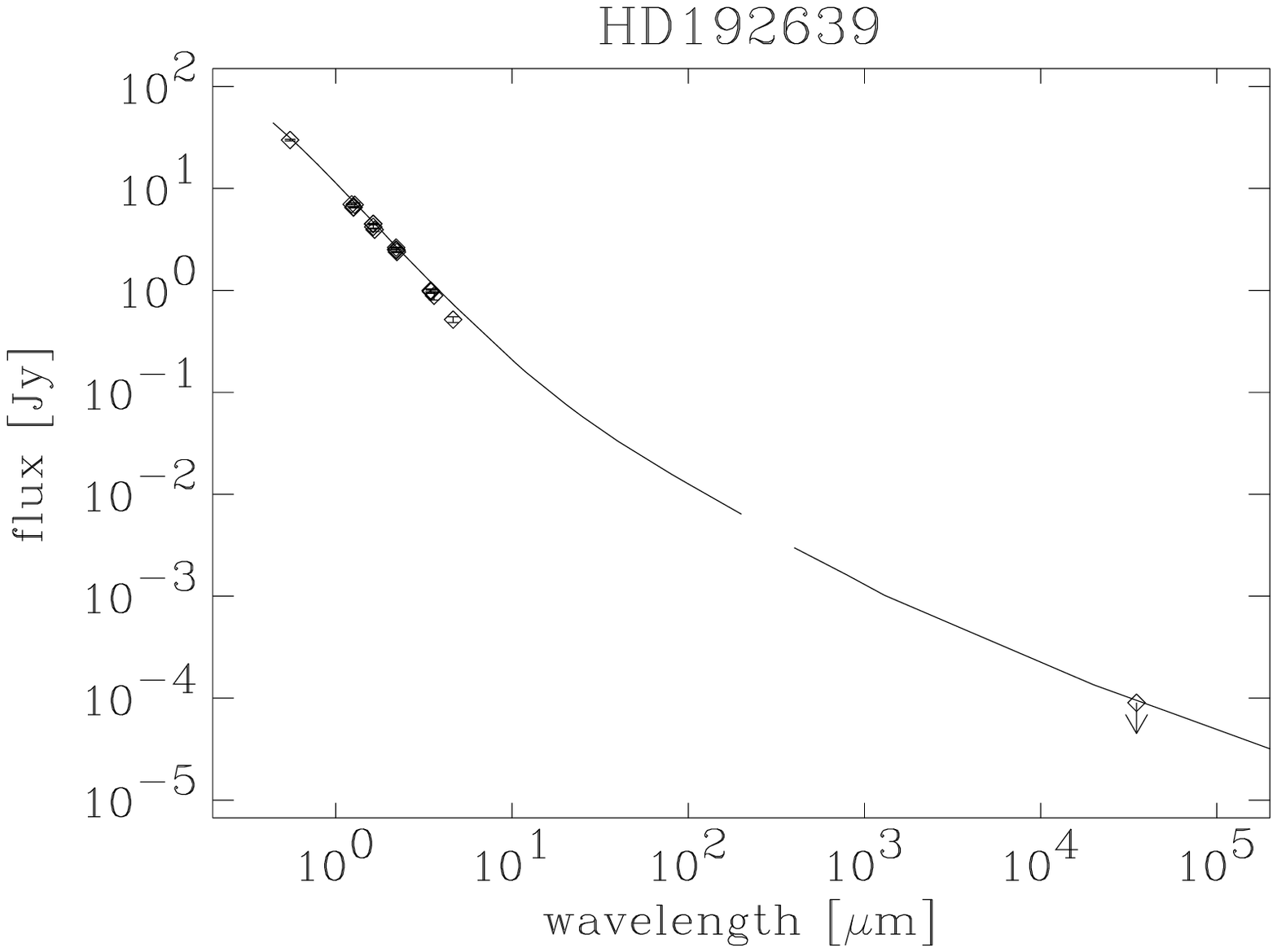}}
\end{minipage}

\begin{minipage}{7.8cm}
   \resizebox{\hsize}{!}
   {\includegraphics{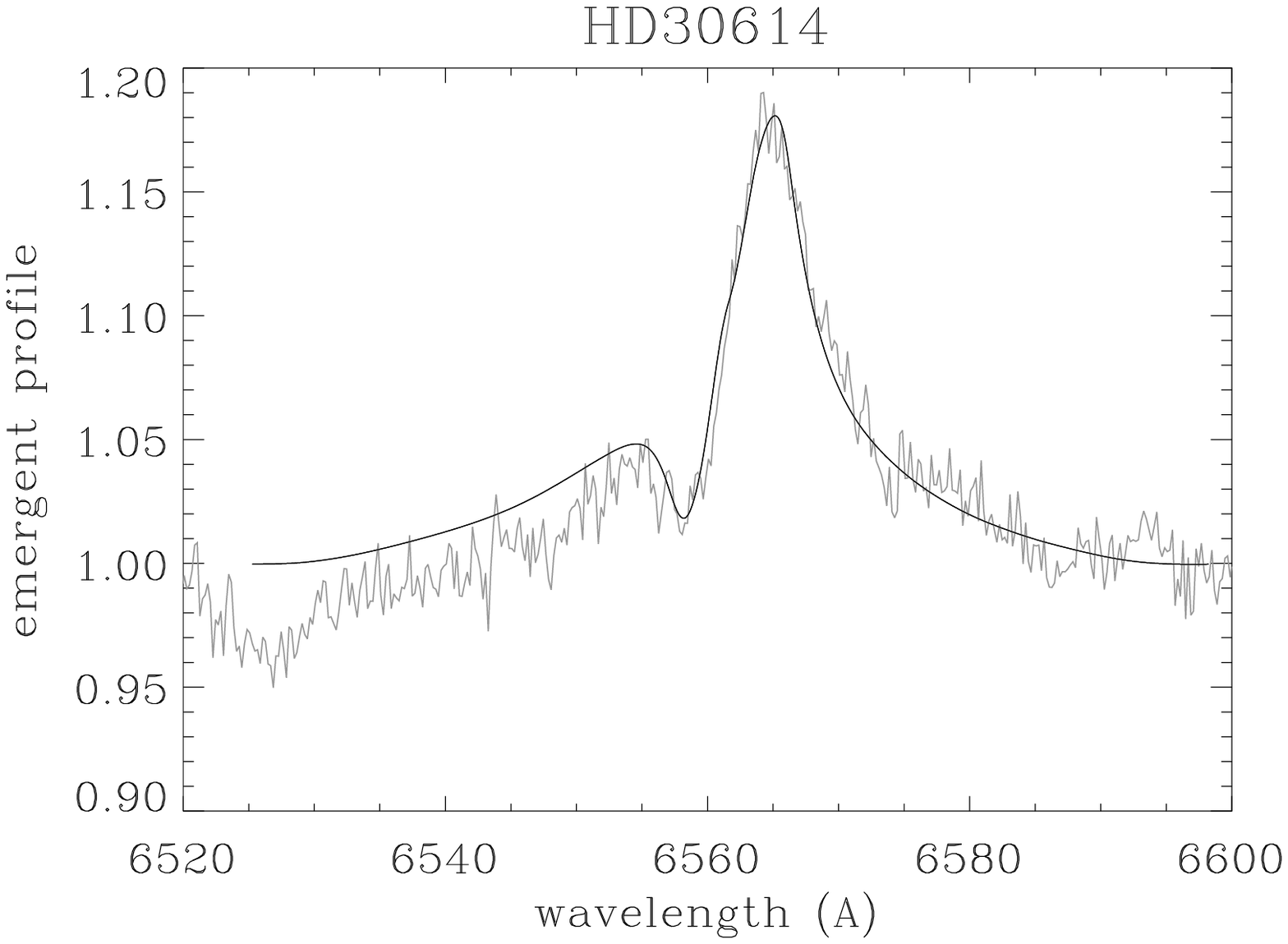}}
\end{minipage}
\begin{minipage}{7.8cm}
   \resizebox{\hsize}{!}
   {\includegraphics{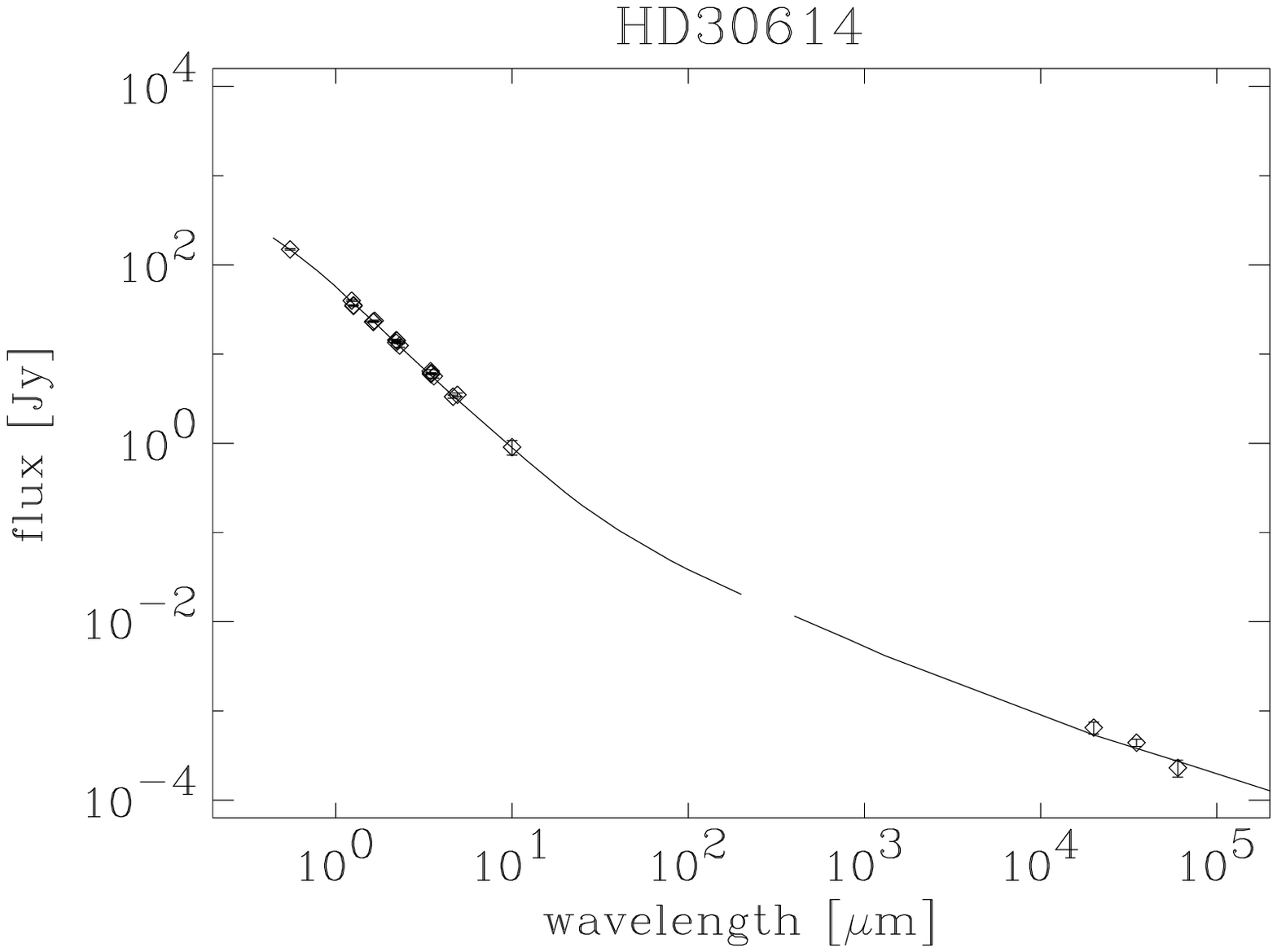}}
\end{minipage}
\end{center}
\caption{As Fig.~\ref{fits_haem_1}.}
\label{fits_haem_2}
\end{figure*}

\begin{figure*}
\begin{center}
\begin{minipage}{7.8cm}
\resizebox{\hsize}{!}
   {\includegraphics{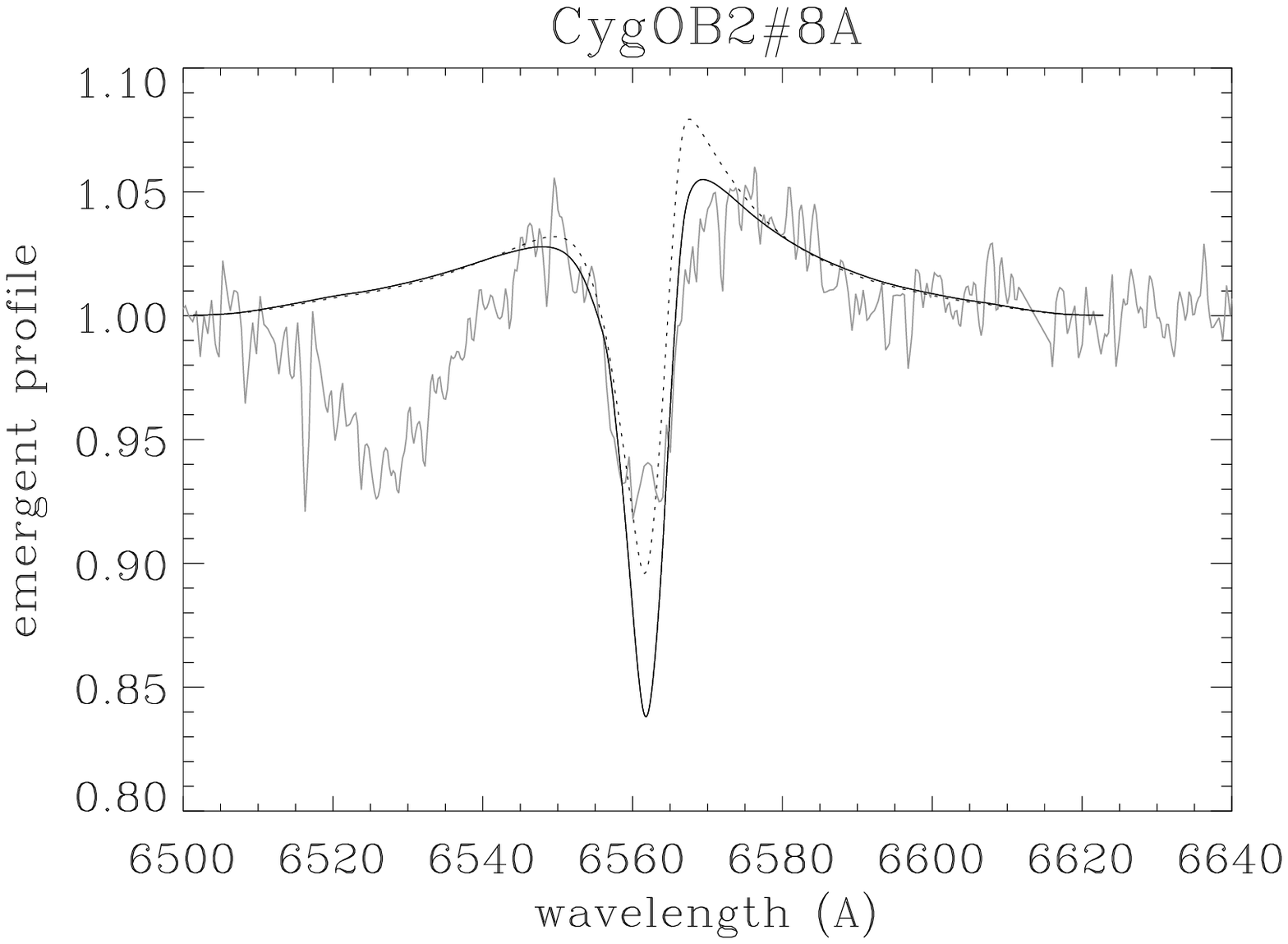}}
\end{minipage}
\begin{minipage}{7.8cm}
   \resizebox{\hsize}{!}
   {\includegraphics{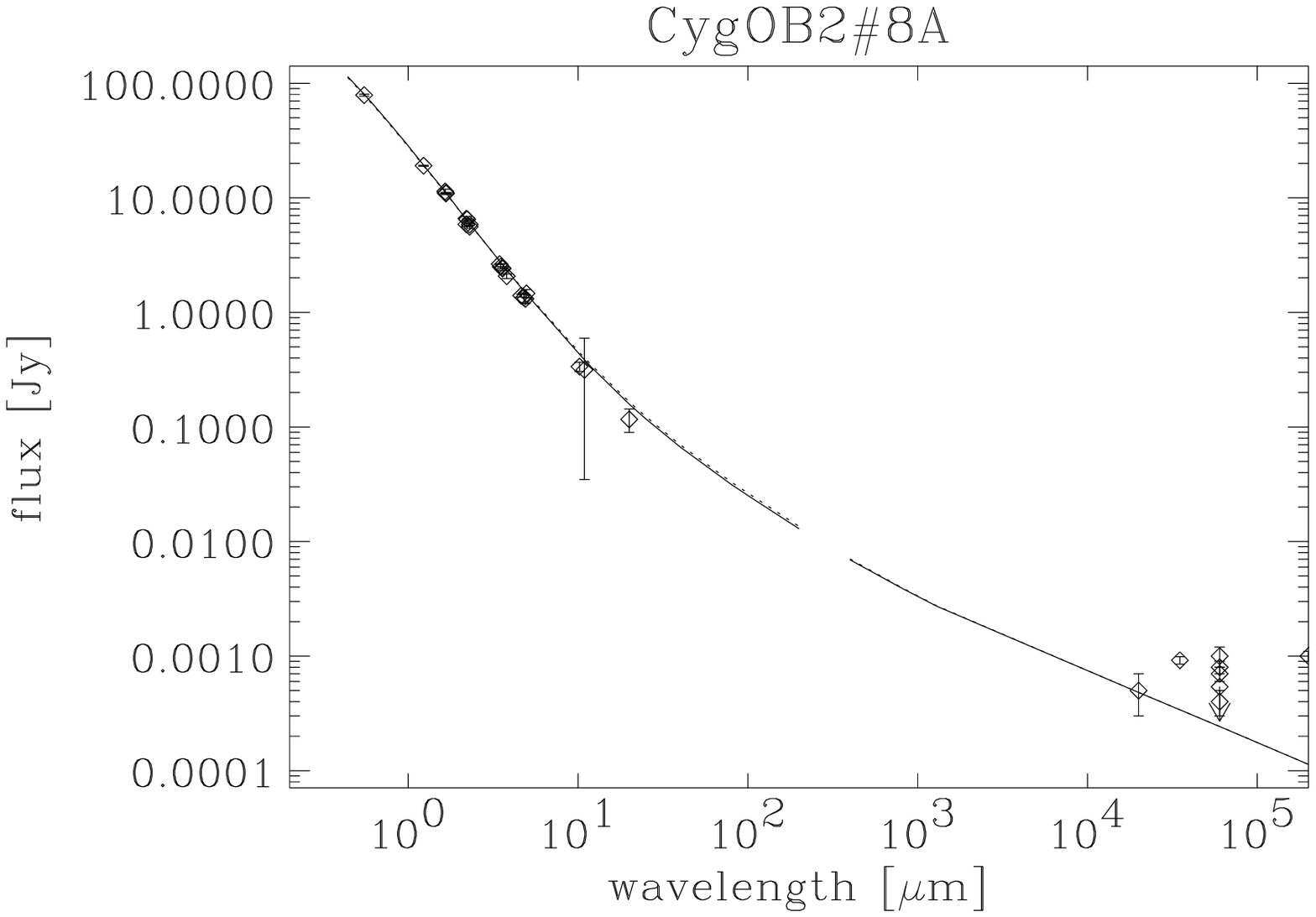}}
\end{minipage}

\begin{minipage}{7.8cm}
\resizebox{\hsize}{!}
   {\includegraphics{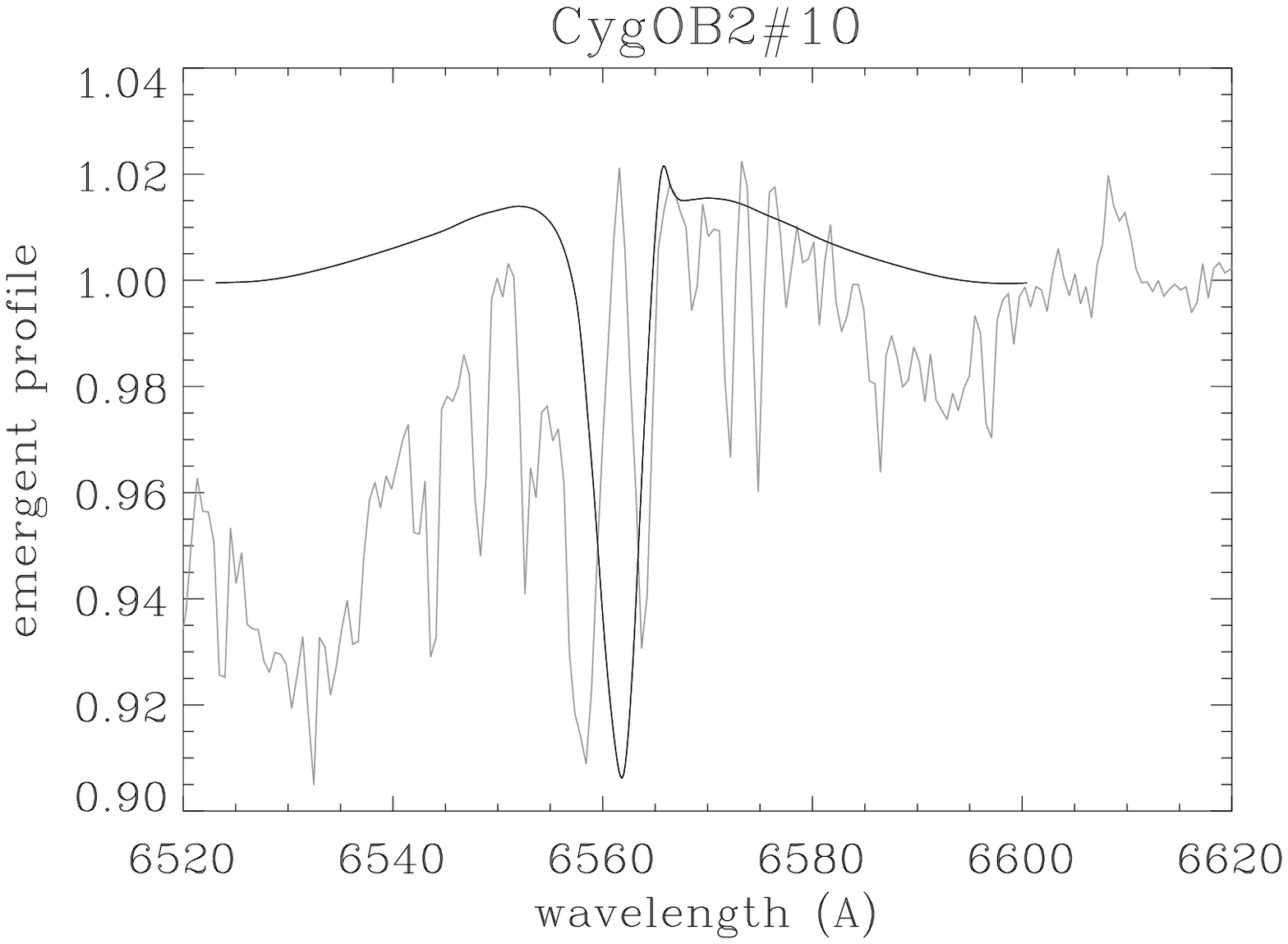}}
\end{minipage}
\begin{minipage}{7.8cm}
   \resizebox{\hsize}{!}
   {\includegraphics{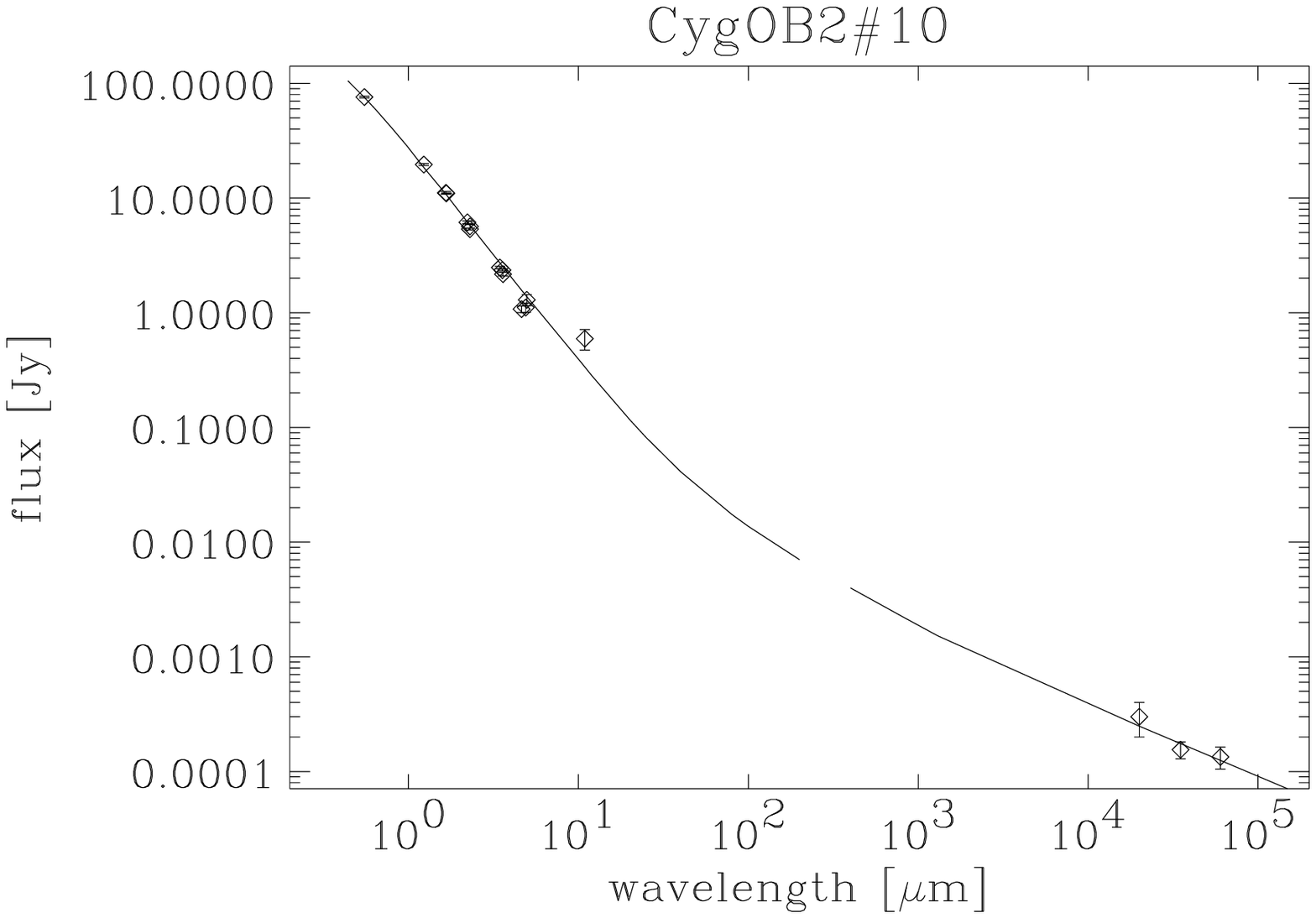}}
\end{minipage}
\end{center}
\caption{As Fig.~\ref{fits_haem_1}, but for objects with ``intermediate''
\Ha-profile type.}
\label{fits_hainter}
\end{figure*}

In the following, we will give, where necessary, some comments on the fits
for the invidual objects. All results have been summarized in
Table~\ref{table_clf}. The fits for objects with \Ha\ in emission are
displayed in Figs.~\ref{fits_haem_1} and \ref{fits_haem_2}, for objects with
``intermediate'' \Ha\ profile types in Fig.~\ref{fits_hainter}, and for
objects with \Ha\ in absorption in Figs.~\ref{fits_haabs_1} and
\ref{fits_haabs_2}.

\subsection{Objects with \Ha\ in emission}

\paragraph{Cyg\,OB2\#7.} 

For the hottest object in our sample, only upper limits for the radio fluxes
are available. The derived mass-loss rate is consequently an upper limit as
well (and the clumping factors corresponding lower limits), and based on the
{\it assumption} that this star is a thermal emitter. By means of our
regression (Eq.~\ref{herec_para}), helium is predicted to remain doubly
ionized throughout the entire wind (this is the only object in our sample
for which this is so), whereas specific models within our grid (located in
the relevant parameter range) indicate that helium might still recombine in
the outermost, radio-emitting region. Thus we have derived two solutions for
this object, both for an ionized and a recombined radio regime.\footnote{The
IR fluxes have been synthesized with doubly ionized helium in both cases,
since they form well below the radio photosphere.} 

For the doubly ionized solution, we derive a (maximum) mass-loss rate of
2.8\Mdu. The lower wind is strongly clumped to a similar degree in regions 2
and 3 ($\fin=10$ and $\fmid=8{\ldots}12$, respectively. The lower value for
$\fmid$ results in a good fit of the 10~$\mu{\rm m}$ flux, but slightly too
narrow wings of \Ha, whereas with $\fmid=12$ we can fit these wings
perfectly, but somewhat overestimate the 10~$\mu{\rm m}$ flux. As for the unclumped
models (\citealt{mokiem05}), the absorption trough cannot be fitted well by
models with $\beta \le 0.9$ (nebular emission?), though the wings are nicely
matched. If we assume, on the other hand, that the trough is refilled by the
wind alone, the complete profile can be reproduced with $\beta \approx 1$
and $\fin=8,\fmid=10{\ldots}12$, respectively. From the shape of the trough
we derive $\rin \la 1.1$, otherwise it becomes too narrow or too deep.

The alternative solution with helium recombined {\it in the radio region}
yields a considerable larger mass-loss rate, \Mdot = 4\Mdu, since we have
adopted a large helium content, \YHe=0.21 (compare with the case of
$\zeta$~Pup; see Sect.~\ref{proto}). All clumping properties scale
accordingly, and the best solution (for $\beta$=0.9) is obtained with
$\fin=5$ and $\fmid=4{\ldots}6$. Since the 10~$\mu{\rm m}$ flux indicates
that helium is not completely ionized, even in the outermost IR photosphere
(otherwise it would lie somewhat higher), we prefer the recombined model for
our final solution (see Table~\ref{table_clf}). In the corresponding fit
diagram, we have indicated both possibilities though (solid: recombined;
dotted: ionized).

\paragraph{HD\,190429A.} For this object, there are two measurements at 3.5
cm which are considerably different, namely 200 $\mu$J (our observations)
and 280 $\mu$J from \citet{scu98}. As is obvious from the fit diagram, the 6
cm flux (our measurement) is consistent with the unpublished 3.5 cm value
provided by Scuderi et al., whereas it lies too high with respect to our 3.5
cm measurements. Thus, either the star is strongly variable, or a
non-thermal emitter, or the errors estimated for our observations are too
optimistic. Note that the 0.7 cm measurement (upper limit) is consistent
with our 3.5 cm flux. A ``wrong'' assumption concerning the He recombination
cannot explain this dilemma: if the ionization degree was higher than
predicted, the 0.7 cm flux would be most affected and would lie at a level
higher than actually observed.

On the assumption that we see thermal emission and that the discrepancy is
due to measurement problems, the maximum mass-loss rate is constrained to
lie between 7.5 (dotted) and 9.5\Mdu\ (solid), and both limits have been
indicated in Table~\ref{table_clf}. By adjustment of the clumping factors,
we obtain a perfect fit for \Ha. If the 0.7 cm flux is not much lower than
its upper limit, $\fout$ must be lower than, or equal to, 2. The only other
discrepancy found for this object concerns the 4.63~$\mu$m measurement from
\citet{castor83}, which cannot be matched by any of our models.

\paragraph{HD\,15570} can be fitted without any problems, and the only
complication arises because of the large error bars attributed to the {\sc
scuba} fluxes. Since, for the corresponding wavelength, it is not completely
clear whether He is already recombined or not, we have investigated both
possibilities. In the recombined case (which is consistent with our
predictions: recombination at 6.3 \Rstar, 1.3mm radiation becoming optically
thick at 9.5 \Rstar), the wind must be significantly clumped in region 4
($\fout = 5{\ldots}20$); larger values can be excluded from the \Ha\ wings.
If, on the other hand, the wind is not recombined in the 1.3\,mm-forming
region, a value of $\fout=1$ is still consistent with the limit of the {\sc
scuba} data. The clumping in region 3 ($\fmid = 6$) is well-constrained from
the \Ha\ line wings, though a lower value, $\fmid = 4$, results when we
force the 10~$\mu$m flux to be matched. In the latter case then, \Ha\ becomes
a bit too narrow.

Note that the two measurements at 3.5 cm almost overlap (but not completely,
indicating a certain variability), and we have forced our solution to comply
with their average value. In the fit diagram, we have plotted three
solutions which are consistent with the error bars for the {\sc scuba}
measurements: $\fout = 5$ (dotted), $\fout = 13$ (solid) and $\fout=20$
(dashed). To find even closer constraints on the outer wind clumping
requires lower error bars. Additional far-IR observations (though being
important as consistency checks) will not help to improve this uncertainty,
since the far-IR is insensitive to any reasonable variation of $\fout$ for
this object.

\paragraph{HD\,66811} has been already discussed in some detail; see 
Sect.~\ref{proto}.

\paragraph{HD\,14947.} The 3.5 cm flux is well determined (with some
variability), whereas only upper limits are available at 2 and 6 cm. The
resulting mass-loss rate is \Mdot\ = 8{\ldots}12~\Mdu, and \Ha\ can be
perfectly fitted, with rather low clumping factors in the lower wind. In the
fit diagram and Table~\ref{table_clf}, we have indicated the intermediate
solution with \Mdot\ = 10\Mdu\ and clumping factors $\fin$ = 3.1 and $\fmid$
= 2.5. 

\paragraph{Cyg\,OB2\#11} has similar clumping properties to HD\,14947, and
the maximum mass-loss rate can be derived to within small errors: \Mdot\ =
5\,$\pm$\,0.5 \Mdu. From \Ha, the potentially unclumped region must be
located within $\rin \la 1.2$. From the line wings, $\fmid$ is somewhat
larger than $\fin$, and $\fout$ might be tightly constrained if far-IR
observations were available.  Problems for this object concern the blue side
of the \Ha\ emission being predicted as too narrow, and the 10.9~$\mu$m flux
(\citealt{leitherer82}), which cannot be matched by any of our models.

\paragraph{HD\,210839.} Though the error bars for the {\sc scuba} fluxes are
smaller than for HD\,15570, two different, barely overlapping fluxes have been
measured, which might introduce a twofold solution for region 4, though the
maximum mass-loss rate is well defined. 

A first solution (solid) can be derived for the lower {\sc scuba}
measurement, with constant clumping in the lower wind, $\fin$ = $\fmid$ =
6.5 until $\rout$ = 10, and no clumping in the outer part. The upper {\sc
scuba} measurement can be fitted by additional clumping in region 4, with
$\fout$ = 5{\ldots}20, but in this case the 0.7 cm flux appears as too
large. 

A second, slightly better solution (which is indicated in
Table~\ref{table_clf}) can be found if one assumes constant clumping (again
with $\fcl = 6.5$) until $r < 4$, and a larger clumping factor of $\fcl =
10$ until $r < 15$. With $\fout$ = 1, the lower 1.3\,mm flux is matched
(dotted), whereas with $\fout$ = 8 the upper one can be fitted (dashed). As
before, however, the 0.7 cm flux is then predicted as too large. For all
solutions, \Ha\ is perfectly reproduced, and a value of $\rin \la 1.2$ can
be constrained from its trough.

\paragraph{HD\,192639.} Only one radio measurement is available, and only as
an upper limit. Adopting this value and assuming thermal emission, the
maximum mass-loss rate can be restricted to \Mdot\ $\la$ 3\Mdu, with
constant clumping factors, $\fin = \fmid = 3.5$, in the lower wind, and
$\rin \la 1.1$. For all our simulations, the observed 4.63~$\mu$m flux
(taken from \citealt{castor83}) is smaller than synthesized, though better
reproduced than for HD\,190420A, and independent of the ionization
equilibrium for helium.

\paragraph{HD\,30614} is perfectly matched, both in the radio and in \Ha,
with a moderate degree of clumping in the inner and intermediate wind.

\subsection{Objects with ``intermediate'' \Ha\ profile type}

\paragraph{Cyg\,OB2\#8A} is a confirmed non-thermal radio emitter
(\citealt{bieging89}). In order to obtain at least an estimate, as low a
maximum mass-loss rate as possible has been adopted (from the 2 cm flux),
although this might still be even smaller, of course. With $\beta$ = 0.74
(taken from the optical analysis using homogeneous models,
\citealt{mokiem05}), the wings of \Ha\ are fitted best, whereas the
absorption becomes too deep. A value of $\beta$ = 0.85 (dotted) improves the
trough, but the emission then becomes too large. The 20~$\mu$m flux
indicates that our prediction for the recombination radius of helium might
be erroneous, and a completely recombined model (which at these temperatures
is rather improbable) can indeed fit this measurement. Only low clumping
factors are required to fit \Ha, though higher values would be necessary if
the mass-loss rate were lower. Note that with $\fmid=2.0$ the wings of \Ha\
are nicely matched, but the 10~$\mu$m flux is slightly overestimated.
With $\fmid = 1$, on the other hand, the latter problem can be cured, at
the expense of \Ha.

\paragraph{Cyg\,OB2\#10} can be fitted accounting for weak clumping in the
lower wind ($\fin=1.4$, $\fmid=1.8$) if $\beta$ is left at its nominal value
of 1.05. $\rin$ must be $\la 1.2$, and clumping effects are seen only in the
inner wind. The observed 10~$\mu$m flux is larger than predicted, which
cannot be corrected for by a non-recombined wind, as the temperature is too
low for such a scenario.

\subsection{Objects with \Ha\ in absorption}

\begin{figure*}
\begin{center}
\begin{minipage}{7.8cm}
\resizebox{\hsize}{!}
   {\includegraphics{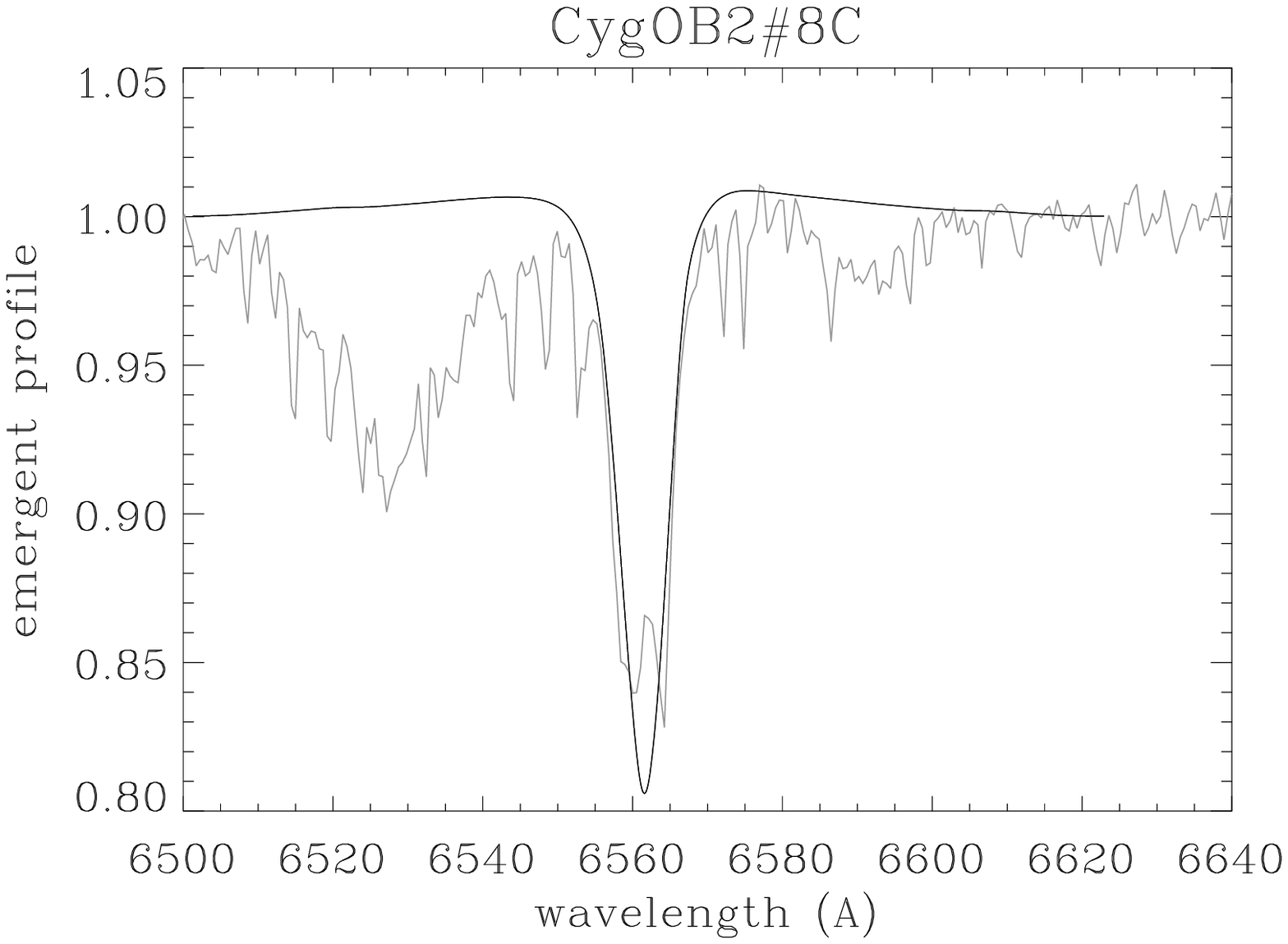}}
\end{minipage}
\begin{minipage}{7.8cm}
   \resizebox{\hsize}{!}
   {\includegraphics{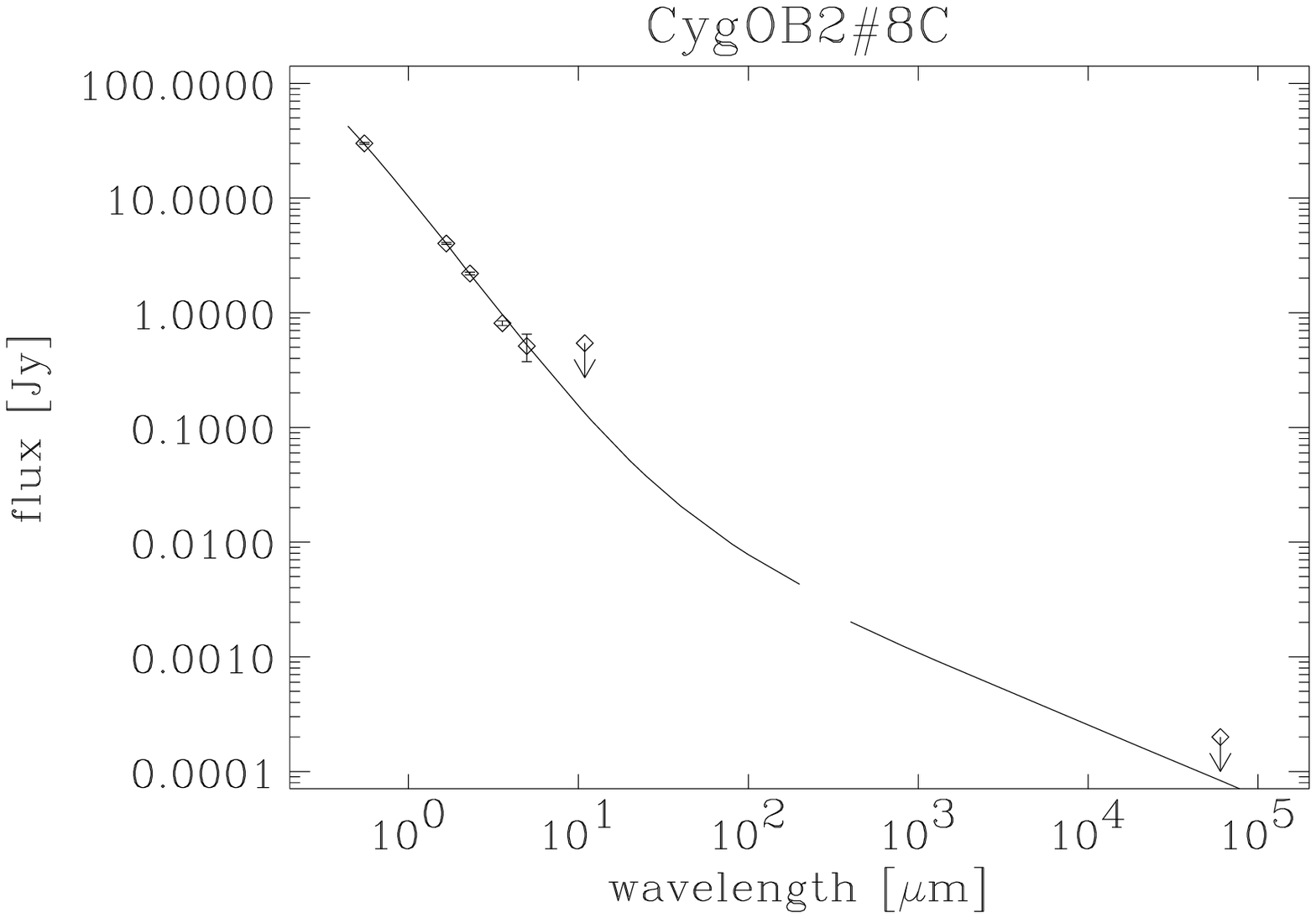}}
\end{minipage}

\begin{minipage}{7.8cm}
   \resizebox{\hsize}{!}
   {\includegraphics{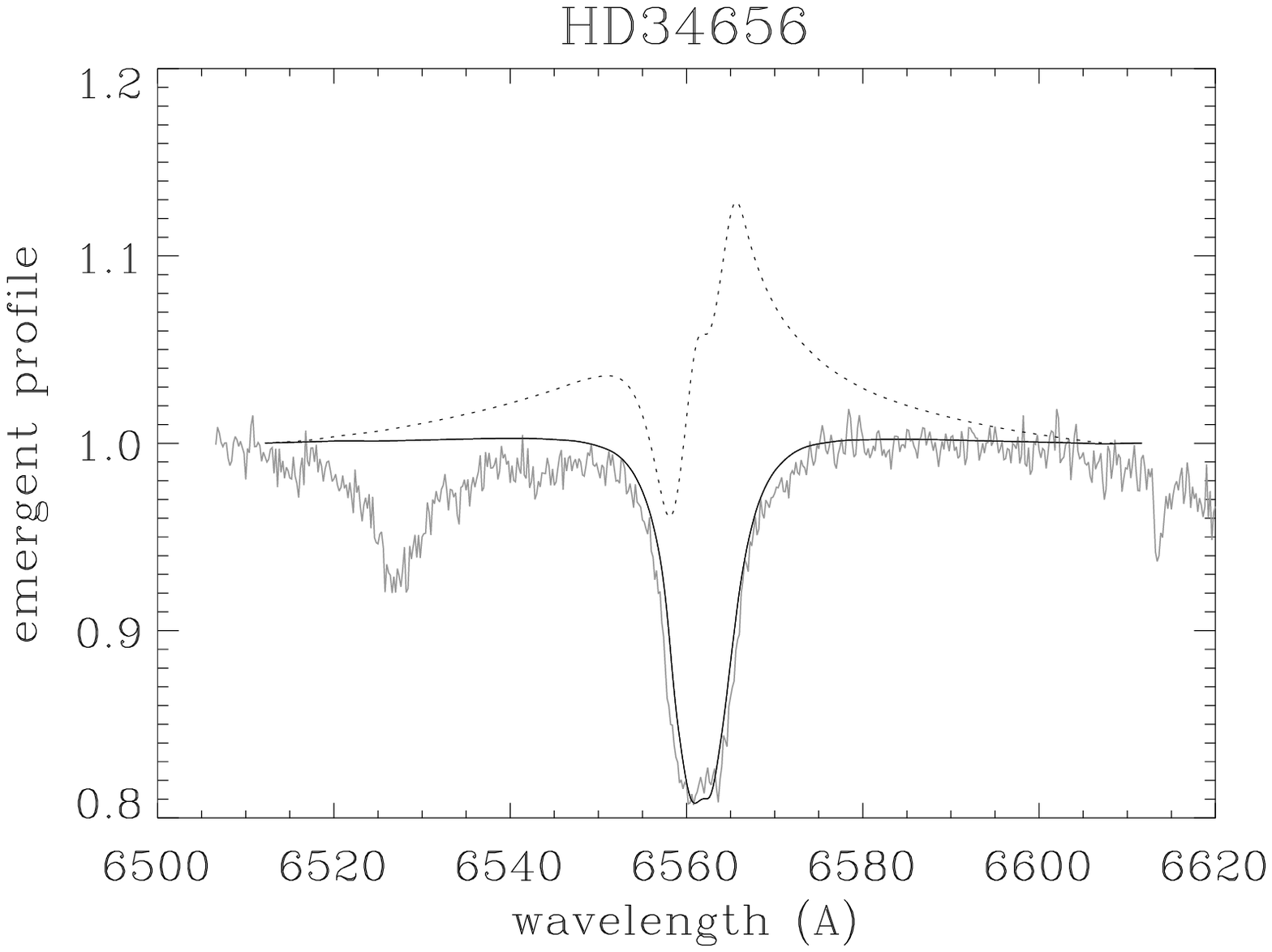}}
\end{minipage}
\begin{minipage}{7.8cm}
   \resizebox{\hsize}{!}
   {\includegraphics{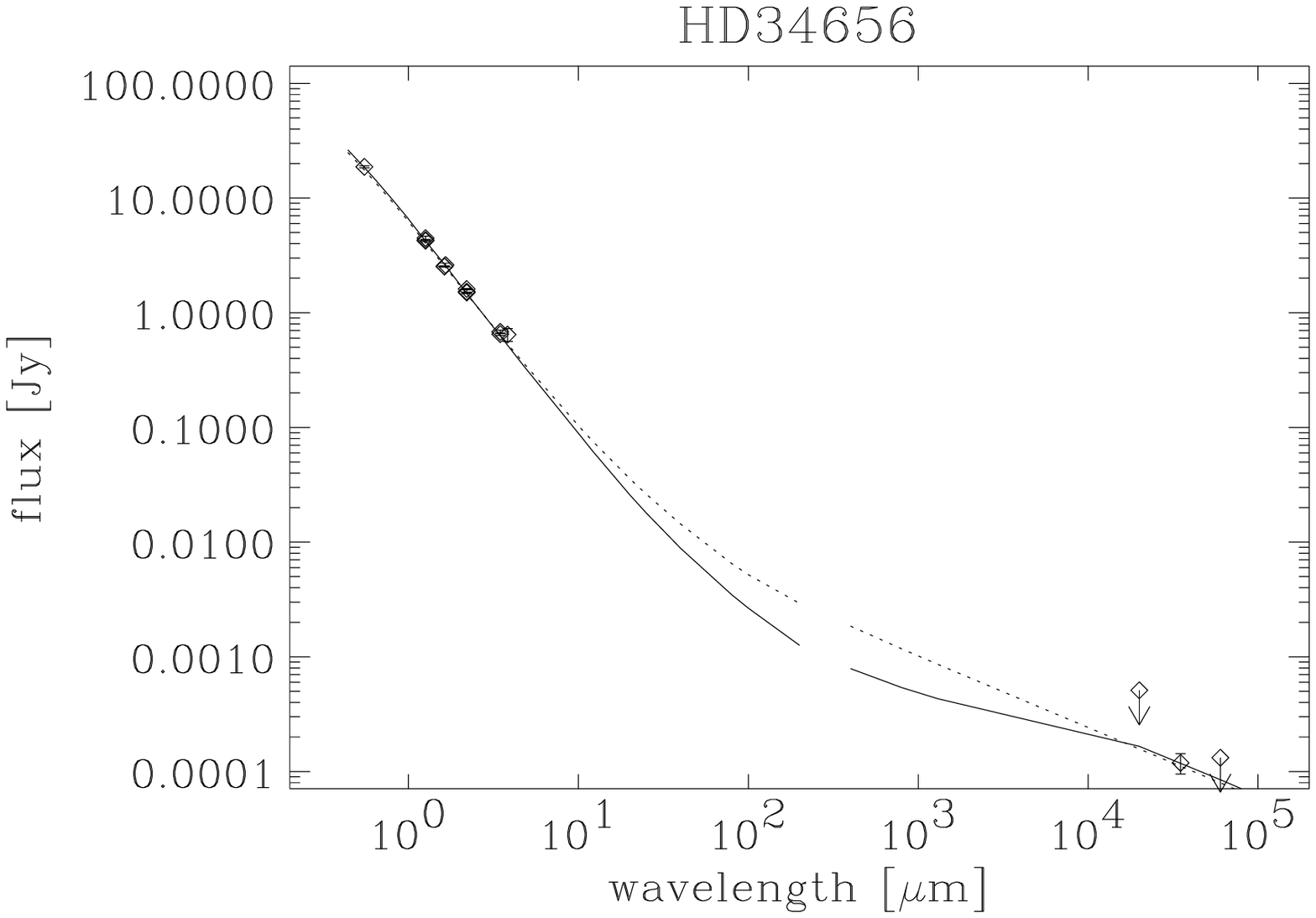}}
\end{minipage}

\begin{minipage}{7.8cm}
   \resizebox{\hsize}{!}
   {\includegraphics{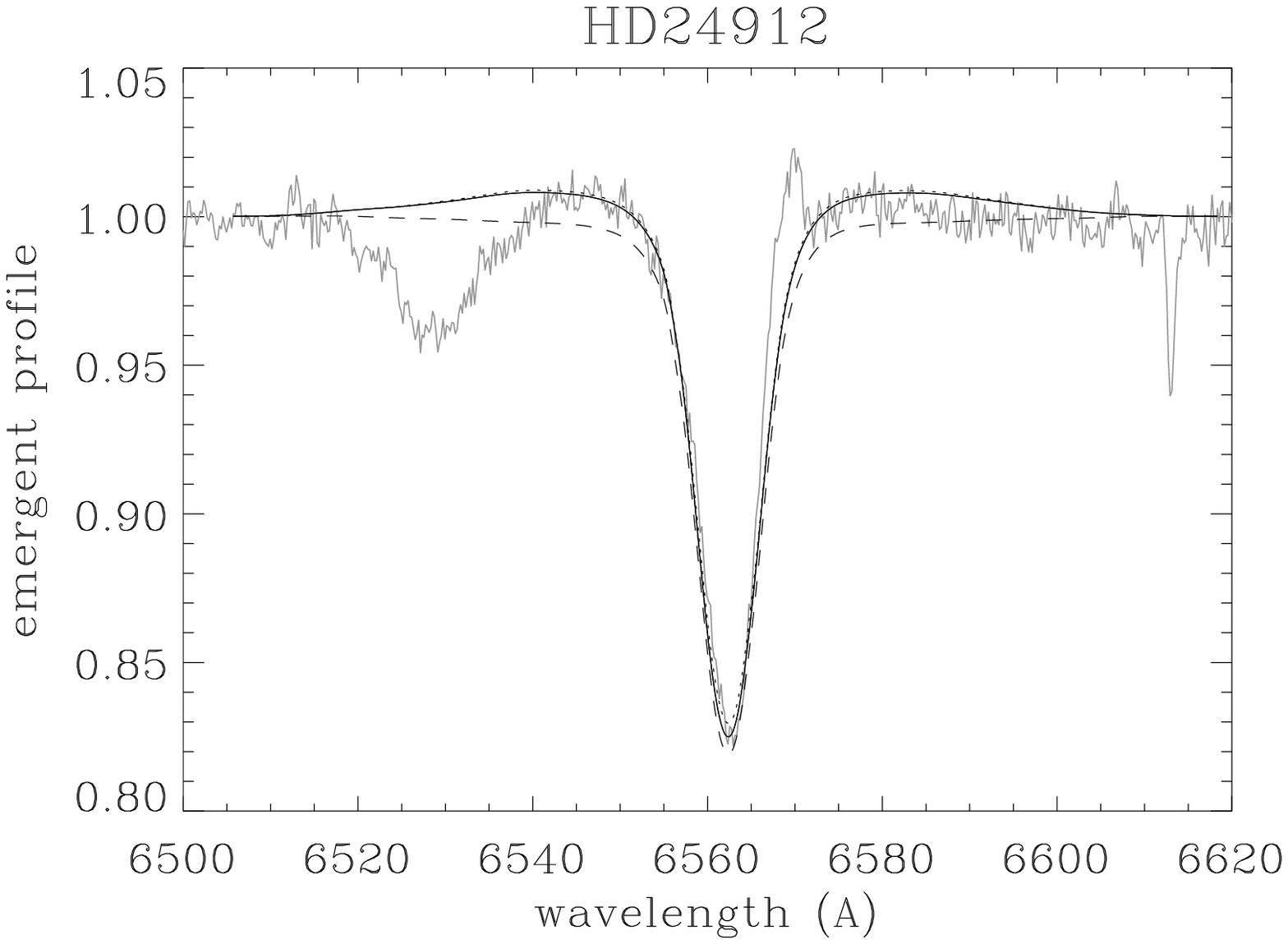}}
\end{minipage}
\begin{minipage}{7.8cm}
   \resizebox{\hsize}{!}
   {\includegraphics{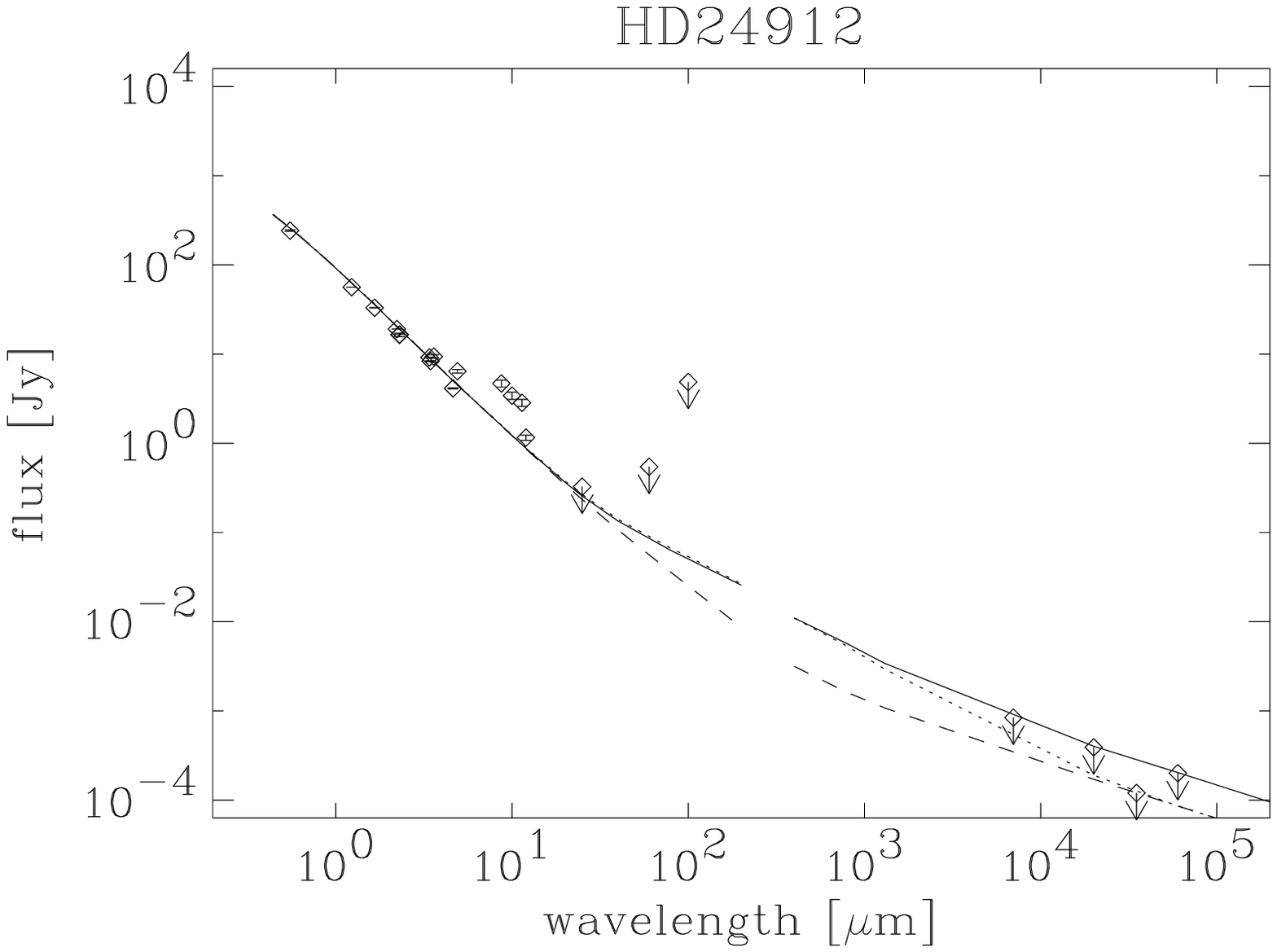}}
\end{minipage}

\begin{minipage}{7.8cm}
   \resizebox{\hsize}{!}
   {\includegraphics{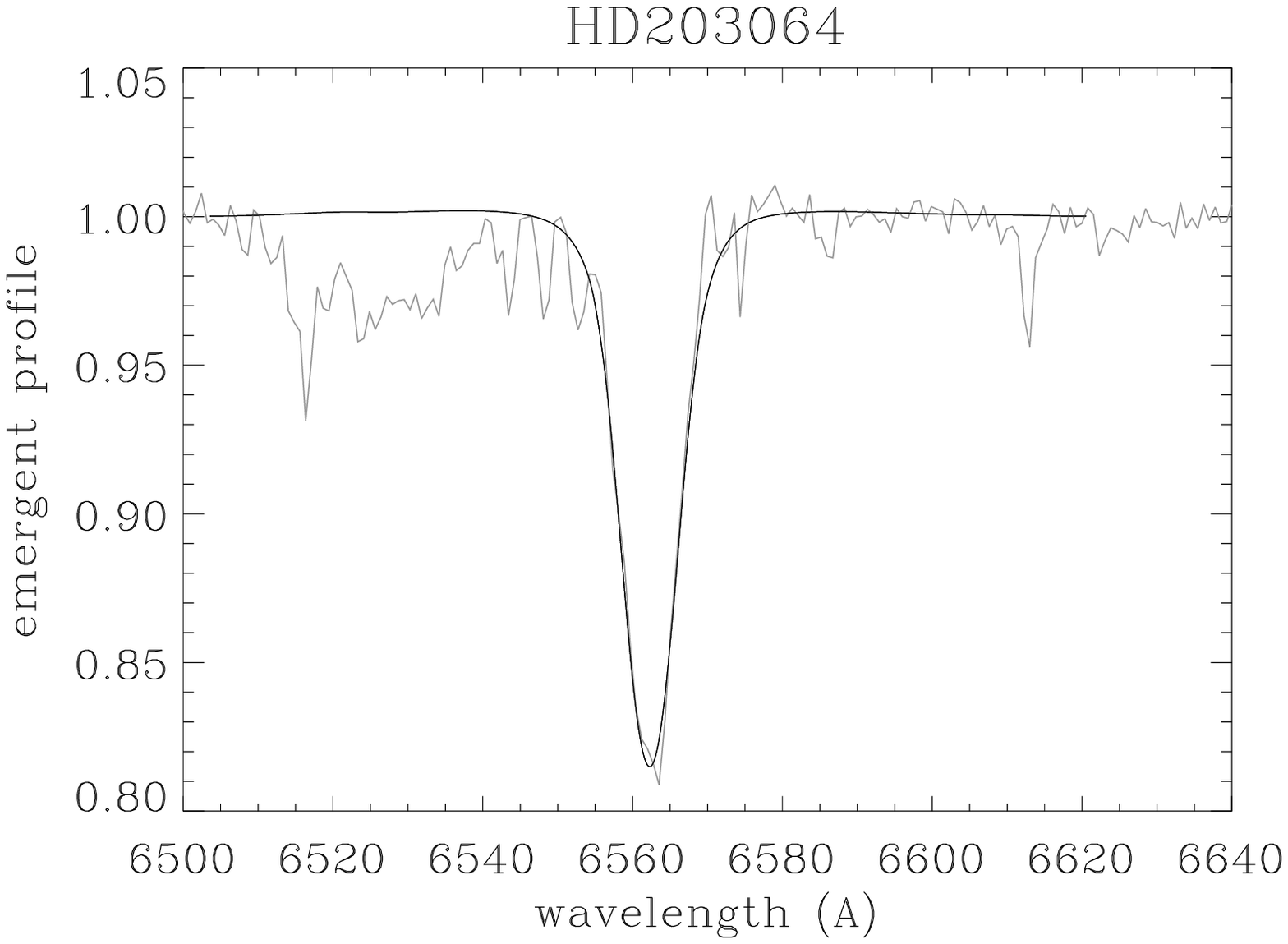}}
\end{minipage}
\begin{minipage}{7.8cm}
   \resizebox{\hsize}{!}
   {\includegraphics{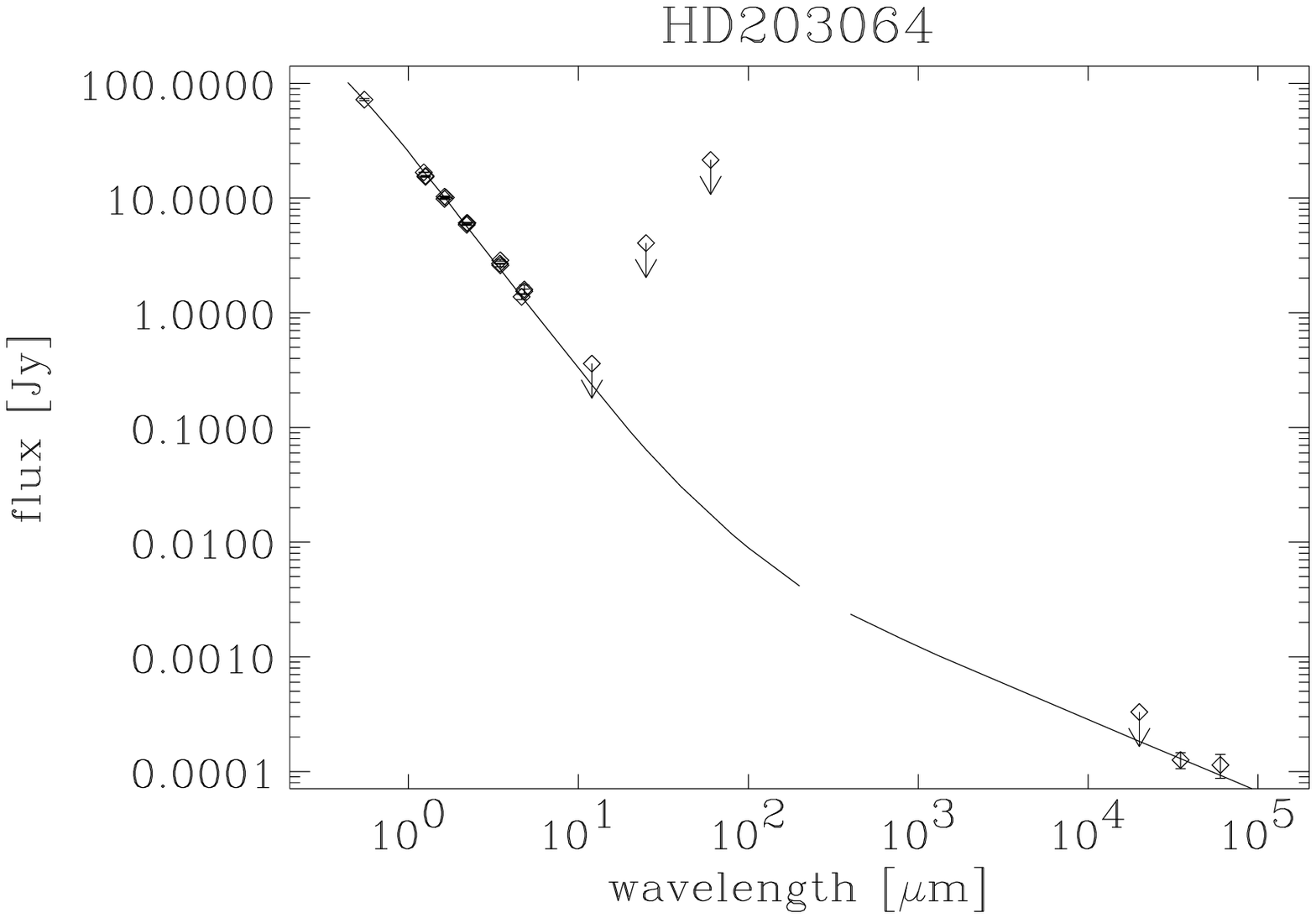}}
\end{minipage}
\end{center}
\caption{As Fig.~\ref{fits_haem_1}, but for objects with \Ha\ in absorption.}
\label{fits_haabs_1}
\end{figure*}

\begin{figure*}
\begin{center}
\begin{minipage}{7.8cm}
   \resizebox{\hsize}{!}
   {\includegraphics{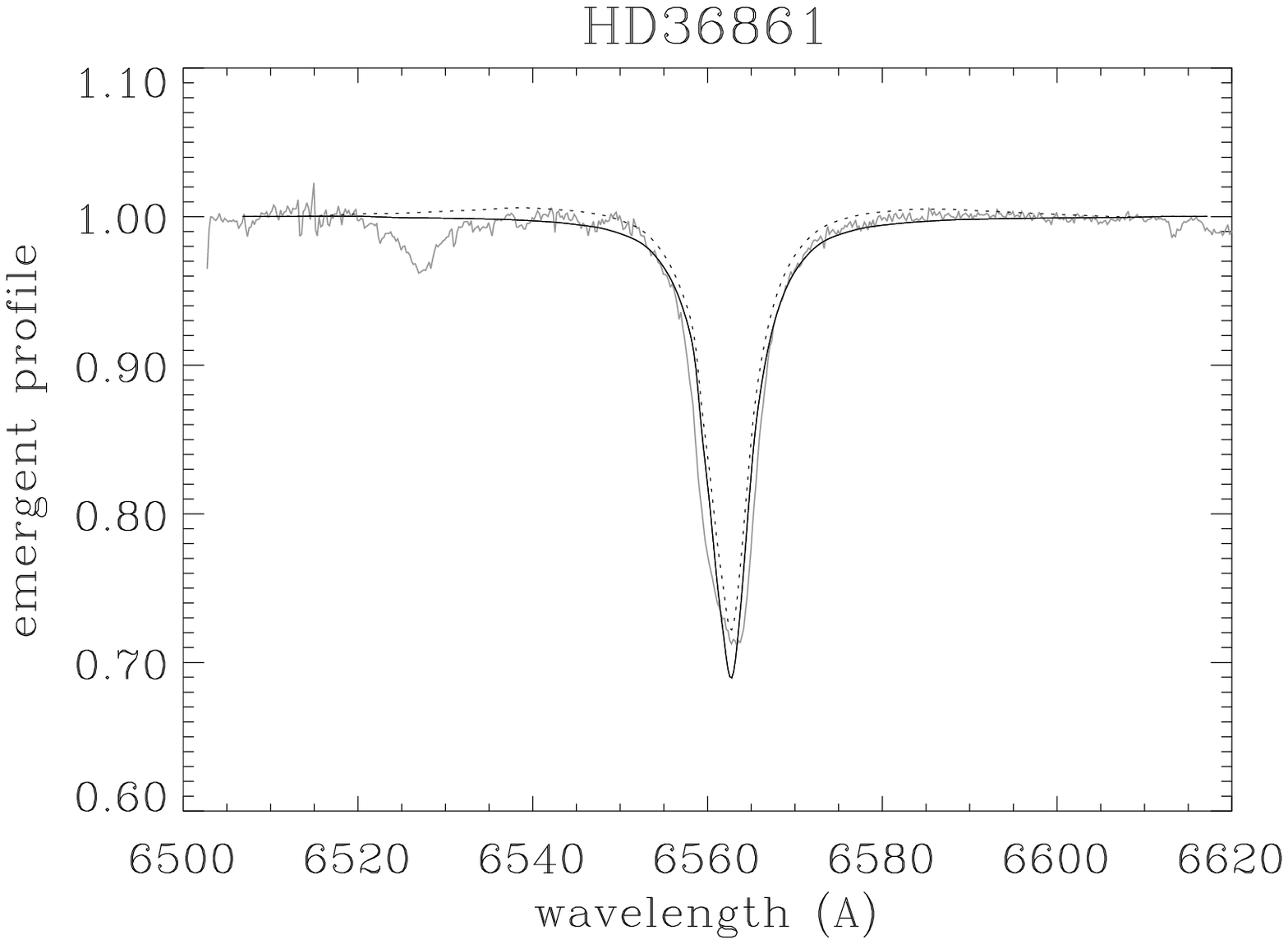}}
\end{minipage}
\begin{minipage}{7.8cm}
   \resizebox{\hsize}{!}
   {\includegraphics{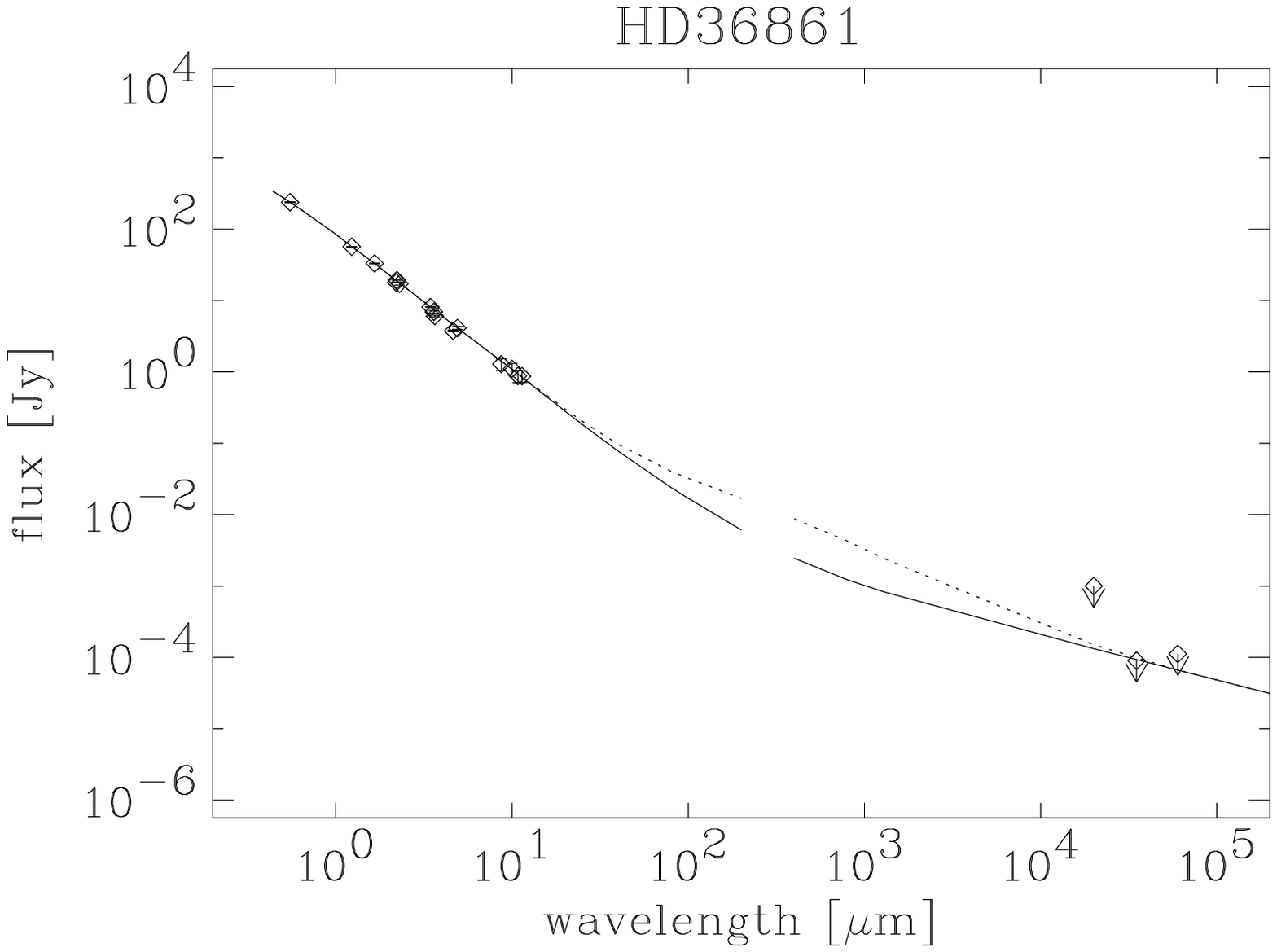}}
\end{minipage}

\begin{minipage}{7.8cm}
   \resizebox{\hsize}{!}
   {\includegraphics{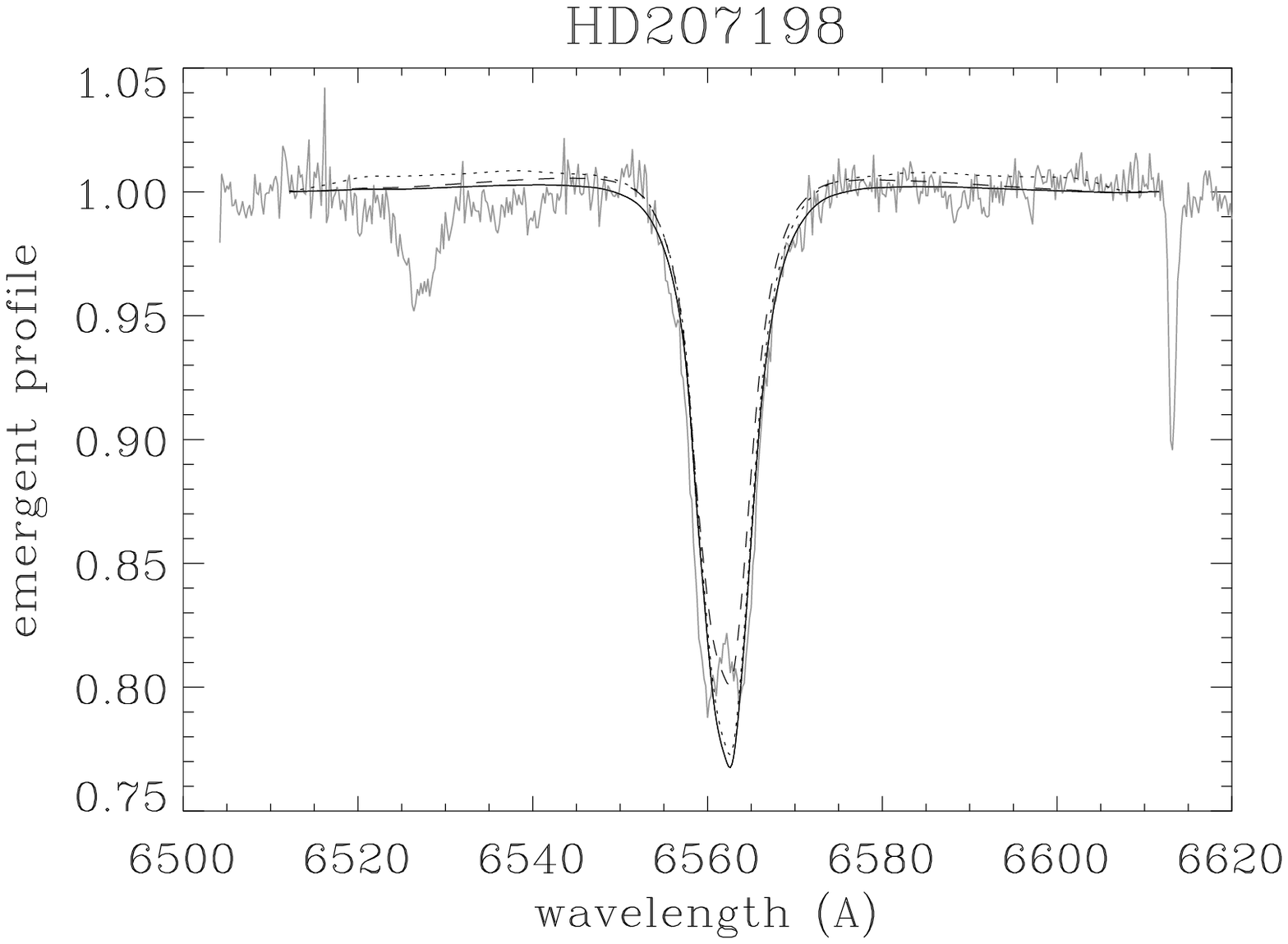}}
\end{minipage}
\begin{minipage}{7.8cm}
   \resizebox{\hsize}{!}
   {\includegraphics{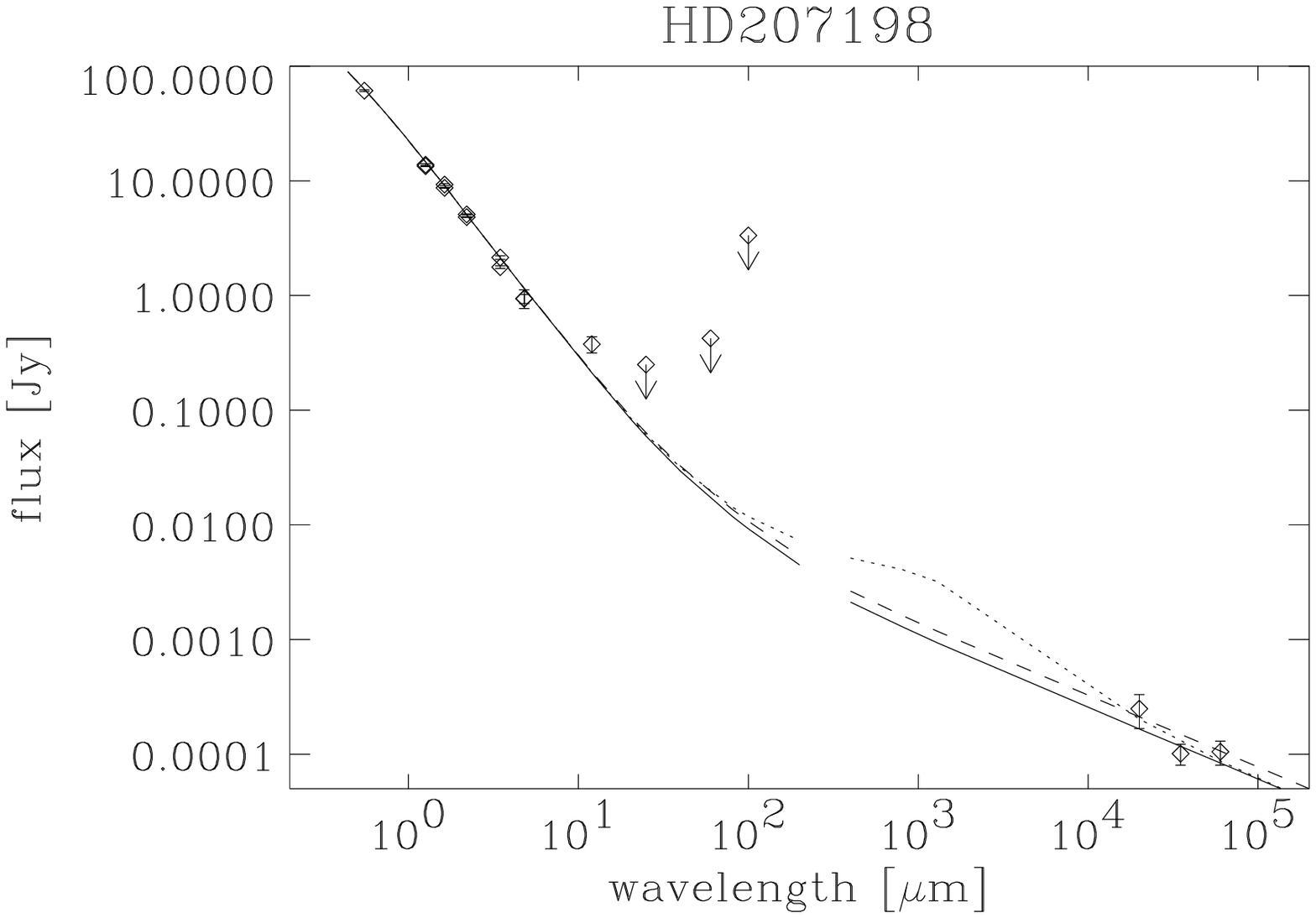}}
\end{minipage}

\begin{minipage}{7.8cm}
   \resizebox{\hsize}{!}
   {\includegraphics{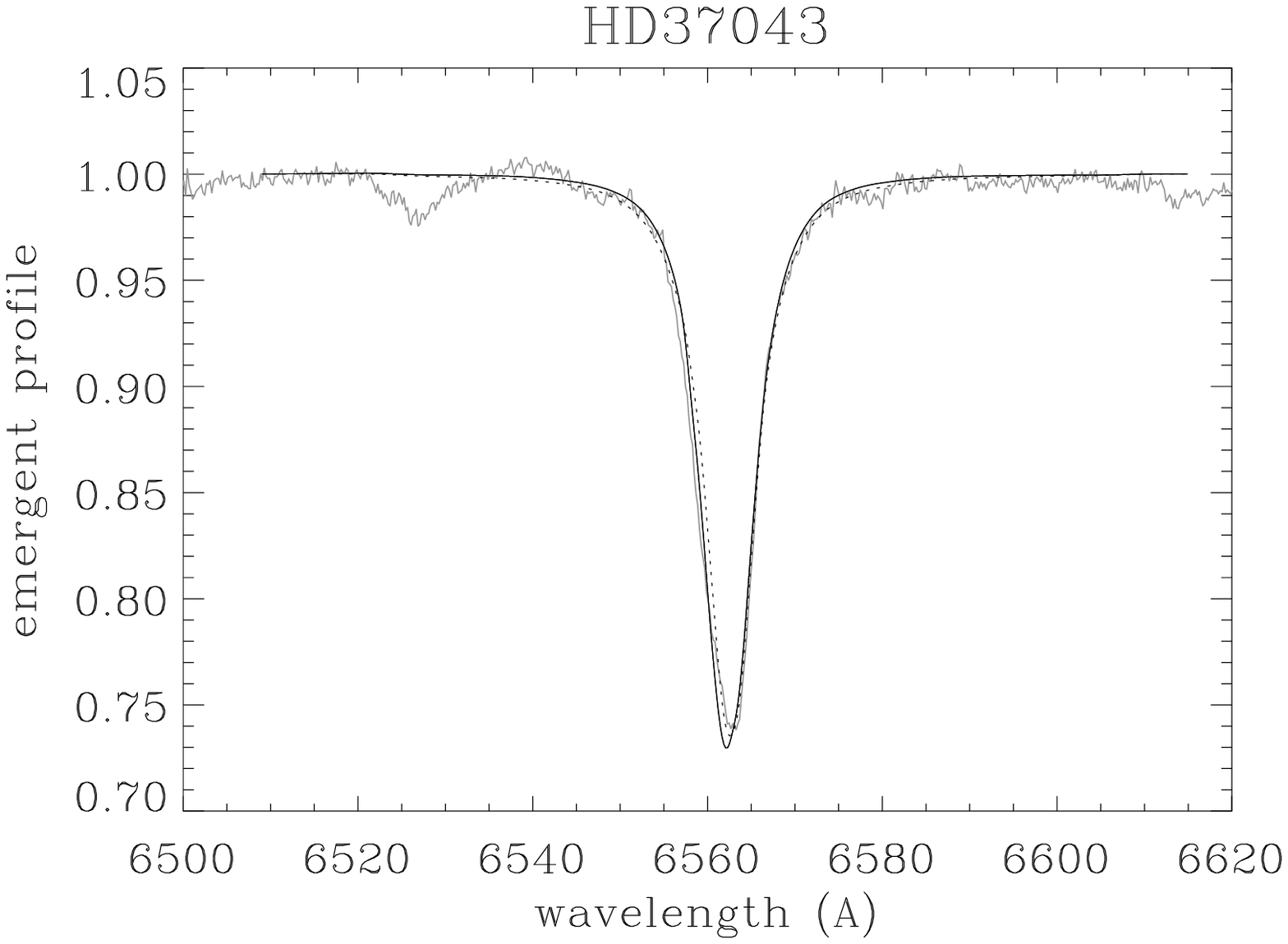}}
\end{minipage}
\begin{minipage}{7.8cm}
   \resizebox{\hsize}{!}
   {\includegraphics{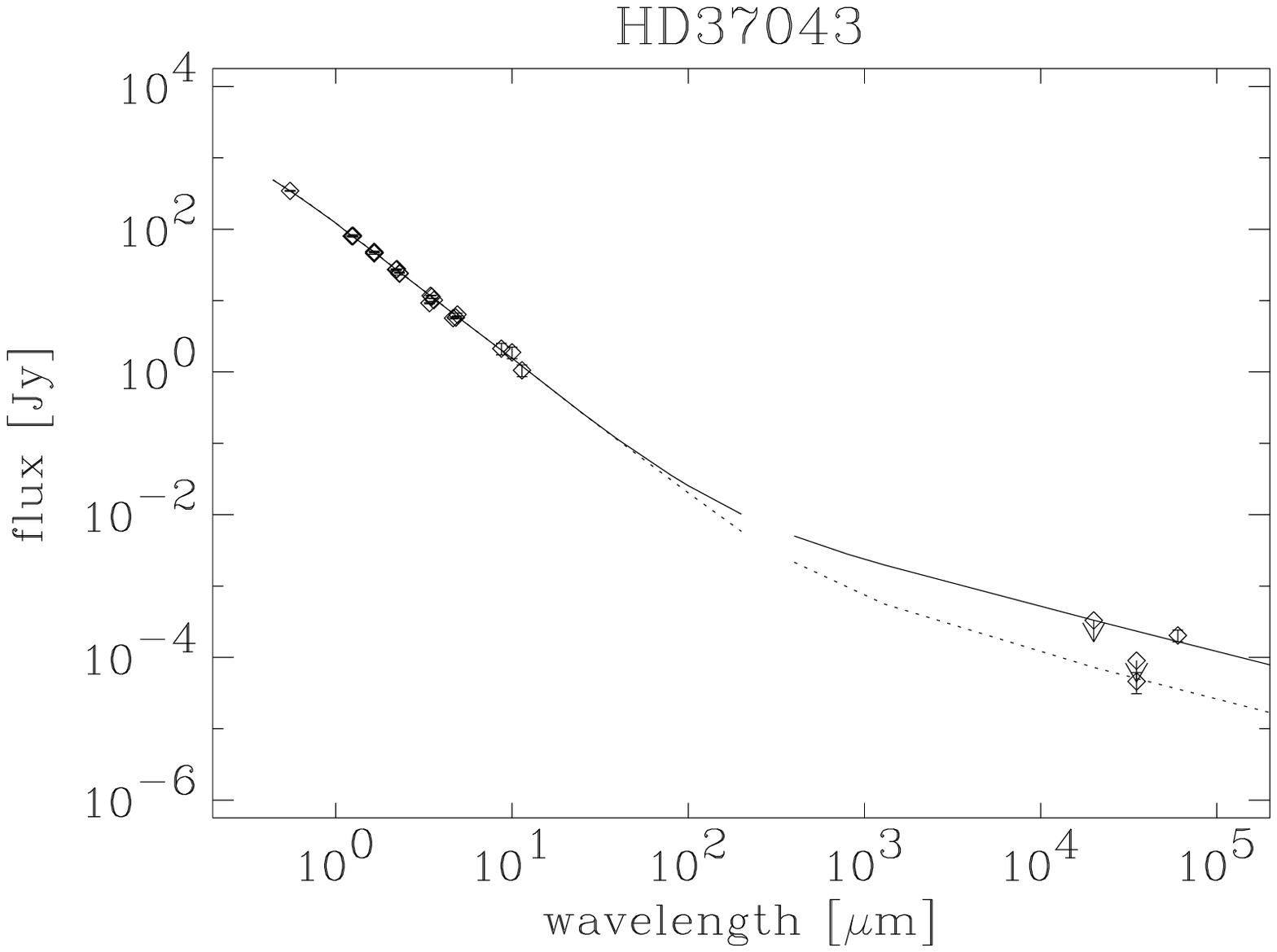}}
\end{minipage}
\end{center}
\caption{As Fig.~\ref{fits_haabs_1}.}
\label{fits_haabs_2}
\end{figure*}

For all objects with \Ha\ in absorption, we have used $\rout=10$, since due
to the lower wind density, the IR and radio emission is formed at smaller
distances from the star (cf. Sect.~\ref{proto}). E.g., for HD\,36861, the
wind becomes optically thick at 2~cm only for $r \la 10 \Rstare$.

\paragraph{Cyg\,OB2\#8C} remains rather unconstrained by our analysis, since
only one upper limit in the radio range is available (at 6 cm), and this
upper limit yields a mass-loss rate larger than the one derived from \Ha\
(\Mdot = 3.5\Mdu\ for $\beta =1$). Thus, the largest possible mass-loss rate
has been adopted from this value, and the only definite statement concerns
$\fmid$ being similar to $\fin$.

\paragraph{HD\,34656} is the only object within our sample where the radio
mass-loss rate (if thermal emission) is definitely larger than the \Ha\
mass-loss rate.\footnote{For HD\,209975, both mass-loss rates overlap within
the errors.} Unfortunately, only one measurement (at 3.5 cm) provides a hard
number, whereas the 2 and 6 cm measurements yield upper limits only. Thus,
non-thermal emission cannot be excluded, at least to some extent, if one
compares the 3.5 and 6 cm fluxes (Fig.~\ref{fits_haabs_1}). Besides being a
non-thermal emitter, there are two other possibilities: either the 3.5 cm
measurements are somewhat corrupted (i.e., can be regarded as upper limits
only), or the outer wind is more heavily clumped than the inner one. In the
latter case, the maximum mass-loss rate results from \Ha\ instead of from
the radio, and corresponds to 3\Mdu\ for $\beta = 1$, which is the lowest
possible value such that $\beta$ remains consistent with our data (wings of
\Ha). Note that a value of $\beta = 1.1$ and \Mdot\ = 2.6\Mdu, as derived in
Paper~I, gives a slightly better solution. Since the (thermal) radio
mass-loss rate corresponds to a value of 7\Mdu\ (and would result in an
emission profile for \Ha; cf. the dotted solution), clumping factors of
$\ffar = (\Mdote_{\rm radio}/\Mdote_{\rm H\alpha})^2 \approx 6$ are
necessary to obtain a simultaneous fit, with $\fmid$ = $\fin$ being
well-constrained.

\paragraph{HD\,24912.} All radio fluxes measured for $\xi$~Per are upper
limits. Our fit diagram shows that the 0.7, 2 and 6 cm limits, if taken at
face value, are consistent with thermal emission and a maximum mass-loss
rate of 2.3\Mdu, which is very close to the value provided by 
\citet{repo04}, using unclumped models. When accounting additionally for the
3.5 cm flux, one derives a maximum mass-loss rate of 1.2\Mdu. In the
following we will consider both possibilities. 

The solution with larger \Mdot\ requires weak clumping in the lowermost wind
($\fin=2.1$), and additional clumping in region 3 ($\fmid$ = 5), if the
small emission humps on the red and blue side of the \Ha\ absorption are due
to clumping and not to other processes (see below). The maximum value of
$\fout$ is restricted by $\fout \la 2$, otherwise the radio band becomes
affected and the maximum mass-loss rate must be decreased.

The lower \Mdot\ solution (which is consistent with {\it all} radio
measurements) requires considerable clumping in the lower wind. Assuming a
``standard value'' of $\beta= 0.9$ (which has been used for most of the
following objects as well, but see Sect.~\ref{errors}), $\fin=8$ and $\fmid
\la 20..25$, the humps can be explained by clumping. Furthermore, the
unclumped region (if any) can be constrained by $\rin \la 1.1$. For this
solution, $\fout \la 3$, otherwise \Mdot\ is even lower. In our fit
diagrams, we have plotted the high \Mdot\ solution (solid), the low \Mdot\
solution (dotted) and the low \Mdot\ solution with $\fmid=1$, which does not
fit the emission humps in \Ha. At least this uncertainty might be resolved
if future far-IR measurements become available.

Let us comment finally on the strong excess measured in the mid-IR, between
8.7 to 11.4~$\mu$m (taken from \citealt{gehrz74}), which is in stark 
contrast to the 12 $\mu$m IRAS data from \citet{beichman88}. This
discrepancy (see also Sect.~\ref{summary}) cannot be due to a wrong flux
calibration, since measurements from the same source have been used also for
HD\,30614 and HD\,36861, without any apparent problems. Thus, $\xi$~Per is
either strongly variable in the mid-IR, or the mid-IR excess is due to
another physical process (e.g., co-rotating interaction zones, see
\citealt{deJong01}, or a wind compressed equatorial region). The latter
interpretation in particular is consistent with the red and blue emission
humps observed in \Ha, which have also been seen in \HeII\ 4686
(\citealt{h92}, Fig.~4).

\paragraph{HD\,203064.} For the standard value of $\beta =0.9$, a model with
all clumping factors being unity is consistent with the observations. Values
for $\fmid \ga 2$ can be excluded.

\paragraph{HD\,36861.} All radio measurements provide only upper limits, and
we have indicated a model with a consistent maximum mass-loss rate, \Mdot\ =
0.4\Mdu. For this value and $\beta$ = 0.9, the innermost clumping is weak
again, $\fin$ = 2{\ldots}4, and $\fmid$ must be lower than 20 (from the
wings of \Ha). Solutions with $\fout > 2$ are no longer consistent with the
adopted mass-loss rate. In Fig.~\ref{fits_haabs_2}, we have indicated the
solutions with minimum ($\fin = 2, \fmid =1$, solid) and maximum ($\fin = 4,
\fmid =20$, dotted) clumping. Note that only far-IR or mm observations will
help to disentangle this uncertainty.

\paragraph{HD\,207198} has well-defined radio measurements, and an unclumped
wind with \Mdot\ = 1.0{\ldots}1.2\Mdu\ (for $\beta = 0.9$) matches all
observational constraints ($\fmid \la 2$). The 2 cm flux can be reproduced with
$\fout = 10$ (and $<$15); larger values are excluded by the
\Ha\ wings. Displayed are the solutions for an unclumped wind at \Mdot\ =
1.0\Mdu\ (solid), a wind with additional clumping in the outer region ($\fout
= 10$, dotted) and a homogeneous wind with \Mdot\ = 1.2\Mdu\ (dashed).  

\paragraph{HD\,37043.} By inspection of the measured radio fluxes, this star
is either a non-thermal emitter (SB2!), or the 6 cm flux is erroneous.
At 3.5 cm, we have two measurements which are consistent. To obtain more
conclusive results, one needs to re-observe this star in the radio range.

Nevertheless, we present two solutions: an upper one discarding the 3.5 cm
data and being consistent with the upper limit at 2 cm, and a lower, more
likely one (which would also be an upper limit if the object were
a non-thermal emitter), discarding the 6 cm measurement.

In the first case, \Mdot\ = 0.8\Mdu, and a smooth wind is consistent with
the observations. In the second case, \Mdot\ = 0.25\Mdu, and the wind is
strongly clumped at least in the lowermost wind, with $\fin = 12$ for $\rin
= 1{\ldots}1.05$ to $\rmid \la 1.3$. Due to the low density, clumping in
other regions has a very low impact on the model fluxes, and we can exclude
only values $\fmid > 20$ and $\fout > 10$ (otherwise the maximum mass-loss
rate must be lower). Plotted are the ``smooth'' solution with the upper value
for \Mdot\ (solid), and the lower \Mdot\ solution, which is strongly clumped
(dotted).

\paragraph{HD\,209975} has already been discussed in Sect.~\ref{proto}.
\end{document}